\documentclass{aa}

\usepackage{txfonts}
\usepackage{natbib}
\usepackage{graphics}
\usepackage{array}

\def\kmsmpc{~km~s$^{-1}$~Mpc$^{-1}$}

\def\ergs{ergs~s$^{-1}$}
\def\ergscm{ergs~s$^{-1}$ cm$^{-2}$}
\def\ergshz{ergs~s$^{-1}$ Hz$^{-1}$}
\def\msunyr{M$_{\odot}$~yr$^{-1}$}
\def\halpha{\ifmmode {\rm H{\alpha}} \else $\rm H{\alpha}$\fi}
\def\hbeta{\ifmmode {\rm H{\beta}} \else $\rm H{\beta}$\fi}
\def\o2{[O\,{\sc ii}]}
\def\oii{[O\,{\sc ii}] $\lambda$3727}

\def\oiii{[O\,{\sc iii}] $\lambda\lambda$4959,5007}

\def\ntwo{[N\,{\sc ii}]}
\def\nii{[N\,{\sc ii}] $\lambda$6584}

\def\niia{[N\,{\sc ii}] $\lambda$6548}

\def\stwo{[S\,{\sc ii}]}
\def\sii{[S\,{\sc ii}] $\lambda\lambda$6717,6731}
\newcommand{\msun}{\,{\rm M_\odot}}
\newcommand{\kms}{~$km~s^{-1}$}
\newcommand{\vf}{velocity field}
\newcommand{\vfs}{velocity fields}
\newcommand{\vd}{velocity dispersion}
\newcommand{\vdm}{velocity dispersion map}
\newcommand{\vdms}{velocity dispersion maps}
\newcommand{\Ha}{H$\alpha$}
\def\degr{\hbox{$^\circ$}}

\def\jnl@style{\it}
\def\aaref@jnl#1{{\jnl@style#1}}
\def\aaref@jnl#1{{\jnl@style#1}}

\def\aj{\aaref@jnl{AJ}}                   % Astronomical Journal
\def\araa{\aaref@jnl{ARA\&A}}             % Annual Review of Astron and Astrophys
\def\apj{\aaref@jnl{ApJ}}                 % Astrophysical Journal
\def\apjl{\aaref@jnl{ApJ}}                % Astrophysical Journal, Letters
\def\apjs{\aaref@jnl{ApJS}}               % Astrophysical Journal, Supplement
\def\ao{\aaref@jnl{Appl.~Opt.}}           % Applied Optics
\def\apss{\aaref@jnl{Ap\&SS}}             % Astrophysics and Space Science
\def\aap{\aaref@jnl{A\&A}}                % Astronomy and Astrophysics
\def\aapr{\aaref@jnl{A\&A~Rev.}}          % Astronomy and Astrophysics Reviews
\def\aaps{\aaref@jnl{A\&AS}}              % Astronomy and Astrophysics, Supplement
\def\azh{\aaref@jnl{AZh}}                 % Astronomicheskii Zhurnal
\def\baas{\aaref@jnl{BAAS}}               % Bulletin of the AAS
\def\jrasc{\aaref@jnl{JRASC}}             % Journal of the RAS of Canada
\def\memras{\aaref@jnl{MmRAS}}            % Memoirs of the RAS
\def\mnras{\aaref@jnl{MNRAS}}             % Monthly Notices of the RAS
\def\pra{\aaref@jnl{Phys.~Rev.~A}}        % Physical Review A: General Physics
\def\prb{\aaref@jnl{Phys.~Rev.~B}}        % Physical Review B: Solid State
\def\prc{\aaref@jnl{Phys.~Rev.~C}}        % Physical Review C
\def\prd{\aaref@jnl{Phys.~Rev.~D}}        % Physical Review D
\def\pre{\aaref@jnl{Phys.~Rev.~E}}        % Physical Review E
\def\prl{\aaref@jnl{Phys.~Rev.~Lett.}}    % Physical Review Letters
\def\pasp{\aaref@jnl{PASP}}               % Publications of the ASP
\def\pasj{\aaref@jnl{PASJ}}               % Publications of the ASJ
\def\qjras{\aaref@jnl{QJRAS}}             % Quarterly Journal of the RAS
\def\skytel{\aaref@jnl{S\&T}}             % Sky and Telescope
\def\solphys{\aaref@jnl{Sol.~Phys.}}      % Solar Physics
\def\sovast{\aaref@jnl{Soviet~Ast.}}      % Soviet Astronomy
\def\ssr{\aaref@jnl{Space~Sci.~Rev.}}     % Space Science Reviews
\def\zap{\aaref@jnl{ZAp}}                 % Zeitschrift fuer Astrophysik
\def\nat{\aaref@jnl{Nature}}              % Nature
\def\iaucirc{\aaref@jnl{IAU~Circ.}}       % IAU Cirulars
\def\aplett{\aaref@jnl{Astrophys.~Lett.}} % Astrophysics Letters
\def\apspr{\aaref@jnl{Astrophys.~Space~Phys.~Res.}}
\def\bain{\aaref@jnl{Bull.~Astron.~Inst.~Netherlands}} 
\def\fcp{\aaref@jnl{Fund.~Cosmic~Phys.}}  % Fundamental Cosmic Physics
\def\gca{\aaref@jnl{Geochim.~Cosmochim.~Acta}}   % Geochimica Cosmochimica Acta
\def\grl{\aaref@jnl{Geophys.~Res.~Lett.}} % Geophysics Research Letters
\def\jcp{\aaref@jnl{J.~Chem.~Phys.}}      % Journal of Chemical Physics
\def\jgr{\aaref@jnl{J.~Geophys.~Res.}}    % Journal of Geophysics Research
\def\jqsrt{\aaref@jnl{J.~Quant.~Spec.~Radiat.~Transf.}}
\def\memsai{\aaref@jnl{Mem.~Soc.~Astron.~Italiana}}
\def\nphysa{\aaref@jnl{Nucl.~Phys.~A}}   % Nuclear Physics A
\def\physrep{\aaref@jnl{Phys.~Rep.}}   % Physics Reports
\def\physscr{\aaref@jnl{Phys.~Scr}}   % Physica Scripta
\def\planss{\aaref@jnl{Planet.~Space~Sci.}}   % Planetary Space Science
\def\procspie{\aaref@jnl{Proc.~SPIE}}   % Proceedings of the SPIE

\begin{document}

%TC: proposition de titre, qui sera coherent avec celui du papier de Julien: .....: the mass-metallicity relation at $z\sim 1.4$
%\title{3D KINEMATICS AND TULLY-FISHER RELATIONS FROM $1.3 \leq  z  \leq 1.6$ LATE-TYPE VVDS GALAXIES}
   \title{Integral field spectroscopy with SINFONI of VVDS galaxies\thanks{Based on observations collected at the European Southern Observatory (ESO) Very Large Telescope, Paranal, Chile, as part of the Programs 75.A-0318 and 78.A-0177}}
%   \title{Integral field spectroscopy with SINFONI of VVDS galaxies\thanks{Based on public data collected at the European Southern Observatory (ESO) Very Large Telescope, Paranal, Chile, as part of the Programs 75.A-0318 and 78.A-0177}}
   \subtitle{I. Galaxy dynamics and mass assembly at $1.2 < z < 1.6$}
%   \subtitle{I. Kinematical properties and the Tully-Fisher relation at $1.2 < z < 1.6$}
%
\author{B. Epinat
	\inst{1,2}\thanks{E-mail: benoit.epinat@ast.obs-mip.fr}
	\and T. Contini\inst{1}
	\and O. Le F\`evre\inst{2}
	\and D. Vergani\inst{3}
	\and B. Garilli\inst{4}
	\and P. Amram\inst{2}
	\and J. Queyrel\inst{1}
%	\and M. Kissler-Patig\inst{5}
	\and L. Tasca\inst{4}
	\and L. Tresse\inst{2}
%	\and M. Lemoine-Busserolle\inst{6}
%	\and F. Lamareille\inst{1}
	}
\institute{Laboratoire d'Astrophysique de Toulouse-Tarbes, Universit\'e de Toulouse, CNRS, 14 Avenue \'Edouard Belin, F-31400 Toulouse, France
	\and Laboratoire d'Astrophysique de Marseille, Universit\'e de Provence, CNRS, 38 rue Fr\'ed\'eric Joliot-Curie, F-13388 Marseille Cedex 13, France
	\and INAF-Osservatorio Astronomico di Bologna, via Ranzani 1, I-40127, Bologna, Italy
	\and IASF-INAF, Via Bassini 15, I-20133, Milano, Italy
%	\and ESO, Karl-Schwarzschild-Str.2, D-85748 Garching b. M\"unchen, Germany
%	\and Oxford Physics, University of Oxford, Keble Road, Oxford, OX1\,3RH, UK
	}
%
%\author[M. Lemoine-Busserolle et al. ]{M. Lemoine-Busserolle$^{1}$ \thanks{mrlb@astro.ox.ac.uk}, B. Epinat$^{2,3}$, T. Contini$^{2}$, F. Lamareille$^{2}$, D. Vergani$^{4}$,
%\newauthor B. Garilli$^{5}$, O. Le F\`evre$^{3}$, M. Kissler-Patig$^{6}$, J. Queyrel$^{2}$, P. Amram$^{3}$, L. Tasca$^{2,5},$
%\newauthor L. Tresse$^{3}$ \\
%and A. N. Other$^{2}$\footnotemark[1]\thanks{This file has been amended to highlight the proper use of \LaTeXe\ code with the class file. These changes are for illustrative purposes and do not reflect the original paper by A. V. Raveendran.}\\
%$^{1}$\,Oxford Physics, University of Oxford, Keble Road, Oxford, OX1\,3RH, UK\\
%$^{1}$\,Institute of Astronomy, University of Cambridge, Madingley Road, Cambridge, CB3\,0HA, UK\\
%$^{2}$\,Laboratoire d'Astrophysique de Toulouse-Tarbes, Universit\'e de Toulouse, CNRS, 14 Avenue \'Edouard Belin, F-31400 Toulouse, France\\
%$^{3}$\,Laboratoire d'Astrophysique de Marseille (UMR6110), Universit\'e de Provence, CNRS, 38 rue Fr\'ed\'eric Joliot-Curie,\\
%F-13388 Marseille Cedex 13, France\\
%$^{4}$\,INAF-Osservatorio Astronomico di Bologna, via Ranzani 1, I-40127, Bologna, Italy\\
%$^{5}$\,IASF-INAF, Via Bassini 15, I-20133, Milano, Italy\\
%$^{6}$\,ESO, Karl-Schwarzschild-Str.2, D-85748 Garching b. M\"unchen, Germany}

%\date{Accepted ??. Received ??}

%\pagerange{\pageref{firstpage}--\pageref{lastpage}} \pubyear{2009}

\label{firstpage}

\abstract
{Identifying the main processes of galaxy assembly at high redshifts is still a major issue to understand galaxy formation and evolution at early epochs in the history of the Universe.}
%{The Mass Assembly Survey with SINFONI in VIMOS VLT Deep Survey (MASSIV) is undertaken to map the dynamical evolution of star-forming galaxies in using integral field spectroscopy within a purely I-band selected redshift survey.}
{This work aims to provide a first insight into the dynamics and mass assembly of galaxies at redshifts $1.2 < z < 1.6$, the early epoch just before the sharp decrease of the cosmic star formation rate.}
%{We analyse the dynamics of nine star forming galaxies at $1.2<z<1.6$, the early epoch just before the sharp decrease of the global star formation rate.}
%{We present new results on the dynamics of star forming galaxies observed at the peak of star formation.}
{We use the near-infrared integral field spectrograph SINFONI on the ESO-VLT under 0.65\arcsec~seeing to obtain spatially resolved spectroscopy on nine emission line galaxies with $1.2 \leq z \leq 1.6$
%purely flux selected
from the VIMOS VLT Deep Survey. We derive the velocity fields and velocity dispersions on kpc scales using the H$\alpha$ emission line.}
{Out of the nine star-forming galaxies, we find that galaxies distribute in three groups: two galaxies can be well reproduced by a rotating disk, three systems can be classified as major mergers and four galaxies show disturbed dynamics and high velocity dispersion. We argue that there is evidence for hierarchical mass assembly from major merger, with most massive galaxies with $M>10^{11}\msun$ subject to at least one major merger over a 3 Gyr period as well as for continuous accretion feeding strong star formation.}
%{Out of the nine star-forming galaxies, we identify three major merger systems, two massive rotating disks with dynamical masses higher than $10^{11}\msun$ and four disturbed rotating disks. Star formation rates are high with an average of 55\msunyr. No Seyfert-2 have been found in our sample.}
{These results point towards a galaxy formation and assembly scenario which involves several processes, possibly acting in parallel, with major mergers and continuous gas accretion playing a major role. Well controlled samples representative of the bulk of the galaxy population at this key cosmic time are necessary to make further progress.}

%{We find that galaxies distribute in three equivalently populated groups: galaxies which can be well reproduced by a rotating disk, major mergers and galaxies with disturbed dynamics and high velocity dispersion. These results point towards a galaxy formation and assembly scenario which involves major mergers and several disk collapsing processes.
%% demonstrate that the galaxy formation and assembly scenario is not monolithic but rather proceeds with a number of physical processes involving major mergers, disk collapse 1, disk collapse 2.
%Well controlled samples representative of the bulk of the galaxy population at this  key cosmic time are necessary to make further progress.}
%% We are currently  expanding this sample from a new Large Program.}
%
%%We present pioneer results on a ongoing SINFONI Large program obtain two-dimensional (2D) information on the rest-frame optical emission lines of spectroscopically confirmed late-type selected galaxies with $1.3 \leq  z  \leq 1.6$ from VVDS, using near-infrared integral field spectroscopy with SINFONI.
%

% -- technique: integral field spectroscopy}

%\input{abstract_aa_v4.4.tex}

\maketitle

\section{Introduction}

In the current paradigm, galaxies 
are formed in dark matter haloes which grow along
cosmic time via hierarchical assembly of smaller units. 
While in this picture halo-halo merging is the main physical process driving the
assembly of dark matter haloes, it is unclear 
how this merging process directly affects the build-up of galaxies, and how
other physical processes play a role.

Considerable efforts have been invested to 
understand galaxy evolution in the past two decades.
We now have a global, but as yet incomplete, picture
of the evolution of the parameters describing
the main galaxy population, with e.g. the global star formation
%rate reaching a peak at redshifts 1 to 2.5 (e.g. \citealp{Tresse:2007,Rodighiero:2009,Hopkins:2006}).
rate reaching a peak at redshifts 1 to 2.5 (e.g. \citealp{Tresse:2007,Hopkins:2006}).
A clear result now emerging is that early type galaxies
have experienced a major growth in stellar mass density
from a redshift $z\sim2$ to $z\sim1$, while late-type 
galaxies have been growing their stellar mass more
slowly in this period (e.g. \citealp{Arnouts:2007,Bundy:2005}). It is also claimed
that galaxy evolution proceeds via a downsizing
scenario, whereby the most massive galaxies assemble
their mass first \citep{Cowie:1996}.

However, much remains to be done to understand how
galaxies of different types have been built
along cosmic time. Several key questions are still subject to considerable debate: 
(i) How were galaxies assembled in details?
(ii) When and at what rate did galaxies of different masses form?
(iii) What is the connection between bulge and disk formation?
(iv) What is the link between the various high-z galaxy populations, 
their origin and their subsequent fates?
(v) What is the connection between star formation and the 
AGN phenomenon ?

Several physical processes have been identified to be
contributing to the evolution of galaxies, but their relative
contributions, main epoch of action, and associated 
timescales, remain poorly constrained.
Merging is identified as an important contributor
through direct evidence of on-going events, or indirect 
fossil remnants (e.g. \citealp{Conselice:2003,Lotz:2008}),
and the  merger rate and its evolution is now demonstrated
to be strongly dependent on the luminosity or stellar
mass of the galaxies involved (e.g. \citealp{de-Ravel:2009}).
%mass of the galaxies involved (e.g. \citealp{de-Raveletal:2008}).
In addition, the effect of the competing processes of cooling, angular momentum exchange, feedback from star formation and 
AGNs, or gas accretion, are to be further investigated \citep{Somerville:2008}.

Until recently, the favoured picture was that the star  formation and stellar mass assembly of massive galaxies was a direct consequence of major merging events at early cosmic epochs \citep{Glazebrook:2004}.
However, recent observations of individual
galaxies have revealed that large and massive disks with
strong star formation seem to be already in place at redshifts 
$z\sim2$, without apparent signs for major merging events 
\citep{Forster-Schreiber:2006,Wright:2007,Wright:2009,Law:2007,Law:2009,Genzel:2006,Genzel:2008}. This has
%\citep{Forster-Schreiberetal:2006,Law:2007,Genzeletal:2008}. This has
led to suggest that the majority of star forming galaxies 
is fed by gas via cold flows along streams of the 
cosmic web (including minor merger), 
continuously fueling 
the star formation \citep{Dekel:2009}. 
This scenario is further supported by the tight correlation between stellar mass and star formation rate in high redshift star forming galaxies (e.g. \citealp{Daddi:2007,Noeske:2007, Elbaz:2007}).
Dynamical processes internal to 
galaxies can then drive secular evolution of disks and the
formation of bulges and spheroids.

The detailed knowledge of the kinematics of a galaxy
%'s velocity field 
on kiloparsec scales is needed 
to identify its dynamical state (disk in rotation, spheroid, major marger event or more complex kinematics).
%recognize the organized rotation velocity field of 
%a disk, to identify a spheroid, a major merger event, or 
%more complex kinematics.
The integral field technique enables one to compute accurate total dynamical masses, 
to trace the spatial distribution of stars and gas, 
as well as to evaluate
the contribution of stellar populations including stellar 
initial mass function (IMF), or gas metallicity.
The velocity field of spiral/disk galaxies,
can be used to put important constraints on total masses 
and hence on dark matter halo masses, thought
to be an important driver of galaxy evolution. 

At low redshifts, the knowledge of 2D velocity fields has 
proved to be a powerful kinematical tool to investigate the 
properties of nearby galaxies \citep{Ostlin:2001,Veilleux:2001,Swinbank:2003,Mendes-de-Oliveira:2003,Garrido:2004,Epinat:2008b,Epinat:2008a}. The velocity fields of 
galaxies at intermediate redshift ($0.4<z<0.75$) have been investigated using FLAMES/GIRAFFE at VLT \citep{Flores:2006,
Puech:2006,Yang:2008,Neichel:2008,Puech:2008}.
%Puech:2006,Yangetal:2008,Neicheletal:2008,Puechetal:2008}.\\
In their sample of 65 galaxies, \citet{Yang:2008} 
found about 32\% of relaxed rotating disks. These rotating disks 
produce a Tully-Fisher relation \citep{Tully:1977} which has apparently not evolved in slope and scatter since z=0.6. They however detect an evolution of the K-band Tully-Fisher relation zero point that, if interpreted as an evolution of the K-band luminosity, would imply a brightening of 0.66$\pm$0.14 mag between $z\sim0.6$ and $z=0$. They suggest that the large scatter found in previously reported Tully-Fisher relations at moderate redshifts are produced by the numerous (65\%) galaxies with perturbed or complex kinematics.

At higher redshifts the SINS survey has pioneered the
detailed observations of galaxies at $z\sim2$ \citep{Forster-Schreiber:2006,Genzel:2006,Genzel:2008,Shapiro:2008,Cresci:2009}, demonstrating that near infrared
%detailed observations of galaxies at $z\sim2$ \citep{Forster-Schreiberetal:2006,Genzeletal:2006,Shapiroetal:2008,Genzeletal:2008}, demonstrating that near infrared
integral field spectroscopy is the best tool to
securely measure the kinematical properties at high redshifts $z>1$.
These authors have found that galaxies have an extremely large velocity
dispersion ($\sigma\sim80$\kms) as compared to their rotational velocity.
They favor the hypothesis of early buildup of central disks and bulges
by secular evolution in gas-rich disks. Fast turbulent speeds in the
gaseous component imply the formation of massive clumps. Gas-rich
primordial disks may evolve through a clumpy phase into bright
early-type disk galaxies with a massive exponential disk, a classical
bulge and possibly a central black hole \citep{Noguchi:1999,Immeli:2004a,Immeli:2004b,Elmegreen:2005,Bournaud:2007,Bournaud:2008}.
%(Noguchi, 1999,ApJ, 514, 77;Immeli et al 2004, AA, 413, 547;Immeli et al 2004, ApJ, 611,20;Elmegreen et al 2005, ApJ 634,101; Bournaud Elmegreen Elmegreen 2007 ApJ 670, 237; Bournaud et al  2008, AA 486, 741).
These massive clumps are suggested from NACO high resolution deep imaging and \Ha\ maps from adaptive optics SINFONI observation in two galaxies of the SINS sample \citep{Genzel:2008}.
%due to the lack of deep imaging addressing the question of their formation. 
%These massive clumps are nevertheless not confirmed in the SINS sample due to the lack of deep imaging addressing the question of their formation. 

%However, 
%On the other hand, 
The number of galaxies observed at or around the 
peak in star-formation at $z\sim 1-2.5$ is still relatively small,
and detailed observations of 
a large volume complete sample of galaxies selected 
from an homogeneous set of selection criteria are needed to clarify the respective contributions of merging and continuous gas feed by accretion. 

%However, 
Assembling large samples of galaxies is not an easy task as we need to 
extract from large spectroscopic surveys,
selected with well controlled criteria, those galaxies 
with accurate spectroscopic redshifts for which 
the H$\alpha$ line emission falls outside bright OH sky 
lines. The VVDS \citep{Le-Fevre:2005} survey has been designed to map the evolution of galaxies, large scale
%lines. The VVDS \citep{Le-Fevreetal:2005} is one such survey designed to map the evolution of galaxies, large scale
structures and AGNs from the spectroscopic redshift 
measurement of tens of thousands of objects.
% down to a magnitude $I_{AB}=24.75$. 
This dedicated program has successfully crossed the ``redshift desert'' $1 < z < 3$, providing for the 
first time a complete magnitude-limited sample of $\sim800$ galaxy
redshifts in the ``Ultra-deep'' ($I_{AB}\leq24.75$) sample
(Le F\`evre et al., in preparation), $\sim$ 11\,000 galaxy 
redshifts in the ``deep'' ($I_{AB}\leq24$) sample
\citep{Le-Fevre:2005}, and $\sim$ 40\,000 in the ``wide'' ($I_{AB}\leq22.5$) sample \citep{Garilli:2008}.
%\citep{Le-Fevreetal:2005}, and $\sim$ 40\,000 in the ``wide'' ($I_{AB}\leq22.5$) sample \citep{Garillietal:2008}. 

% "the results for the $z \sim 3$ galaxy sample": je pense
% qu'il est pr�f�rable de ne pas en parler ici, c'est 
% sans logique avec $1<z<1.8$, il faut en parler dans 
% la discussion fin du papier

We are using VVDS to select targets for a large 
observing program at the ESO-VLT aimed at probing the mass assembly and metallicity evolution of a
representative sample of galaxies at $1 < z < 2$, a crucial epoch corresponding to the peak of cosmic star formation activity. 
The main goal of the
MASSIV\footnote{\texttt{www.ast.obs-mip.fr/massiv/}} (Mass Assembly Survey with SINFONI in VVDS) project is to obtain a detailed
description of the mix of dynamical types
% (rotating disks, spheroids, and mergers) 
at this epoch and to follow the evolution of
fundamental scaling relations, such as the Tully-Fisher or the
mass-metallicity relations, and therefore constrain galaxy 
evolution scenarios. 

In this paper we present the first results obtained by MASSIV
focusing on the kinematics
%velocity fields, kinematics, and dynamical masses 
of nine galaxies  with
$1.2 \leq z \leq 1.6$ observed during the MASSIV pilot program. 
A companion paper \citep{Queyrel:2009} 
is devoted to the mass-metallicity relation of galaxies at 
these redshifts using the same data. 
In (\S2) we describe our observations and data reductions. In (\S3) we present the morphological, physical, kinematical and dynamical measurements of the galaxies of our sample. In (\S4 ) we classify the galaxies exploiting the full kinematical information and discuss the Tully-Fisher relation. Our discussion and conclusions are provided in (\S5). Appendix \ref{individual} contains detailed individual comments for each galaxy.
%In (\S2) we describe our observations and data reductions. In (\S3) we address the physicals properties of the galaxies of our sample, and attempt to classify them.
%%In (\S4 ) we provide and %discuss the B and K band absolute TF %relations, {\bf etc..}
%In (\S4 )... (a preciser)...
%Our conclusions are summarised in (\S5). 
We assume a cosmology with $\Omega_{m} = 0.3$, $\Omega_{\Lambda} = 0.7$ and $H_{0}$ = 70 \kmsmpc~
throughout.
%%%%%%%%%%%%%%%%%%%%%%%%%%%%%%%%%%%%%%%%%%%%%%%%%%%%%%%%%%%%%%%%%
%
\section{Data and Observations}
\subsection{Sample selection}
\begin{figure}
%\vspace{-2.2cm}
\begin{center}
\resizebox{1.0\columnwidth}{!}{\includegraphics{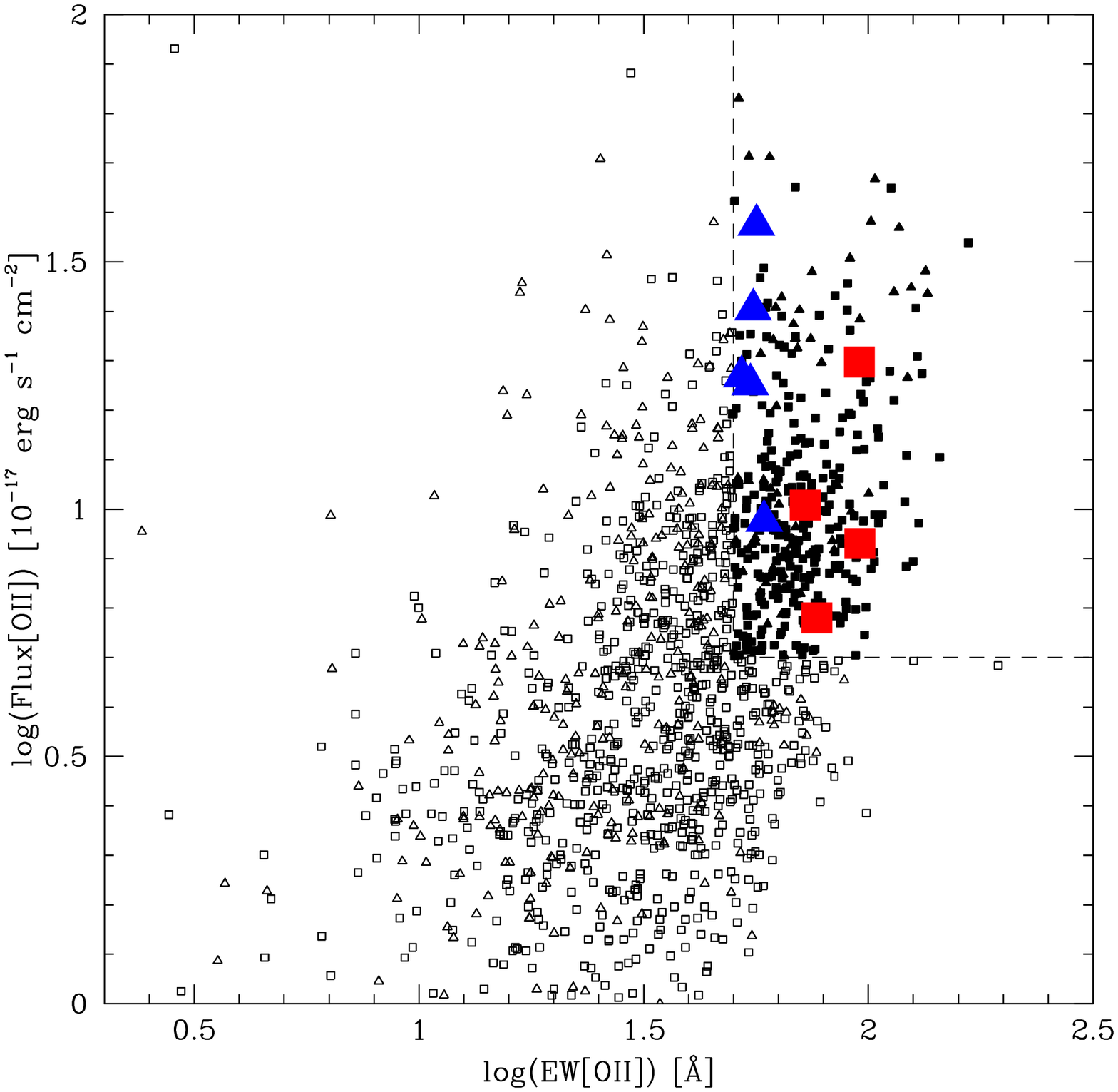}}
\end{center}
%\vspace{1.8cm}
\caption{Selection of late-type $1 < z < 1.54$ star-forming galaxies with secure redshift 
and measured \oii\AA\ emission line in the VVDS-22h ($I_{\rm AB}\leq22.5$, small triangles) 
and VVDS-02h ($I_{\rm AB}\leq24$, small squares) fields for SINFONI follow-up 
observations. Among these galaxies we selected those showing the strongest \oii\AA\ 
emission line (filled symbols: EW $> 50$ \AA\ and flux $> 5\times 10^{-17}$ erg~s$^{-1}$cm$^{-2}$, 
measured on VIMOS spectra) for H$\alpha$ to be easily detected 
in the near-IR. The nine galaxies studied in this paper are indicated as large blue triangles (VVDS-22h field) 
and large red squares (VVDS-02h field).
%For higher redshift galaxies ($z > 1.4$), [OII]3727\AA\ 
%is not detected anymore in VIMOS optical spectra. For these galaxies, 
%we used the photometric $UBVRIK$ SED to select late-type star-forming galaxies.
}
\label{criteria}
\end{figure}

\begin{table*}
\caption{Table of observations}
\label{obs} 
\begin{center}
\begin{tabular}{ccccccccc}
\hline
VVDS ID & $\alpha$ & $\delta$ &          z &          $I_{\rm AB}$ &        Field &   t$_{exp}$ &  Seeing  & Run ID \\
 & J2000 & J2000 &   & mag &   &  hours & \arcsec & \\
(1) & (2) & (3) & (4) & (5)& (6) & (7) & (8) & (9) \\
\hline
020116027 & 02:25:51.133 & -04:45:04.48 &     1.5259 &     22.875 & VVDS-02h   &        1.67 &       0.65 & 075.A-0318(A) \\
020182331 & 02:26:44.260 & -04:35:51.89 &     1.2286 &     22.729 & VVDS-02h   &           3 &       0.72 & 078.A-0177(A) \\
020147106 & 02:26:45.386 & -04:40:47.39 &     1.5174 &     22.502 & VVDS-02h   &           2 &       0.74 & 075.A-0318(A) \\
020261328 & 02:27:11.049 & -04:25:31.60 &     1.5291 &     23.897 & VVDS-02h   &           1 &       0.61 & 075.A-0318(A) \\
220596913 & 22:14:29.184 & +00:22:18.89 &     1.2667 &     21.841 & VVDS-22h   &        1.75 &       0.47 & 075.A-0318(A) \\
220584167 & 22:15:23.038 & +00:18:47.01 &     1.4637 &     22.036 & VVDS-22h   &        1.75 &       0.74 & 075.A-0318(A) \\
220544103 & 22:15:25.708 & +00:06:39.53 &     1.3970 &     22.469 & VVDS-22h   &           1 &       0.69 & 075.A-0318(A) \\
220015726 & 22:15:42.455 & +00:29:03.59 &     1.3091 &     22.473 & VVDS-22h   &           2 &       0.58 & 075.A-0318(A) \\
220014252 & 22:17:45.690 & +00:28:39.47 &     1.3097 &     22.101 & VVDS-22h   &           2 &       0.70 & 075.A-0318(A) \\
\hline
\end{tabular}

(1) Source VVDS identification number,
(2) and (3) Coordinates,
(4) VVDS spectroscopic redshift,
(5) AB magnitude in I-band,
(6) VVDS-22h wide field ($17.5 \leq I_{\rm AB} \leq 22.5$) and VVDS-02h deep field ($17.5 \leq I_{\rm AB} \leq 24.0$),
(7) Exposure time,
(8) Median seeing of SINFONI observations measured on PSF stars,
(9) ESO program.
\end{center}

\label{runs}
\end{table*}
We have used the VIMOS VLT Deep Survey sample to select galaxies with known spectroscopic redshifts across the peak of star formation activity. The VVDS is a complete magnitude selected sample avoiding the biases linked to \textit{a priori} color selection techniques. This sample offers the advantage of combining a large sample with a robust selection function and secure spectroscopic redshifts. The latter are necessary to engage into long single objects integrations with SINFONI \citep{Eisenhauer:2003,Bonnet:2004} being sure to observe the H$\alpha$ line outside bright OH night-sky emission lines.

%In order to probe mass assembly and chemical evolution along cosmic time and to estimate the distribution of  kinematical types, including the fraction of mergers, rotating disks, and massive spheroids, we need a sample of galaxies with known spectroscopic redshifts going across the peak of star formation activity and large enough  to enable statistical studies of the galaxy population. The best and largest sample available today is undoubtedly  the VVDS dataset. This parent sample gives a combination of robust selection functions and secure spectroscopic redshifts to engage into long single object integration with SINFONI being sure to avoid objects with the key emission lines embedded in the OH sky lines.

From the existing VVDS dataset, we have available today a unique sample of more than 1500 galaxies in the redshift domain $1 < z < 2$ with accurate (to the fourth digit) and secure spectroscopic redshifts (confidence level $>$ 90\%; \citealp{Le-Fevre:2005}).
%From the existing VVDS dataset, we have available today a unique sample of more than 1500 galaxies in the redshift domain $1 < z < 2$ with accurate (to the fourth digit) and secure spectroscopic redshifts (confidence level $>$ 90\%; \citealp{Le-Fevreetal:2005}).
This sample being purely $I$-band limited, it contains both star-forming and passive galaxies distributed over a wide range of 
stellar masses, enabling to easily define
volume limited sub-samples. For the MASSIV program, we have defined a sample of $\sim$ 140 VVDS star-forming galaxies at $1.0 < z < 1.8$
suitable for SINFONI observations. In most of the cases, the galaxies are selected on the basis of their measured intensity of \oii\ emission line in the VIMOS spectrum or, for a
few cases, on their observed photometric $UBVRIK$ spectral energy distribution which is typical of star-forming galaxies. The star formation criteria ensure that the brightest rest-frame optical emission lines, mainly H$\alpha$, \oiii, \nii\ used to probe kinematics and chemical abundances, will be observed with SINFONI in the NIR bands.
%In most of the cases, the galaxies are selected on the basis of their measured intensity of \oii\ emission line in the VIMOS spectrum or, for a few cases, on their observed photometric $UBVRIK$ spectral energy distribution which is typical of star-forming galaxies.

For the pilot observations presented in this paper, the selection has been made on \oii\ emission line only thus restricting the redshift range 
to $1 < z < 1.54$. We have selected galaxies showing the strongest \oii\ emission line (EW $> 50$ \AA\ and flux 
$> 5\times 10^{-17}$ erg~s$^{-1}$cm$^{-2}$) as measured on VIMOS spectra (see Figure \ref{criteria}) for H$\alpha$ to be easily 
detected in the near-IR. Among these candidates, we have further restricted the sample taking 
into account one important observational constraint: the observed wavelength of H$\alpha$ line had to fall at least 10\AA\ 
away from strong OH night-sky lines, to avoid heavy contamination of the galaxy spectrum by sky subtraction residuals. Among the remaining list of promising 
candidates, we have randomly picked up twelve galaxies to carry out these first pilot observations. 
%The nine galaxies studied in this paper have accurate (to the fourth digit) and secure (confidence level $>$ 90\%) spectroscopic redshifts derived from VIMOS spectroscopy. 
%For higher redshift galaxies ($z > 1.4$), [OII]3727\AA\ was not detected in the first-epoch VIMOS optical spectra. 
% For these galaxies, we used the photometric SED to select late-type star-forming galaxies. 
These criteria for selecting late-type star-forming galaxies have been shown to be very efficient. For the
first observing runs, our success rate of selection has been around 85\%: 9 galaxies over 12 observed show strong rest-frame
optical emission lines in SINFONI datacubes, the 3 remaining show a signal too faint to derive reliable velocity maps. Among the nine VVDS star-forming galaxies, five have been selected in the 
VVDS-22h ($17.5 \leq I_{\rm AB} \leq 22.5$) field and four in the VVDS-02h ($17.5 \leq I_{\rm AB} \leq 24.0$) field. 
These targets span a redshift range between $z \sim 1.22$ and $z \sim 1.53$ (see Table \ref{runs}).

%--------------------------------------------------------------------------

\begin{table*}
\caption{Fluxes and luminosities of VVDS galaxies} 
\begin{center}
\begin{tabular}{ccccccc}
\hline
VVDS ID & U$_{AB}$ & L$_{1500}$ & F(\Ha) & L(\Ha) & F([O\,{\sc ii}]) & L([O\,{\sc ii}]) \\
 & mag & $10^{+28}$ \ergshz & $10^{-17}$ \ergscm & $10^{+41}$ \ergs & $10^{-17}$ \ergscm & $10^{+41}$ \ergs \\
(1) & (2) & (3)  & (4) & (5) & (6) & (7) \\
\hline
 020116027 &   24.0 &    5.1 $\pm$ 0.2 &   6 $\pm$ 1  &   9 $\pm$ 2  &  6 $\pm$ 2 &    9 $\pm$ 2 \\
 020182331 &   24.1 &    3.7 $\pm$ 0.1 &  55 $\pm$ 14 &  47 $\pm$ 13 & 20 $\pm$ 2 &   17 $\pm$ 2 \\
 020147106 &   23.4 &   12.9 $\pm$ 0.1 &  22 $\pm$ 6  &  32 $\pm$ 9  &  9 $\pm$ 2 &   12 $\pm$ 3 \\
 020261328 &   24.2 &    4.4 $\pm$ 0.2 &   5 $\pm$ 1  &   7 $\pm$ 2  & 10 $\pm$ 2 &   15 $\pm$ 3 \\
 220596913 &   23.1 &    8.7 $\pm$ 0.4 &  33 $\pm$ 6  &  32 $\pm$ 6  & 19 $\pm$ 2 &   17 $\pm$ 2 \\
 220584167 &   22.9 &   12.5 $\pm$ 0.6 &  48 $\pm$ 10 &  64 $\pm$ 14 & 18 $\pm$ 2 &   24 $\pm$ 3 \\
 220544103 &   23.3 &    9.4 $\pm$ 0.5 & 161 $\pm$ 29 & 192 $\pm$ 35 & 25 $\pm$ 3 &   30 $\pm$ 4 \\
 220015726 &   24.0 &    3.1 $\pm$ 0.4 &  26 $\pm$ 3  &  26 $\pm$ 3  &  9 $\pm$ 2 &   10 $\pm$ 2 \\
 220014252 &   22.7 &   11.2 $\pm$ 0.3 & 212 $\pm$ 27 & 216 $\pm$ 28 & 38 $\pm$ 2 &   38 $\pm$ 2 \\
\hline                                                              
\end{tabular}

(1) Source VVDS identification number,                              
(2) AB magnitude in $U$-band,                                       
(3) UV luminosity,                                                  
(4) SINFONI \Ha\ flux,                                              
(5) \Ha\ luminosity,                                                
(6) VIMOS \oii\ flux,
(7) \oii\ luminosity.
\end{center}
\label{tableflux}
\end{table*}

\begin{table*}
\caption{SFRs and extinctions of VVDS galaxies} 
\begin{center}
\begin{tabular}{ccccccc}
\hline
VVDS ID & SFR$_{UV}$ & SFR$_{\mathrm{H}\alpha}$ & SFR$_{\textup{\scriptsize{[O\,{\sc ii}]}}}$ & SFR$_{SED}$ & $E(B-V)_{UV/H\alpha}$ & $E(B-V)_{SED}$ \\
 & \msunyr & \msunyr & \msunyr & \msunyr & mag & mag\\
(1) & (2) & (3)  & (4) & (5) & (6) & (7) \\
\hline
 020116027 &   7.2 $\pm$ 0.3 &   7 $\pm$ 2  &  13 $\pm$ 4  &   46 $^{+28}_{-21}$  &  0.0 & 0.2\\
 020182331 &   5.2 $\pm$ 0.2 &  38 $\pm$ 10 &  24 $\pm$ 7  &   67 $^{+3}_{-25}$   &  0.2 & 0.3\\
 020147106 &  18.1 $\pm$ 0.2 &  25 $\pm$ 7  &  17 $\pm$ 6  &   55 $^{+6}_{-2}$    &  0.0 & 0.1\\
 020261328 &   6.2 $\pm$ 0.2 &   6 $\pm$ 2  &  21 $\pm$ 7  &    8 $^{+23}_{-5}$   &  0.0 & 0.1\\
 220596913 &  12.1 $\pm$ 0.6 &  24 $\pm$ 5  &  24 $\pm$ 7  &   77 $^{+109}_{-29}$ &  0.1 & 0.1\\
 220584167 &  17.5 $\pm$ 0.8 &  51 $\pm$ 11 &  34 $\pm$ 10 &  158 $^{+113}_{-92}$ &  0.1 & 0.3\\
 220544103 &  13.2 $\pm$ 0.7 & 152 $\pm$ 28 &  42 $\pm$ 13 &   46 $^{+73}_{-15}$  &  0.3 & 0.3\\
 220015726 &   4.3 $\pm$ 0.5 &  21 $\pm$ 3  &  13 $\pm$ 4  &   86 $^{+39}_{-32}$  &  0.2 & 0.2\\
 220014252 &  15.7 $\pm$ 0.4 & 171 $\pm$ 22 &  54 $\pm$ 16 &   84 $^{+121}_{-48}$ &  0.3 & 0.3\\
\hline
\end{tabular}

(1) Source VVDS identification number,
SFRs deduced respectively from (2) UV flux, (3) \Ha\ flux, (4) \oii\ flux and (5) SEDs,
Reddening suffered by the ionised gas respectively computed (6) from the UV/\Ha\ ratio and (7) from SED modeling.
\end{center}
\label{tablesfr}
\end{table*}
%
%_--------------------------------------------------------------------------
\subsection{SINFONI observations and data reduction}
The NIR spectroscopic observations were obtained with the 3D-spectrograph SINFONI at ESO-VLT during two four-nights runs, on September
5-8, 2005 (ESO run 75.A-0318) and on November 12-15, 2006 (ESO run 78.A-0177). SINFONI was used in its seeing-limited mode, with the
0.125\arcsec$\times$0.25\arcsec\ pixel scale leading to a field-of-view of 8\arcsec$\times$8\arcsec, and the H grism providing a 
spectral resolution
$R \sim 4000$. 
%Note also that two galaxies (VVDS220544103 and VVDS220584167) have also been observed with the J grism. 
Conditions were not photometric and the mean seeing measured on PSF stars was around 0.65\arcsec\ (see details in Table \ref{runs}). The PSF stars were observed by SINFONI in $H$-band for each pointing on any target.
Each target was acquired through a blind offset from a nearby
bright star. Each observation was obtained by nodding the position of the galaxy within the 8\arcsec\ $\times$ 8\arcsec\ SINFONI
field-of-view, generally by locating the source in two opposite corners. This observational procedure allows background subtraction
by using frames contiguous in time, but with the galaxy in different locations. Moreover, the target was never located exactly at
the same position on the detector, a minimal dithering of 0.3\arcsec\ was required in order to minimize instrumental artifacts when
the individual observations are aligned and combined together. The total on-source integration times are listed in
Table \ref{runs} and range between 1 and 3 hours. As observations were performed in visitor mode, the integration time 
has been adjusted on the fly depending on the signal to noise achieved for the H$\alpha$ emission line. 

Data reduction has been performed with the ESO-SINFONI pipeline (version 1.7.1, see \citealp{Abuter:2006,Modigliani:2007}). The pipeline subtracts the sky background from the
temporally contiguous frames, flat-fields images, spectrally calibrates each individual observation and then reconstructs the
datacube. Individual cubes were aligned in the spatial direction by relying on the telescope offsets from a nearby bright star and
then combined together. A flux calibration is mandatory in order to derive absolute parameters (e.g. star formation rate) from the
%uncalibrated
 flux measured in emission lines. To this end, each science observation has been immediately followed by the observation of a telluric
standard star.
%From them, we have extracted a 1D spectrum which can be used both to perform the flux calibration of our galaxy cubes, and to correct them from telluric absorption lines. We assume that the standard stars are well fitted by a pure blackbody curve, and that the only absorption features which are present in the standard spectra are due to the atmosphere. Starting from the blackbody temperature of the standard star and its magnitude in $J$-,$H$- or $K$-band, we are able to recover the efficiency spectrum. The galaxy cubes are then divided by this efficiency spectrum in order to be flux calibrated and corrected for atmospheric absorption.
%
\section{Measurements}

\subsection{Stellar masses, star formation rates and extinction}
\label{seds}
\begin{figure*}
\begin{center}
%\vspace{-2.2cm}
\resizebox{2\columnwidth}{!}{\rotatebox{270}{\includegraphics{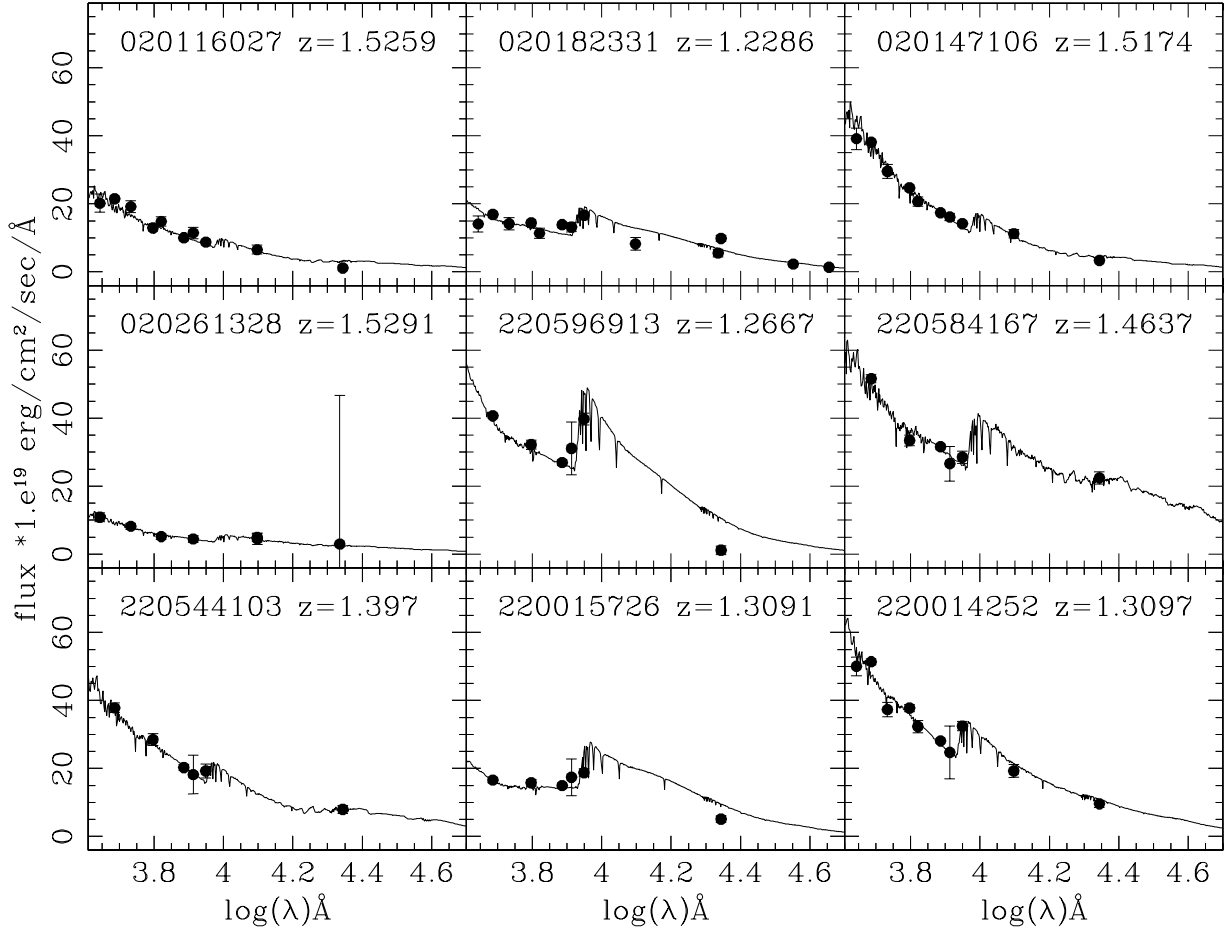}}}
%\resizebox{.67\columnwidth}{!}{\includegraphics{eps/p2}}
%\resizebox{.67\columnwidth}{!}{\includegraphics{eps/fig}}
%\vspace{1.5cm}
\end{center}
\caption{Best SED fit model BC03 (continuous line) and photometric points overplotted for the nine galaxies of the sample.}
\label{sedfit}
\end{figure*}
Stellar mass $M_{star}$ and star formation rate (SFR) estimates for our sample were obtained using the GOSSIP spectral energy distribution (SED) modeling software \citep{Franzetti:2008}. We used as input for the SED fitting the multi-band photometric observations available in VVDS fields, including $BVRI$ data from the CFHT, $UBVRZs$ data from the CFHT Legacy Survey, $J$- and $K$-bands data from SOFI at the NTT and from the UKIDSS survey, and the VVDS-Deep spectra. The photometric and spectroscopic data were fitted with a grid of stellar population models, generated using the BC03 population synthesis code \citep{Bruzual:2003}, assuming a set of ``delayed'' star formation histories (see \citealp{Gavazzi:2002} for details), a \citet{Salpeter:1955} IMF with lower and upper mass cutoffs of respectively 0.1 and 100 $\msun$, a metallicity ranging from 0.02 and 2.5 solar metallicity and a Large Magellanic Cloud reddening law \citep{Pei:1992} with an extinction $E(B-V)$ ranging from 0 to 0.3. The parameters for the best-fitting model for each galaxy are taken as the best fitting values for both the galaxy stellar mass and SFR. On top of the best fitting values, GOSSIP computes also the Probability Distribution Function (PDF), following the method described in \citet{Walcher:2008}. The median of the PDF and its confidence regions are then used to derive a robust estimate of value and error for the  parameter that is to be determined. Stellar masses for the galaxies of this paper are reported in Table \ref{table_mass}. The best SED fit model for each galaxy is presented in Figure \ref{sedfit}.
%
%The stellar masses are estimated thanks to the full optical and near-infrared VVDS photometry available for our targets. In this paper, we have used stellar masses computed by the VVDS team (see \citealp{Pozzetti:2007} and \citealp{Walcher:2008} for details). In brief, the photometric points are compared to those of a library of synthetic stellar population spectra based on CB07 (\citealp{Bruzual:2007}; Charlot \& Bruzual, in preparation) models. The synthetic stellar populations are based on exponential declining star formation histories with the addition of random secondary bursts. The stellar mass is defined as the median of the probability distribution function obtained when comparing each observed galaxy to all models in the library. The results are self-consistently corrected for the effects of age, metallicity and dust. Degeneracy between these secondary parameters are reflected into the error bars on derived quantities.
%

%Additionally the star formation rates, averaged on the last hundred million years, have been derived from the full photometry using the same approach (see \citealp{Lamareille:2009}).% They may then be compared to other estimators (see below).
%They are reported in Table \ref{tablesfr}, together with star formation rates derived from other estimators (see below).
%_--------------------------------------------------------------------------
%
%\subsection{Star formation rates and extinction}
%
%Here we compare 
In addition to the determination from the SED fitting (SFR$_{SED}$), SFRs for the nine star-forming galaxies in our sample can be deduced
%from the SED fitting of the complete photometry (see above), 
from the H$\alpha$ (SFR$_{\mathrm{H}\alpha}$) flux, from the \oii\ flux (SFR$_{\textup{\scriptsize{[O\,{\sc ii}]}}}$) and from the rest-frame UV continuum emission (SFR$_{UV}$).
%, and from the SED fitting of the complete photometry;
As these four estimators are affected differently by dust and star formation history, our results can in principle tell us about the extinction and stellar populations of the galaxies.

The H$\alpha$ recombination line is a direct probe of the young, massive stellar population and therefore provides a nearly instantaneous (i.e. averaged on the last ten million years) measure of the SFR. Moreover, the H$\alpha$ line is less affected by dust extinction compared to e.g. UV estimators. We have calculated  SFR$_{\mathrm{H}\alpha}$ following \citet{Kennicutt:1998}: 

\begin{equation}
\frac{\textup{SFR}_{\mathrm{H}\alpha}}{[\msun \textup{ yr}^{-1}]} = 7.9 \times 10^{-42} \frac{\textup{L}(\mathrm{H}\alpha)}{[\textup{erg s}^{-1}]}
%\textup{SFR } (\msun \textup{ yr}^{-1}) = 7.9 \times 10^{-42} \;L(H\alpha) \:\:(\textup{erg s}^{-1})
\label{sfrha}
\end{equation}
where the \Ha\ luminosity $\textup{L}(\mathrm{H}\alpha)$ has been derived from the \Ha\ flux measured on SINFONI data by \citet{Queyrel:2009}.
%computed from our SINFONI data.

The \oii\ emission line luminosity $\textup{L([O\,{\sc ii}])}$ is not directly coupled to the ionising luminosity and its excitation is sensitive to abundance and the ionization state of the gas. It can however be empirically used as a quantitative SFR tracer. We have computed SFR$_{\textup{\scriptsize{[O\,{\sc ii}]}}}$ following the calibration provided by \citet{Kennicutt:1998}:
\begin{equation}
\frac{\textup{SFR}_{\textup{\scriptsize{[O\,{\sc ii}]}}}}{[\msun \textup{ yr}^{-1}]} = (1.4 \pm 0.4) \times 10^{-41} \frac{\textup{L([O\,{\sc ii}])}}{[\textup{erg s}^{-1}]}
\label{sfrOII}
\end{equation}
where $\textup{L([O\,{\sc ii}])}$ has been derived from the \oii\ flux measured on VIMOS spectra.

Ultraviolet-derived SFRs were calculated from the broadband optical photometry. Given the redshifts of our targets, optical photometry provides indeed an information on the amount of rest-frame UV flux. The UV continuum around 1500 \AA\ is roughly flat if the flux is expressed in frequency units. The mean luminosity inside the optical broad-band filter which corresponds to the rest-frame 1500 \AA\ continuum L$_{1500}$, at the given redshift $z$ of the source, is thus a very good approximation of the level of this continuum. The mean luminosity in frequency units is calculated following this equation:
\begin{equation}
%\frac{L_{1500}} {[\textup{erg~s$^{-1}$~Hz$^{-1}$}]} = 10^{-0.4\times m(AB)} \times 3631\times 10^{-23} \times 4\pi \frac{(D_{L}/[\textup{cm}])^2}{1+z}
%\frac{L_{1500}} {[\textup{erg~s$^{-1}$~Hz$^{-1}$}]} = 10^{-0.4\times m(AB)} \times 3631\times 10^{-23} \times 4\pi
\frac{\textup{L}_{1500}} {[\textup{erg~s$^{-1}$~Hz$^{-1}$}]} = 10^{-0.4~U_{AB}-23} \times 3631 \times 4\pi \frac{\left(\frac{D_{L}}{[\textup{cm}]}\right)^2}{1+z}
%\frac{L_{1500}} {[$\ergshz$]} = 10^{(-0.4~m(AB))} \times 3631\times 10^{-23} \times 4\pi \frac{D_{L(cm)}^2}{1+z}
\end{equation}
$D_{L}$ beeing the luminous distance.

We have then deduced SFR$_{UV}$ following \citet{Kennicutt:1998}: 
\begin{equation}
\frac{\textup{SFR}_{UV}}{[\msun \textup{ yr}^{-1}]} = 1.4 \times 10^{-28}~\frac{\textup{L}_{1500}}{[\textup{erg~s$^{-1}$~Hz$^{-1}$}]}
%\textup{SFR } (\msun \textup{ yr}^{-1}) = 1.4 \times 10^{-28}\;L_{1500} \:\:($\ergshz$ )
\label{sfruv}
\end{equation}
The three equations \ref{sfrha}, \ref{sfrOII} and \ref{sfruv} assume
%The above relation assumes 
a Salpeter IMF with lower and upper mass cutoffs of 0.1 and 100 $\msun$ and
%does 
do not include any effects of attenuation by dust.
% and case B recombination at $T_e$=10,000 K.

Table \ref{tableflux} gives the AB magnitude in $U$-band, UV flux, \Ha\ and \oii\ fluxes and luminosities used to compute the four SFRs stored in Table \ref{tablesfr} together with the reddening values deduced from SED fitting and from the UV/\Ha\ SFR ratio.

% It also assumes that all of the ionizing photons are
%reprocessed into lines, therefore $SFR_{H\alpha} = SFR_{UV}$ provide an estimation of the extinction.
The UV SFR may be much more strongly affected by extinction than the \Ha\ one due to a shorter wavelength emission. Its comparison with the H$\alpha$ SFR may thus provide an estimate of the amount of dust.
Assuming that all of the ionizing photons are reprocessed into lines, therefore $\textup{SFR}_{\mathrm{H}\alpha} = \textup{SFR}_{UV}$, we thus computed the interstellar gas extinction ($E(B-V)$) using the intrinsic starburst flux density $f_i(\lambda)$ via:
\begin{equation}
f_i(\lambda)=f_o(\lambda)~10^{0.4~E(B-V)~k^e(\lambda)}~,
\end{equation}
where the obscuration curve for the stellar continuum, $k^e(\lambda)$, is given by \citet{Calzetti:2000} for star-forming galaxies:
$$k^e(\lambda)=1.17\times(-2.156+1.509/\lambda-0.198/\lambda^2+0.011/\lambda^3)+1.78$$
\begin{equation}
\textup{for~~~}0.12\mu m \le \lambda < 0.63 \mu m
\end{equation}

\begin{figure}
\begin{center}
%\vspace{-2.2cm}
\resizebox{1.0\columnwidth}{!}{\includegraphics{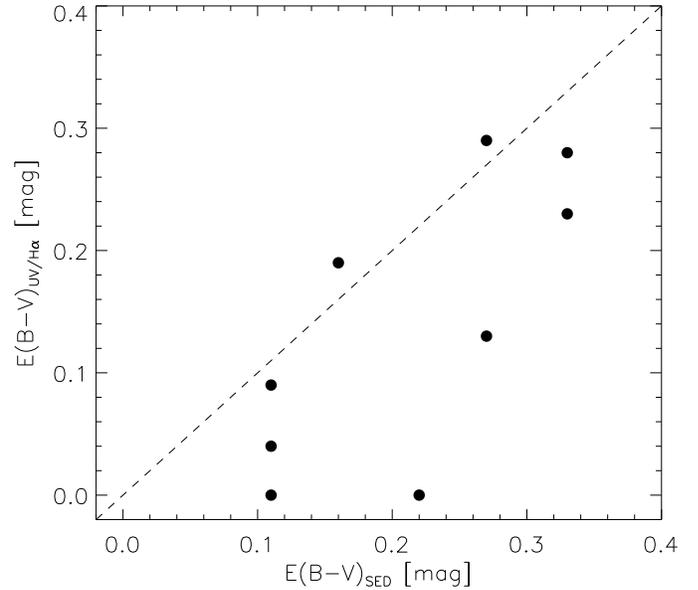}}
%\vspace{1.5cm}
\end{center}
\caption{Comparison of extinctions computed from SED fitting (x-axis) and from the comparison of SFRs deduced from UV and \Ha\ luminosities (y-axis). The dashed line indicates $y=x$.}
\label{extinction}
\end{figure}

The amount of interstellar extinction estimated in our sample (mean value $E(B-V) \sim 0.15$) is typical of star-forming galaxies in the local and intermediate-redshift universe and is in very good agreement with the reddening values derived at similar redshifts in UV-selected galaxy samples by \citet{Erb:2006}.
%but slightly higher than the reddening values derived at similar redshifts in UV-selected galaxy samples ($E(B-V) \sim 0.15$, \citealp{Erb:2006}).
Figure \ref{extinction} shows a good agreement between extinctions derived from SED fitting and from the UV/\Ha\ SFR ratio. However, the SFR-based extinctions are systematically underestimated compared to the SED ones. This may be due to the uncertainties in the absolute flux calibration both in the rest-frame UV and even more important in the SINFONI data.

\begin{figure*}
\begin{center}
%\vspace{-2.2cm}
\resizebox{1.0\columnwidth}{!}{\includegraphics{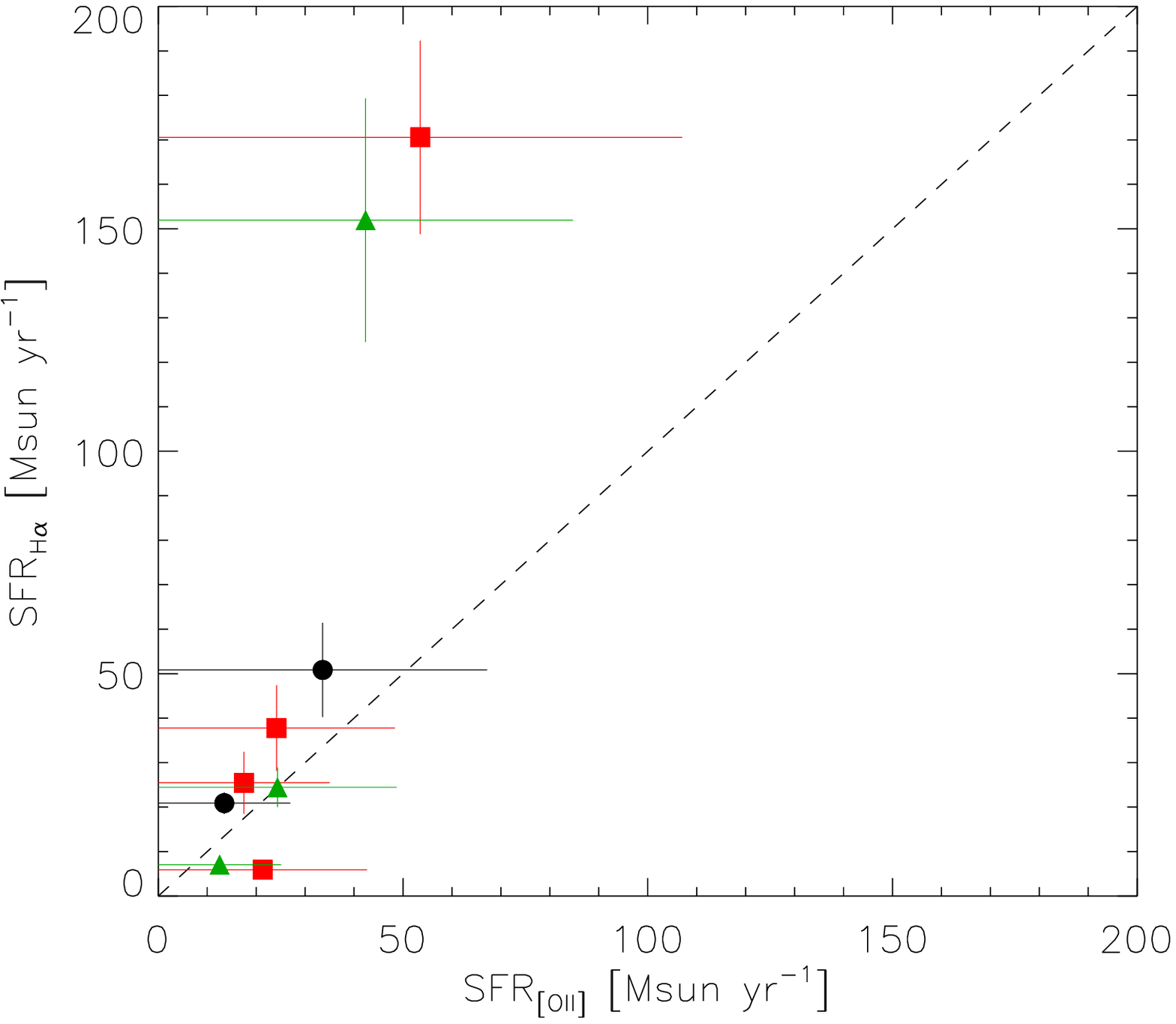}}
\resizebox{1.0\columnwidth}{!}{\includegraphics{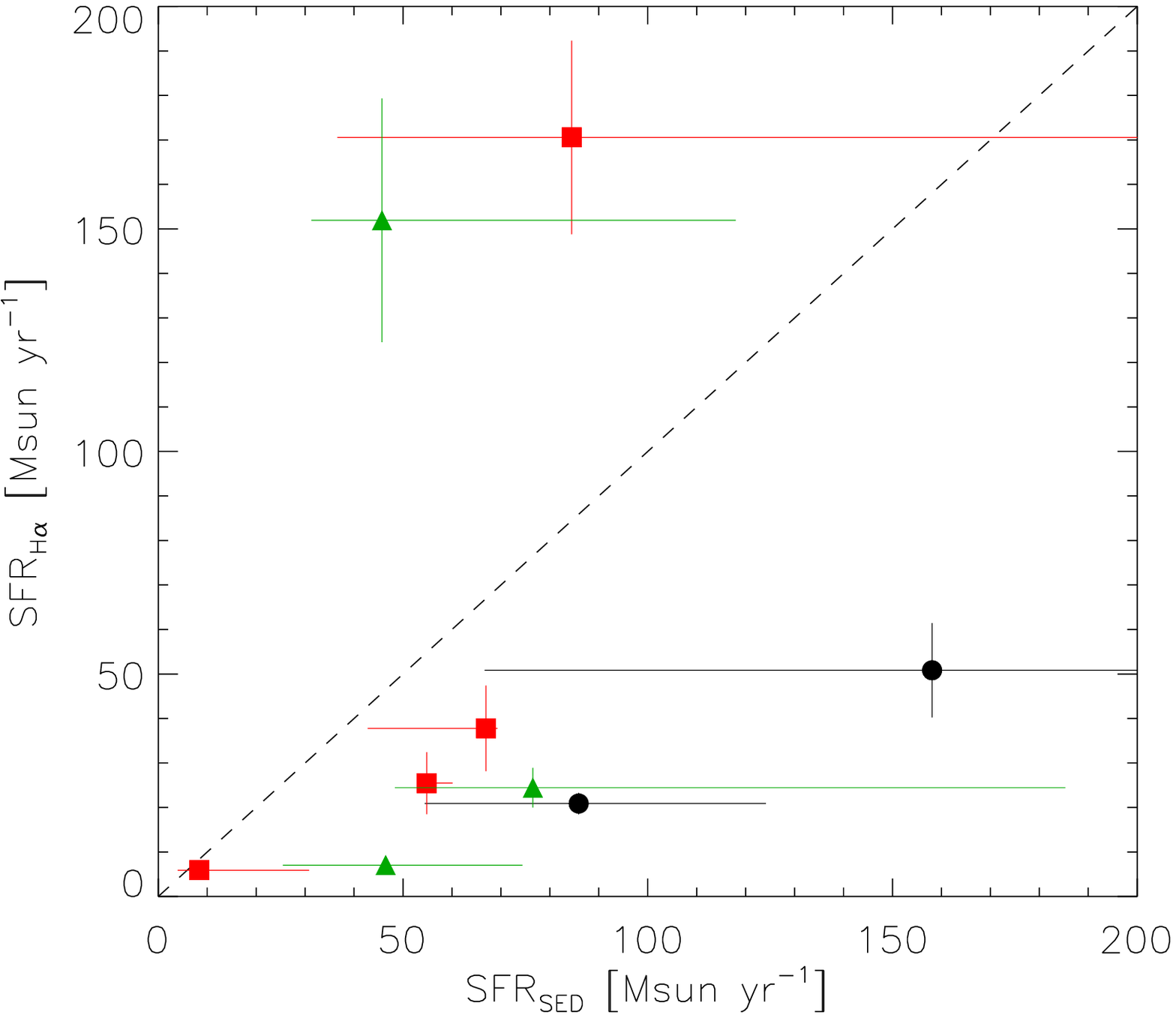}}
%\vspace{1.5cm}
\end{center}
\caption{Comparison of \Ha\ SFR with \o2\ (left) and photometric (right) SFRs. The dashed line indicates $y=x$. Black dots, red squares, green triangles correspond respectively to objects classified (see section \ref{classif}) as rotating disks (RD), perturbed rotators (PR) and merging systems (MS).}
\label{sfrha_vs_sfrsedo2}
\end{figure*}

In Figure \ref{sfrha_vs_sfrsedo2} (left), we compare the SFR derived from \o2\ and \Ha\ emission lines. The agreement is good taking into account that neither \Ha\ nor \o2\ flux has been corrected for reddening. Indeed, \Ha\ based SFR are systematically higher than \o2\ ones because \o2\ is more affected by extinction than \Ha, except for the two galaxies with almost no extinction (see Table \ref{tablesfr}). The fact that the two galaxies with the highest extinction ($E(B-V)\sim 0.3$) show the largest deviation is supporting this interpretation.

The SFR deduced from the SED fitting (SFR$_{SED}$) provides an internal correction for dust extinction but is averaged on a timescale ten times longer than the SFR derived from H$\alpha$. The comparison between these two SFRs may provide information on the ratio between recent and older star formation activity, and thus can give an hint on the presence of recent bursts (see Figure \ref{sfrha_vs_sfrsedo2}, right).
%
%The SFR derived from H$\alpha$ can be compared with the value resulting from the SED fitting (see Figure~\ref{sfrha_vs_sfrsed.eps}).
%
Despite the large uncertainties inherent to the SED-based SFR determination, a good
agreement between the two SFR estimators is seen for the three galaxies
classified as perturbed rotators (see section \ref{classif}) with a H$\alpha$-based SFR value lower than $50$ M$_{\odot}$ yr$^{-1}$. 
The two galaxies classified as rotating disks 
and the two merging systems VVDS020116027 and VVDS220596913 show an instantaneous
H$\alpha$-based SFR (timescale $\sim 10$ Myr) which is lower 
(by a factor 2-4) than the SED-based SFR integrated over a longer 
timescale (a few hundreds Myr). On the contrary, the two galaxies with 
the highest values of H$\alpha$-based SFR ($> 150$ M$_{\odot}$ yr$^{-1}$, 
VVDS220544103 and VVDS220014252) are typical starburst galaxies with 
a high ratio between the instantaneous and integrated SFRs.

% Despite the large uncertainties inherent to the SED-based SFR determination, a good 
% agreement between the two SFR estimators is seen for the four galaxies 
% with the lowest SFR values ($< 50$ M$_{\odot}$ yr$^{-1}$). 
% The two galaxies classified as rotating disks 
% and the merging system VVDS220596913 (see section \ref{classif}) show an instantaneous 
% H$\alpha$-based SFR (timescale $\sim 10$ Myr) which is lower 
% (by a factor 2-4) than the SED-based SFR integrated over a longer 
% timescale (a few hundreds Myr). On the contrary, the two galaxies with 
% the highest values of H$\alpha$-based SFR ($> 150$ M$_{\odot}$ yr$^{-1}$, 
% VVDS220544103 and VVDS220014252) are typical starburst galaxies with 
% a high ratio between the instantaneous and integrated SFRs.

%\begin{figure}
%\begin{center}
%\vspace{-4.5cm}
%\resizebox{1.0\columnwidth}{!}{\includegraphics{eps/ebv_vs_incl_vs_oh.eps}}
%\end{center}
%\caption{}
%\label{ebv_vs_incl_vs_oh.eps}
%\end{figure}

%For the UV star formation rate, we give the name of observed band which is closest to the rest-frame 1500 \AA\ region, together with its actual rest-frame median wavelength.
%
%_----------------------------------------------
\subsection{Morphologies}
\label{morphologies}
For each galaxy of our sample we have estimated the inclination from the $I$-band CFHT images using the GALFIT software \citep{Peng:2002}. We have used CFHT legacy survey images with best seeing ($<$0.65\arcsec) for VVDS-02h field galaxies (Figures \ref{map1} to \ref{map4}) and CFH12K/CFHT \citep{McCracken:2003} images for VVDS-22h field galaxies (Figures \ref{map5} to \ref{map9}). GALFIT convolves a point spread function (PSF) with a model galaxy image based on the initial parameter estimates fitting a \citet{Sersic:1968} profile. Finally, GALFIT converges into a final set of parameters such as the Sersic index, the center, the position angle, and the axial ratio. Residual maps were used to optimize the results. Since the morphological inclination is used as input for kinematical modeling, we have imposed the morphological position angle of the major axis to be equal to the kinematical one (determined in section \ref{model_fitting}). When the morphological position angle is a free parameter, the agreement with the kinematical one is better than 25\degr\ except for galaxies which have a morphology compatible with a galaxy seen face-on within the uncertainties, as it would be the case for local disks \citep{Epinat:2008b,Epinat:2008a}. This agreement is also true for some of the galaxies that we have classified as merging systems (see section \ref{classif} and individual comments in Appendix \ref{individual}).
%This suggests that disks will result from these mergers geometry or that the merger is in an advanced stage.
For mergers, the use of a fit relying on a disk hypothesis is undoubtly less meaningful than for relaxed systems. However, this fitting method enables to study the whole sample in an homogeneous way.
% which are shown in column 9 of Tab.\ref{table:prop}.
\\
\par
%\citet{Epinat:2009} have used toy models in order to study the effects of beam smearing on the inclination determination of disks. 
\citet{Epinat:2009} have shown that observation of high redshift disk galaxies, strongly affected by beam smearing effects, have a disk inclination difficult to retrieve. The uncertainty on the inclination is
the main source of errors in the rotational velocity determinations and thus on the Tully-Fisher relation, on the mass determination and on the mass assembly history. The extreme cases are provided by galaxies appearing face-on or edge-one from broad-band imagery: the radial velocity field of a face-on galaxy may show a velocity gradient indicating that the disk is not face-on; alternatively, an edge-on disk appears thicker and leads to an underestimation of the inclination.
In order to check the disk inclinations determined by GALFIT, we have compared them to the ones computed using a simple correction provided in \citet{Epinat:2009}.
They modeled inclined thin disks having exponential or flat truncated luminosity profiles with scale lengths from 2 to 6 kpc, observed at redshift $z=1.7$ with a seeing of 0.5\arcsec\ and a sampling of 0.125\arcsec. They introduced a parameter $C$ in order to recover the true inclination $i$ from the seeing FWHM ($s$) and the measured major ($a_m$) and minor ($b_m$) axis lenghts (FWHMs) estimated in arcseconds from a 2D gaussian fitting :
\begin{equation}
\cos{i}=\frac{b}{a}=\sqrt{\frac{b_m^2-C^2\times s^2}{a_m^2-C^2\times s^2}}
\end{equation}
where $a$ and $b$ are actual major and minor axis lengths (FWHMs) on the sky plane. The parameter $C$ represents the fraction of the seeing by which both measured minor and major axis are quadratically overestimated.
In \citet{Epinat:2009}, it is shown that $C=1.014 \pm 0.002$ for an exponential luminosity profile and $C=1.015 \pm 0.010$ for a flat and truncated luminosity profile. This means that these smooth luminosity profiles can be satisfactorily described by a gaussian function. We have used $C=1.014$ to confirm the inclinations determined from the GALFIT fitting procedure.
%In \citet{Epinat:2009}, it is shown that $C$ depends on the luminosity profile. We have used the mean value between exponential and flat lights distributions $C=2$ to confirm the inclinations determined from the GALFIT fitting procedure.
%We measured the axis length from the GALFIT fitting procedure which was confirmed by eye measurements using $C=2$ parameter.
We have used the inclination determined from direct axis measurements for some objects for which GALFIT failed to converge into a realistic model.
Except for those galaxies, GALFIT typically gives a disc component with a Sersic index of one and typical associated uncertainties ranging from 0.2 to 0.5.

%indicates face-on disks with thus infinite maximum rotation velocities.
% and measured the axis lengths by eye. 
%This enables to obtain a very good agreement with GALFIT fitting procedure in most of the cases. 
%We used the inclinations determined by eye since for some objects GALFIT gave a 0\degr~inclination that leads to unrealistic infinite maximum velocities when deprojecting.
%Moreover, the measurement of the axis lengths enabled us to estimate an error bar on the morphological inclinations.
\par
In order to evaluate error bars on the inclination, the uncertainty on both major and minor axis lengths has been taken into account. Assuming that the error $\Delta$ is the same for both axes, we have computed upper and lower limits for the inclination reported in Table \ref{table_kin} following:
\begin{eqnarray}
i_{max}=acos\left[ \frac{b-\Delta}{a+\Delta}\right] \\
i_{min}=acos\left[ \frac{b+\Delta}{a-\Delta}\right] 
\end{eqnarray}
The typical uncertainty is $\Delta=0.25$\arcsec\ (about one third the seeing).
\subsection{Kinematics of the ionized gas}
\subsubsection{$H\alpha$ kinematics}
\label{hakin}

\begin{table*}
\caption{Kinematical parameters} 

\begin{center}
\begin{tabular}{ccccccccc}
\hline
VVDS ID & $z_{H\alpha}$ & Inclination & Position Angle & $r_t$ & $V_t$ & Residuals ($V$) & $\chi^2$ & $V_t/r_t$ \\
   & 		   & \degr & \degr & \arcsec & \kms & \kms & & \kms~kpc$^{-1}$ \\
 (1)  & (2) & (3) & (4)  & (5) & (6)  & (7)  & (8)  & (9) \\
\hline
020116027 & 1.53022 & 44 $^{+12}_{-23}$ & 219 $\pm$   8 &  0.12 &  32 &   8 &  2.4 &    30 \\
020182331 & 1.22832 & 52 $^{+ 9}_{-13}$ & 244 $\pm$   6 &  0.55 & 134 &  18 &  1.2 &    29 \\
020147106 & 1.51949 & 37 $^{+15}_{-37}$ & 307 $\pm$  11 &  0.12 &  30 &   3 &  0.7 &    29 \\
020261328 & 1.52891 & 47 $^{+12}_{-21}$ & 192 $\pm$   5 &  0.77 & 195 &  12 &  2.2 &    30 \\
220596913 & 1.26619 & 66 $^{+ 6}_{- 7}$ & 253 $\pm$   5 &  2.80 & 325 &  16 &  3.5 &    13 \\
220584167 & 1.46588 & 49 $^{+ 9}_{-13}$ & 178 $\pm$   1 &  1.24 & 280 &  15 &  6.6 &    26 \\
220544103 & 1.39659 & 61 $^{+ 7}_{-10}$ & 205 $\pm$   7 &  6.73 & 762 &  17 &  2.8 &    13 \\
220015726 & 1.29300 & 25 $^{+21}_{-25}$ & 184 $\pm$   1 &  0.12 & 323 &  16 &  6.8 &   309 \\
220014252 & 1.31014 & 48 $^{+10}_{-15}$ & 137 $\pm$   3 &  0.12 & 103 &  13 &  2.6 &    99 \\
\hline
\label{table_kin}
\end{tabular}

(1) Source VVDS identification number,
(2) Geocentric SINFONI spectroscopic redshift deduced from model fitting,
(3) Morphological inclination from CFHT $I$-band images,
(4) Kinematical Position Angle of the major axis from North to East deduced from model fitting,
(5) Characteristic radius of the kinematical model,
(6) Characteristic velocity of the kinematical model,
(7) Weighted standard deviation of the residual velocity map,
(8) Reduced $\chi^2$ of the fit,
(9) Inner velocity gradient derived from the model.
\end{center}
\end{table*}

The SINFONI data were analysed using routines written under IDL environment \footnote{Interactive Data Language}. Our main goal was to determine the kinematics of the ionised gas using the strongest emission line, \Ha. In order to increase the signal to noise ratio without degrading the spatial resolution, we have performed a sub-seeing spatial gaussian smoothing (FWHM$=$2 pixels) on the data cube (the mean seeing is around 0.65\arcsec\ i.e. 5 pixels). For each galaxy of the sample we fit the spectrum around \Ha~to a single gaussian function in order to characterize the emission line, and a pedestal to characterise the continuum.
To minimize the effects of sky lines residuals on line parameters determination, the sky spectrum has been used in the fitting procedure in order to lower the weight of strong emitting sky lines wavelengths.
% This assumes the noise to be poissonian.
%To minimize as much as possible the sky lines residuals effects on line parameters determination, we used the square root of the sky spectrum in the line fitting procedure as an error spectrum, assuming the noise to be poissonian. In this way, wavelengths with strong emitting sky lines are considered with a lower weight. Line flux, dispersion, central wavelength and the pedestal were then computed. 
From these fitting techniques it was possible, for each source, to recover the line flux map, the \vf\ and the \vdm\ after correcting for instrumental dispersion (around 2.75\AA) measured on sky lines. 2D error maps have also been derived for each quantity from the fitting procedure. These are statistical errors that take into account the error spectrum and that indicate the accuracy of the fit for each parameter. A signal-to-noise ratio map has been computed, the signal beeing the intensity of the fitted line and the noise beeing the standard deviation of the residual spectrum. These maps are presented in Figures \ref{map1} to \ref{map9}. The velocity maps have been masked using several criteria: we have imposed (i) the line width to be larger than the one of the spectral PSF (the major part of the white pixels in the velocity dispersion error maps), (ii) the uncertainty on the velocity to be less than 30\kms\ and (iii) the signal-to-noise ratio to be larger than $\sim2$ ($1.5$ for the faintest object VVDS020182331). We also made a careful visual inspection in the spectrum to remove outer regions validating the criteria but for which the line was associated with strong sky residuals (see Figures \ref{map5} and \ref{map6}).

%The data were analysed using E3D, the Euro3D visualization tool (Sanchez 2004), and a coded routines for kinematics' analysis (e.g., Sanchez et al. 2004, 2005). Our main goal was to determine the kinematics of the ionised gas using the strongest emission line, H$\alpha$. For each galaxy of the sample we fit the H$\alpha$ region to a single gaussian function, in order to characterize the emission line, and a pedestal to characterise the continuum. The flux intensity, FWHM, central wavelength and the pedestal were then fitted. From the result of these fitting technique it was possible to recover the line flux map, the velocity map and the velocity dispersion map ($\sigma$-map) after correction for instrumental dispersion measured on sky lines (see Fig.\ref{map1}, \ref{map2} \& \ref{map3} ). A 2D error map has been also computed by taking into account the uncertainty on the velocity.

\subsubsection{Model fitting}
\label{model_fitting}

%TABLE CINEMATIQUE
%\input{kinematical_table_v4.3.tex}
\begin{table*}
\caption{Mass parameters} 

\begin{center}
\begin{tabular}{ccccccccccccc}
\hline
VVDS ID & $V_{max}$ & $R_{last}$ & $r_{1/2}$ & $\sigma_0$ & $V_{max}/\sigma_0$ & $M_{star}$ & $M_{dyn}$ & $M_{\theta}$ & $M_{\sigma}$ & $M_{\theta}/M_{\sigma}$ & $M_{halo}$ & Class \\
   & \kms  & $kpc$  & $kpc$ & \kms & & $10^{10}\msun$ & $10^{10}\msun$ & $10^{10}\msun$ & $10^{10}\msun$ & & $10^{10}\msun$ & \\
 (1)  & (2) & (3) & (4)  & (5) & (6)  & (7)  & (8)  & (9) & (10) & (11) & (12) & (13) \\
\hline
020116027 &  32 $^{+ 29}_{-  5}$ &  9.7 &  2.9 &  45 $\pm$  13 &   0.7 $^{+  1.2}_{-  0.3}$ &  1.2 $^{+ 1.7}_{- 0.5}$ &  5.5 $^{+ 4.1}_{- 2.7}$ &  0.2 $^{+ 0.6}_{- 0.1}$ &  5.3 $^{+ 3.5}_{- 2.6}$ &   0.0 $^{+  0.3}_{-  0.0}$ &    0.5 & MS \\
020182331 & 134 $^{+ 32}_{- 14}$ &  6.3 &  4.5 &  71 $\pm$  34 &   1.9 $^{+  2.6}_{-  0.8}$ &  5.8 $^{+ 1.0}_{- 1.9}$ &  4.1 $^{+ 3.2}_{- 1.7}$ &  2.6 $^{+ 1.4}_{- 0.6}$ &  1.5 $^{+ 1.8}_{- 1.1}$ &   1.8 $^{+  8.4}_{-  1.1}$ &   43.9 & PR \\
020147106 &  30 $^{+\infty}_{-  7}$ &  9.1 &  2.0 &  80 $\pm$   8 &   0.4 $^{+\infty}_{-  0.1}$ &  1.7 $^{+ 0.1}_{- 0.6}$ & 27.6 $^{+\infty}_{- 5.8}$ &  0.2 $^{+\infty}_{- 0.1}$ & 27.4 $^{+ 6.4}_{- 5.7}$ &   0.0 $^{+\infty}_{-  0.0}$ &    0.4 & PR \\
020261328 & 194 $^{+125}_{- 31}$ &  6.4 &  1.2 &  55 $\pm$  17 &   3.5 $^{+  4.9}_{-  1.3}$ &  0.6 $^{+ 1.8}_{- 0.4}$ & 19.6 $^{+19.8}_{- 9.0}$ &  5.6 $^{+ 9.7}_{- 1.7}$ & 14.0 $^{+10.1}_{- 7.4}$ &   0.4 $^{+  1.9}_{-  0.2}$ &  110.1 & PR \\
220596913 & 177 $^{+ 13}_{-  7}$ & 12.7 &  8.7$^*$ &  76 $\pm$  20 &   2.3 $^{+  1.1}_{-  0.6}$ &  8.5 $^{+ 2.7}_{- 4.1}$ & 13.0 $^{+ 3.7}_{- 2.5}$ &  9.3 $^{+ 1.4}_{- 0.7}$ &  3.7 $^{+ 2.3}_{- 1.7}$ &   2.5 $^{+  2.9}_{-  1.1}$ &   98.7 & MS \\
220584167 & 280 $^{+ 81}_{- 31}$ & 10.9 &  7.9$^*$ &  47 $\pm$  22 &   5.9 $^{+  8.5}_{-  2.3}$ & 12.1 $^{+10.9}_{- 5.0}$ & 21.1 $^{+14.5}_{- 5.0}$ & 20.0 $^{+13.2}_{- 4.3}$ &  1.1 $^{+ 1.3}_{- 0.8}$ &  18.1 $^{+ 90.7}_{- 11.6}$ &  346.1 & RD \\
220544103 & 146 $^{+ 19}_{-  9}$ & 10.9 &  5.5 &  70 $\pm$  17 &   2.1 $^{+  1.1}_{-  0.5}$ &  5.1 $^{+ 2.3}_{- 3.1}$ & 10.3 $^{+ 4.2}_{- 2.8}$ &  5.4 $^{+ 1.5}_{- 0.7}$ &  4.9 $^{+ 2.7}_{- 2.1}$ &   1.1 $^{+  1.4}_{-  0.5}$ &   51.0 & MS \\
220015726 & 323 $^{+\infty}_{-135}$ &  6.4 &  2.7 &  38 $\pm$  25 &   8.5 $^{+\infty}_{-  5.5}$ &  6.2 $^{+ 5.7}_{- 1.6}$ & 16.7 $^{+\infty}_{-11.3}$ & 15.5 $^{+\infty}_{-10.3}$ &  1.2 $^{+ 2.1}_{- 1.0}$ &  13.2 $^{+\infty}_{- 11.6}$ &  591.3 & RD \\
220014252 & 103 $^{+ 39}_{- 13}$ &  9.2 &  4.4 &  92 $\pm$  19 &   1.1 $^{+  0.8}_{-  0.3}$ &  6.1 $^{+ 1.0}_{- 3.6}$ & 10.2 $^{+ 5.7}_{- 3.5}$ &  2.3 $^{+ 2.1}_{- 0.6}$ &  7.9 $^{+ 3.7}_{- 3.0}$ &   0.3 $^{+  0.6}_{-  0.1}$ &   19.4 & PR \\
\hline
\label{table_mass}
\end{tabular}

(1) Source VVDS identification number,
(2) Maximum velocity deduced from model fitting,
(3) Maximum radius of the kinematical maps,
(4) Half light radius corrected for the beam smearing (an asterisk indicates that it has been computed from the $I$-band image),
(5) ``1/error''-weighted mean local velocity dispersion,
(6) Ratio of the maximum rotation velocity (2) over the local velocity dispersion (5),
(7) Stellar mass,
(8) Total dynamical mass,
(9) Rotation mass,
(10) Dispersion  mass,
(11) Ratio of the dynamical masses,
(12) Halo  mass,
(13) Kinematical classification (RD: Rotating Disk, PR: Perturbed Rotation, MS: Merging System).
\end{center}
\end{table*}

To analyze the velocity fields of high redshift galaxies, four toy rotation curve models used in the literature (an exponential disk by \citealp{Forster-Schreiber:2006,Cresci:2009}, an isothermal disk by \citealp{Spano:2008}, an arctangent function by \citealp{Puech:2008} and a flat rotation curve by \citealp{Wright:2007,Wright:2009} described hereafter) have been tested on a sub-sample of around 150 data cubes of nearby galaxies part of the GHASP sample which have been projected at high redshift \citep{Epinat:2009}. These models consist of two parameters rotation curves that have various shapes, mainly the plateau (decreasing, increasing or flat) and the presence or not of an inner velocity bump. By comparing the parameters determined from high resolution and projected data, \citet{Epinat:2009} have shown that the simple model used by \citet{Wright:2007,Wright:2009} leads to the best estimation of the kinematical parameters in average.
%To help in the kinematical classification, in addition to this set of three maps, a fit with a simple rotating disk was attempted for all sources on their velocity field using the associated velocity error map. 
%
The high resolution rotation curve (velocity $V$ as a function of the radius $r$) of this model is given by:
\begin{equation}
V(r)=V_t\frac{r}{r_t}
\end{equation}
when $r \le r_t$ and 
\begin{equation}
V(r)=V_t
\end{equation}
when $r>r_{t}$.

In the present work, this fit was attempted for all sources on their velocity field using the associated velocity error map
in order to evalutate which ones are compatible with rotating disks.
%using the residual maps.
For objects that we have classified as merging systems (section \ref{classif}), fitting rotating disks on substructures has been attempted and discussed in Appendix \ref{individual}.
%we fit a rotating disk model to the velocity field, using the velocity error map, for all sources. 
The high resolution model is convolved with the seeing measured on the PSF stars and by the 2-pixels gaussian smoothing applied on the data cubes. The parameters of the model are the position angle of the major axis (measured from North to East), the inclination, the center, the systemic velocity, the plateau of the rotation curve $V_t$ and the radius $r_{t}$ at which the plateau is reached.

%TC: d'ou l'interet d'indiquer la valeur du seeing pour chaque galaxie

%This model assumes a thin disk in rotation. As the inclination is hardly constrained, we tried a series of fixed inclination : 20, 30, 40, 50, 60, 70, 80 and find the model with the smallest $\chi^2$.

This model fitting procedure described in details in \citet{Epinat:2009} is based on a $\chi^2$ minimization. One main concern for fitting the data is that the actual high resolution line flux distribution is unknown.
%Since the flux is faint, we excluded deconvolution processes for recovering the high resolution distribution. Moreover, from local samples of galaxies, it does not seem obvious to make flux distribution models. Thus, in
We have assumed that the observed line flux is representative of the high resolution one.
Furthermore, due to projection effects, kinematical models are affected by a degeneracy between the rotation velocity $V(r)$ and the inclination. Even for local resolved galaxies, this degeneracy leads to large error bars on kinematical inclinations \citep{Epinat:2008b,Epinat:2008a}. Due to the low sampling, this degeneracy can not be ruled out for high redshift galaxies. We have fixed the inclination to the morphological value determined from $I$-band continuum images as described in section \ref{morphologies}.

From the best fit model, we have computed a model velocity dispersion map. This model \vdm~only includes the enlargement of the profiles due to the unresolved velocity gradient because of beam smearing effects (see \citealp{Epinat:2009} for analytical computation). It does not include enlargement due to the physics of the gas. Thus by subtracting quadratically the model to the measured \vdm, we are able to recover the local velocity dispersion noted hereafter $\sigma_0$.
% when the model is a good description of the reality.
%
%The features of that modeled velocity dispersion map are only related to the unresolved velocity shear of the velocity field due to beam smearing effects, and not to the physics of the gas. Thus by subtracting quadratically that model to the measured velocity dispersion map, we are able recover the local velocity dispersion when the model is a good description of the reality.
In order to take into account the uncertainty on the \vdm\ while computing the mean \vd, we have used a weight inversely proportional to the error. The same procedure has been applied while computing the mean velocity residual.

The results of this fitting procedure are presented in Table \ref{table_kin}. It contains the redshifts determined from SINFONI data, the morphological inclination used for the kinematical model, the kinematical position angle, the rotation curve parameters $r_t$ and $V_t$, the mean velocity residual, the reduced $\chi^2$ of the fit and the inner velocity gradient computed by the models. The center has been fixed to match the maximum of the flux distribution except for VVDS220584167 and VVDS220014252 (see Appendix \ref{individual}) since the determination of the kinematical center is severaly affected by the low spatial resolution as underlined in \citet{Epinat:2009}. Since no clear turnover is reached in the \vfs\ and due to a symmetrization process, the center could be offset by more than 0.5\arcsec when let free, which is unrealistic. This biases the maximum velocity determination since the extent appears to be larger on the one side than on the other. The bias depends on the offset. It also changes the value of the geocentric spectroscopic redshift to the value corresponding to the center. However, the position angle of the major axis remains a stable parameter.
%For VVDS2200596913, we made several attemps described in section \ref{individual}. For VVDS220584167, as the observation is affected from a strong sky line residual (in particular the South side), we fix the center to match the kinematical center of the Nothern side (using isovelocities). The center of VVDS220014252 has been fixed to match the center of the \Ha\ line flux outer isophotes rather than the flux peak as explained in section \ref{individual}.
The centers used are displayed in Figures \ref{map1} to \ref{map9}. The maximum velocity suggested by the models is sytematically higher than the maximum velocity deduced from the maps computed as
\begin{equation}
V_{max,~map}=\frac{\textup{v}_{max}-\textup{v}_{min}}{2\times\sin{i}}
\label{vmaxmap}
\end{equation}
where $\textup{v}_{max}$ and $\textup{v}_{min}$ are the maximum and minimum values of the \vf\ and $i$ is the inclination (see Table \ref{table_tf}). Indeed, the maximum velocity of a rotating disk is observed along the major axis. Due to beam smearing effects, the velocities measured along the major axis are lowered by the contribution of off-axis regions. The beam smearing may also affect the velocity gradients observed in merging systems \vfs. For a given sampling, this effect is more pronounced for small galaxy.

\subsubsection{Dynamical masses}

%TABLE_MASSES
%\input{mass_table_v4.3.tex}

To follow mass assembly accross the cosmic time, it is worth comparing dynamical masses and stellar masses even if uncertainties for both are rather large.
%We aim at computing masses on a statistical sample in order to be able to follow the evolution of both stellar and dark or gaseous components.

Given our high resolution 3D data on rotating disk candidates, the most robust mass estimate is the enclosed mass estimate.
However a mass estimate requires several assumptions. First of all, the system has to be relaxed. The geometry of the system also has important consequences. If we consider that the mass is principally contained in spheroidal haloes, the enclosed mass $M_{\theta}$ at radius $R$ can be expressed as follow:
\begin{equation}
M_{\theta} = \frac{V^2 R}{G}
%M = \frac{V_{max}^2 R_{last}}{G}
\label{encl_mass}
\end{equation}
$V$ beeing the velocity at radius $R$.
%$R_{last}$ beeing the radius of the last point for which a velocity measurement is done, and $V_{max}$ beeing the maximum velocity reached within $R_{last}$ derived from the model fitting.% (see \ref{table_mass}).

%We may consider as well that disks may already be formed and preponderant. The enclosed mass for a highly flattened spheroid follow the equation \citep{Wright:2009}:
%\begin{equation}
%M = \frac{2V_{max}^2 R_{last}}{\pi G}
%\end{equation}
%However, a refined kinematical model taking into account the disk thickness (via the velocity dispersion for instance) would enable to correct the maximum velocity \citep{Binney:2008}.

For some of the galaxies of our sample, there is no strong evidence for rotation and their kinematics are compatible with slowly rotating spherical systems, since the velocity gradient is lower than the velocity dispersion. For spherical, bound and dynamically relaxed systems with random motions, the virial mass is more appropriate than the enclosed mass:
%We computed the virial mass for the whole sample from the intrinsic \vd~$\sigma_0$ rather than from the \vd~derived from integrated profiles, in order to separate the rotation from random motions:
\begin{equation}
M_{virial}=\frac{C\times\sigma^2~r_{1/2}}{G}
\end{equation}
The parameter $C$ depends on the mass distribution and the geometry of the system, $r_{1/2}$ is the half light radius and $\sigma$ is the mean random velocity. \citet{Binney:2008} suggest that $C=2.25$ is an average value of known galactic mass distribution models.

\begin{figure}
%\vspace{-2cm}
\begin{center}
%\resizebox{1.0\columnwidth}{!}{\includegraphics{eps/mdyn_vs_mstel.eps}}
\resizebox{1.0\columnwidth}{!}{\includegraphics{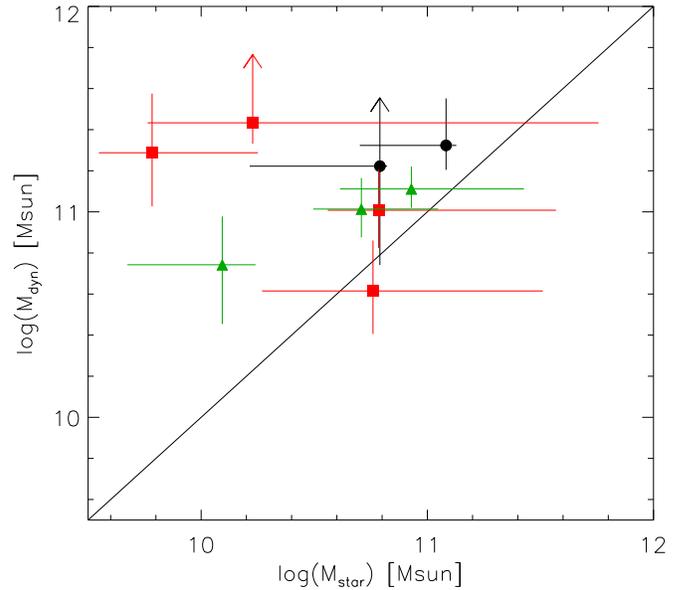}}
\end{center}
%\vspace{2cm}
\caption{Comparison between total dynamical (y-axis) and stellar (x-axis) masses in log units. Black dots, red squares, green triangles correspond respectively to objects classified (see section \ref{classif}) as rotating disks (RD), perturbed rotators (PR) and merging systems (MS).
% The maximum velocity has been determined from the rotating disk fit modeling (filled symboled) and directly from the \vf~(open symbols). 
The solid line represents $y=x$.}
\label{mdyn_mstel}
\end{figure}

\begin{figure}
%\vspace{-2cm}
\begin{center}
\resizebox{1.0\columnwidth}{!}{\includegraphics{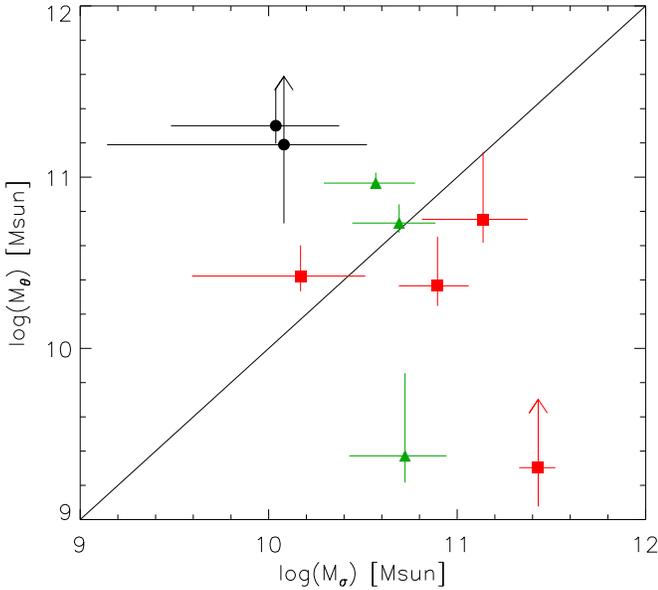}}
%\resizebox{1.0\columnwidth}{!}{\includegraphics{eps/mrotmsig_vmsig}}
\end{center}
%\vspace{2cm}
\caption{Comparison between rotation (y-axis) and dispersion (x-axis) masses in log units. Black dots, red squares, green triangles correspond respectively to objects classified (see section \ref{classif}) as rotating disks (RD), perturbed rotators (PR) and merging systems (MS).
% The maximum velocity has been determined from the rotating disk fit modeling (filled symboled) and directly from the \vf~(open symbols). 
The solid line represents $y=x$.}
\label{mrot_msig}
\end{figure}

Since our objects clearly show both rotation and dispersion motions, one might take into account both components to compute the total dynamical mass. To do that, we have to apply an ``asymmetric drift correction'' term \citep{meurer:1996} which involves radial gradients of the gas surface density, the gaseous \vd, and the disk scale height. To simplifiy, we assume that the disk scale height radial gradient is null as well as the dispersion one since we do not see clear radial dependency in our data.
%with the accuracy of our data. 
Thus, assuming that the gas is in dynamical equilibrium and that the gas velocity ellipsoid is isotropic, the total dynamical mass can be expressed as:
\begin{equation}
M_{dyn}=M_{\theta}+M_{\sigma}= \frac{V^2~R}{G} +  \frac{\sigma^2~R^3}{G~h^2}
\label{massdyntot}
\end{equation}
where $h$ is the gas surface density disk scale length when described by a gaussian function and $\sigma$ is the local \vd~at radius $R$. We note the two terms of equation \ref{massdyntot} respectively $M_{\theta}$ and $M_{\sigma}$ since they respectively refer to rotation and dispersion masses.

The first term $M_{\theta}$ is equal to the enclosed mass within radius $R$. The second term $M_{\sigma}$ is similar to the virial mass. Indeed, the gas surface density is well described by a gaussian function with $r_{1/2}=h\sqrt{2\ln{2}}$. The virial mass is thus equal to the asymmetric drift correction term at
\begin{equation}
R=(C/2\ln{2})^{1/3}~r_{1/2} \sim 1.2~r_{1/2}
\end{equation}

In Table \ref{table_mass}, we have computed $M_{dyn}$, $M_{\theta}$ and $M_{\sigma}$ at $R_{last}$, radius of the last point for which a velocity measurement is done.
%, and $V_{max}$ beeing the maximum velocity reached within $R_{last}$ derived from the model fitting.% (see \ref{table_mass}).
Since the actual rotation velocity amplitude mainly depends on the true inclination of the system, we have used the uncertainties on the inclination to compute upper and lower limits for the rotation mass computed from equation \ref{encl_mass}. We have also used the maximum velocity $V_{max}$ derived from our kinematical modeling within the radius $R_{last}$ ($V_t$ when $R_{last}>r_t$).
Since we assumed the \vd\ to be constant, we used the ``1/error''-weighted mean local \vd\ corrected for beam smearing effects $\sigma_0$ to compute the dispersion mass.
The half light radius $r_{1/2}$ has been determined for each galaxy from a two dimension gaussian fitting on line flux maps provided in Figures \ref{map1} to \ref{map9}. The half light radius measured on $I$-band CFHT images is in average about $1.1$ the \Ha\ one.
% of about 10\% in average.
For two galaxies (VVDS220584167 and VVDS220596913), since the fit on the \Ha\ line flux maps did not converge, we thus used the half light radius measured on CFHT $I$-band images divided by $1.1$.
%The half light radius $r_{1/2}$ has been determined for each galaxy from a two dimension gaussian fitting on I-band CFHT images provided in Figures \ref{map1} and \ref{map2}.
For each galaxy, $r_{1/2}$ has been corrected for the beam smearing using a quadratical subtraction of half the seeing disk. All these parameters are stored in Table \ref{table_mass} together with the dynamical classification defined in section \ref{classif} and two indicators of dynamical support ($V_{max}/\sigma_0$ and $M_{\theta}/M_{\sigma}$) on which the classification partly relies. $R_{last}$ and $r_{1/2}$ beeing approximatively linearly correlated, these two dynamical support indicators are directly linked since:
\begin{equation}
\frac{M_{\theta}}{M_{\sigma}}=2\ln{2} \left(\frac{R_{last}}{r_{1/2}}\right)^2 \left(\frac{V_{max}}{\sigma_0}\right)^2
\label{mtms_vs}
\end{equation}
The use of the mass estimates instead of $V_{max}/\sigma_0$ ratio enables to know which is the main contribution.

%\textbf{For galaxies classified as mergers (see section \ref{classif}), this method to compute the mass is more approximative than for rotating disks in particular because of both the rotation velocity and the extent measurements for the rotation mass component $M_{\theta}$ and because of the half light radius measurement for the dispersion mass component $M_{\sigma}$.
%
For galaxies classified as mergers (see section \ref{classif}), this method to compute the mass is more approximative than for rotating disks. Indeed, the rotation velocity and the extent measurements used to compute the rotation mass component $M_{\theta}$ as well as the half light radius measurement used to compute the dispersion mass component $M_{\sigma}$ are derived using disk hypothesis.
However, since it is difficult to extract unambiguously the components, this method gives a trend (see Appendix \ref{individual} for individual details).
% together with $K$-band and $B$-band absolute and apparent magnitudes (see section \ref{tullyfisher} for more details on these magnitudes).
%together with the maximum radius $R_{last}$, the maximum rotation velocity $V_{max}$, the mean \vd~corrected from beam smearing effects ($\sigma_0$), the stellar, virial and enclosed masses ($M_{star}$, $M_{virial}$ and $M_{enclosed}$) and finally $K$-band and $B$-band absolute and apparent magnitudes (see section \ref{tullyfisher} for more details on these magnitudes).

As shown in Figure \ref{mdyn_mstel}, the total dynamical mass is higher or equal than the stellar mass within the uncertainties.
%, except for VVDS220544103.
The difference between dynamical and stellar masses could be even larger if for some galaxies the turnover in the rotation curve is not reached or if the extent of the stellar distribution is much larger than the gas one.
This trend shows that in most of the cases, the gravitational potential is mainly due to gas and dark matter.

%Perturbed rotators (dynamical classes are defined in section \ref{classif}) have stellar masses around $10^{10}\msun$ and dynamical masses larger by a factor around five.
Rotating disks (dynamical classes are defined in section \ref{classif}) have both large dynamical and stellar masses (around $10^{11}\msun$) while mergers and perturbed rotators present a larger mass range. The comparison between dynamical and stellar masses is discussed for each galaxy in Appendix \ref{individual}.

In Figure \ref{mrot_msig} comparing rotation and dispersion masses, rotation-dominated systems are clearly distinguished from dispersion-dominated ones. Indeed, on the one hand, for the two galaxies classified as rotating disks (VVDS220584167 and VVDS220015726), the rotation mass is higher than the dispersion mass by a factor larger than 13 showing that their gravitational support is rotation dominated. On the other hand, VVDS020116027 and VVDS020147106 are clearly dominated by random motions ($M_{\sigma}/M_{\theta} > 10$) even if inclination effects may lead to slightly higher values. The five remaining galaxies with $M_{\theta}/M_{\sigma}\sim 1$ show both rotation and dispersion gravitational supports.

%For VVDS020116027, VVDS020147106 and VVDS220014252, the dispersion mass is higher than the rotation mass. For two of them this can be due to a wrong inclination determination (VVDS020116027, VVDS020147106) as suggested by the upper errors on the rotation mass. 
%For the two galaxies classified as rotating disks (VVDS220584167 and VVDS220015726), the rotation mass is higher than the dispersion mass by a factor larger than 13 showing that their gravitational support is rotation dominated.
% This underlines our dynamical classifiaction.

%Assuming a virialized motion around the morphological galaxy center, such as a circularly rotating disk, the total mass within $R$ for which the maximum rotational velocity $V_{max}$ is reached, is roughly described by:
%\begin{equation}
%M_{vir} = V_{max}^2 R/G
%\end{equation}
%Nowadays, at high redshift, the dynamical mass derived from spatially resolved emission line can only yield the mass within the largest radius for which a velocity is measured.
% In the context of our data, for each galaxy the maximum velocity

%XXX AJOUTER UNE ESTIMATION DE LA MASSE DE gas?

\subsubsection{Halo masses}

\begin{figure*}
%\vspace{-6.7cm}
\begin{center}
\resizebox{2.0\columnwidth}{!}{\includegraphics{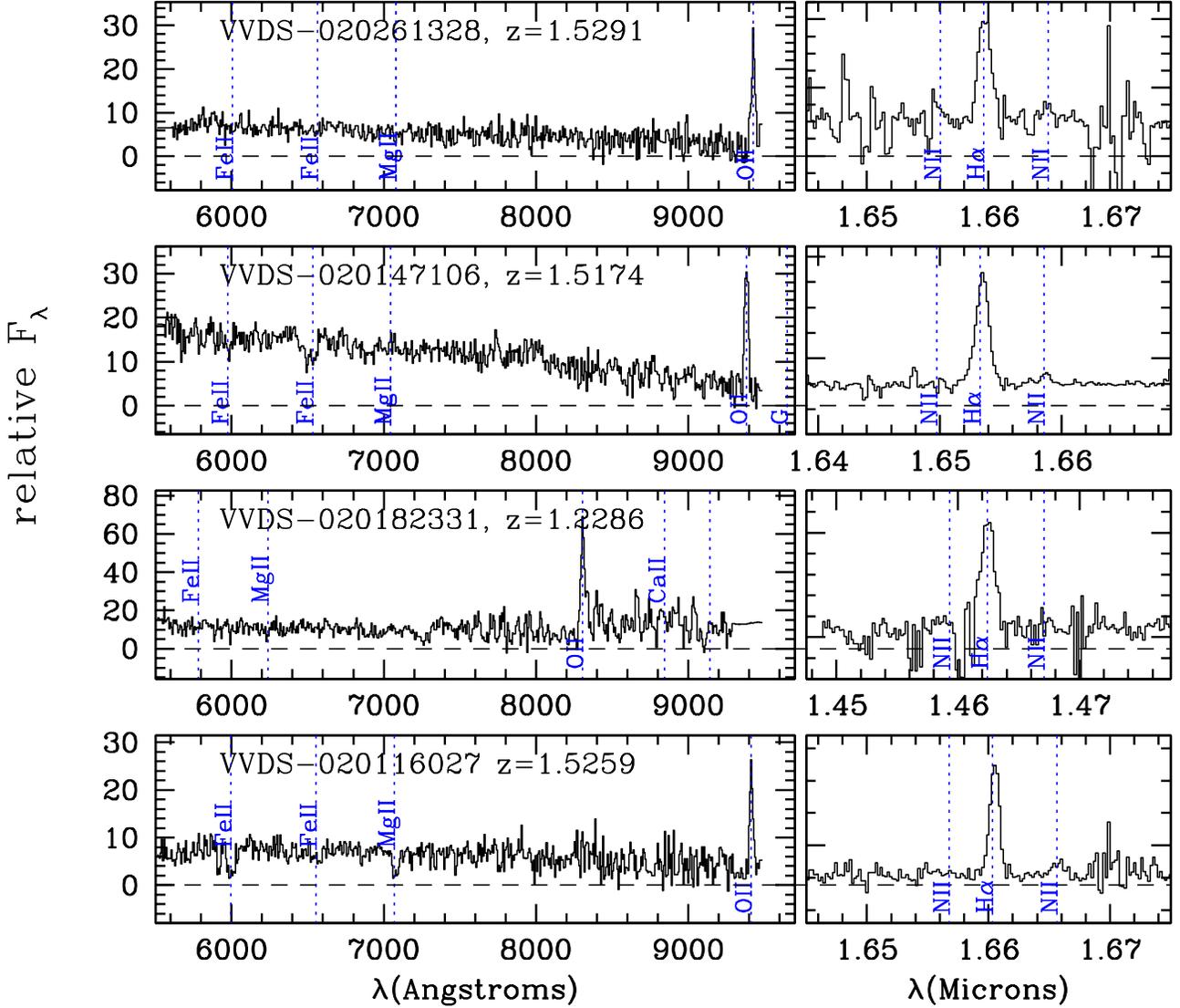}}
\end{center}
%\vspace{3.5cm}
\caption{VIMOS optical spectra from which the original redshifts were derived (left) and SINFONI (right) 1D spectra around the \Ha~line for the four VVDS-02h field galaxies. SINFONI spectra have been produced summing spectra of all spatial resolution elements with detected \Ha~emission. The main emission and absorption lines are marked and labeled for each sources. The line wavelengths are computed using the redshifts deduced from SINFONI data (Table \ref{table_kin}). }
\label{profils1}
\end{figure*}
\begin{figure*}
%\vspace{-5cm}
\begin{center}
\resizebox{2.0\columnwidth}{!}{\includegraphics{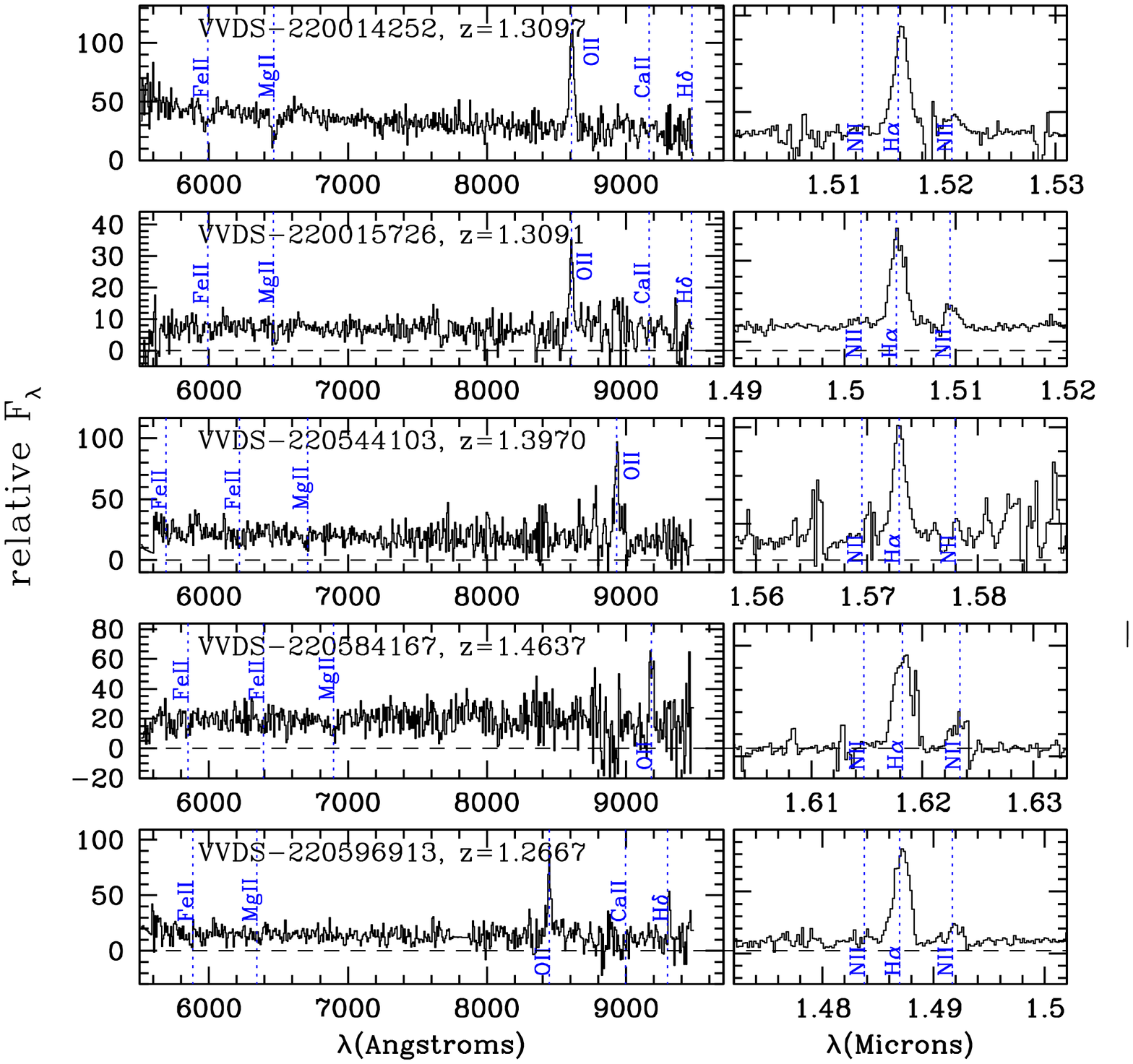}}
\end{center}
%\vspace{4.cm}
\caption{Similar as Figure \ref{profils1}  for the five VVDS-22h field galaxies.}
\label{profils2}
\end{figure*}

We have estimated the total halo mass ($M_{halo}$) using a spherical virialized collapse model \citep{Peebles:1980,White:1991,Mo:2002}. We assumed that the plateau has been reached in our observations and that the plateau velocity $V_{max}$ traces the modeled halo circular velocity.
\begin{equation}
M_{halo}=0.1~H_0^{-1}~G^{-1}~\Omega_m^{-0.5}~(1+z)^{-1.5}~V_{max}^3
\end{equation}
where $H_0$ is the local Hubble constant, $\Omega_m$ is the local matter density, $G$ is the universal gravitational constant and $z$ is the redshift of the galaxy.

In the cases where our models indicate that the plateau is not reached within the observation, we have used the maximum velocity derived from our kinematical modeling within the radius $R_{last}$ instead of the plateau velocity of the model $V_t$ since the plateau is not constrained correctly. The values are available in Table \ref{table_mass}.

If we assume that the length of the halo is the size that a galaxy should have in order that its enclosed mass is equal to the halo mass, then for our sample, the halo length is $113\pm73$. It is larger for fast rotating disks since:
\begin{equation}
R_{halo}=0.1~H_0^{-1}~\Omega_m^{-0.5}~(1+z)^{-1.5}~V_{max}
\end{equation}

%Typical halo masses at the last HI radius for massive local galaxies is around $25\times 10^{10}\msun$ (from galaxies larger than 18 kpc in \citealp{Blais-ouellette:2001,Spano:2008}).
%
These total halo masses have no physical meaning for galaxies which are not gravitationally supported by rotation. For galaxies having a ratio $V_{max}/\sigma_0$ larger than unity, the total halo masses range from 0.2 to 6.0$\times10^{12}\msun$. These values are much larger than the measured dynamical masses. At redshift $z=1.5$, a galaxy with the same maximum rotation velocity than a $z=0$ galaxy is supposed to have a total halo mass lower by a factor $\sim4$. Our rotation dominated sources have $M_{\theta}/M_{halo}\sim0.049\pm0.015$ in average. This value cannot be directly compared to the fraction of baryonic matter to total matter $\Omega_b/\Omega_m=0.171$ \citep{Komatsu:2009} because $M_{\theta}$ may contain a noticeable fraction of dark matter. Indeed, for comparison, typical halo masses measured from mass model derived from rotation curves at the last HI radius for nearby galaxies larger than 18 kpc is around $25\times 10^{10}\msun$ \citep{Blais-ouellette:2001,Spano:2008}. Taking into account that the values computed for local galaxies are measured, in average, at a radius twice as large as the sample of high redshift galaxies, the total cosmological halo mass for high redshift galaxies are still much larger than the total halo mass measured for local galaxies from rotation curves.

%XXX It would be intersting to compare to simulations
%XXX Compare with results of \citet{Forster-Schreiber:2006,Wright:2009}.
%XXX I almost copied/pasted the paragraph of \citet{Wright:2009}...

\subsection{Search for type 2 AGN and metallicity}
\begin{figure}
%\vspace{-4.cm}
%\vspace{-2.2cm}
\begin{center}
\resizebox{1.0\columnwidth}{!}{\includegraphics{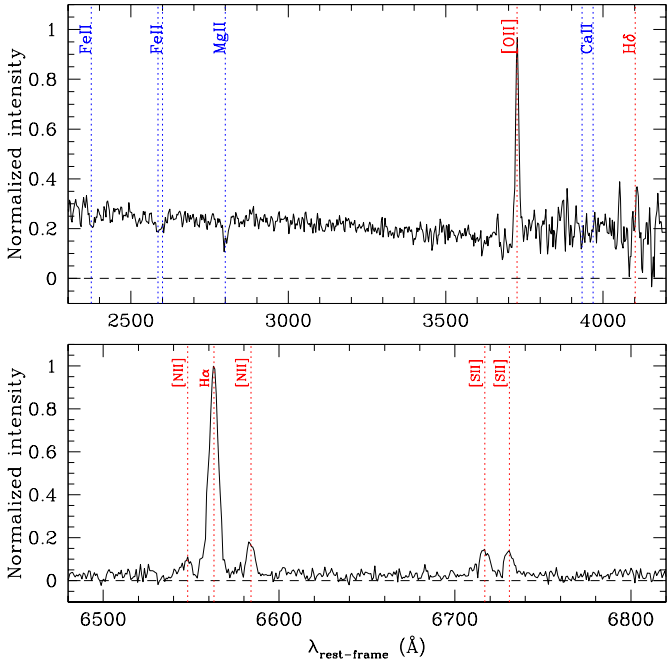}}
%\resizebox{1.0\columnwidth}{!}{\includegraphics{eps/ha_rest.ps}}
\end{center}
%\vspace{1.8cm}
%\vspace{0cm}
\caption{Combined spectrum of the nine VVDS galaxies at $1.2 < z < 1.6$
observed with VIMOS (top) and SINFONI (bottom), in arbitrary flux units and rest frame wavelengths. The position of the main emission and absorption lines are indicated in red and blue respectively.}
%Combined spectrum of the 9 sources observed, in arbitrary flux units and rest frame wavelengths.}
\label{ha_rest}
\end{figure}

Along with the VIMOS optical spectra, we show in Figures \ref{profils1} and \ref{profils2} the integrated 1D SINFONI spectra 
around the H$\alpha$ line, resulting from the sum of all spatial resolution elements in the datacubes at the  position of the 
sources.
VIMOS and SINFONI combined spectra of the nine observed galaxies are shown in Figure \ref{ha_rest}. 
These are median spectra that have been computed using respectively [O\,{\sc ii}] flux and \Ha~flux to normalize each individual spectrum.
%It has been computed using \Ha~flux to normalize each spectrum, and computing the median spectra.
%average spectrum with a rejection.
One can clearly identify \niia\ and \nii\ emission lines, as well as the [S\,{\sc ii}] doublet at 6717 
and 6731\AA\ in the SINFONI combined spectrum. The low \nii/H$\alpha$ and \sii/H$\alpha$ ratios 
($= 0.13$ and $0.23$ respectively) indicate the lack of AGN contribution in our sample. We further performed a careful 
inspection of the spatial distribution of the \nii/H$\alpha$ ratio for each galaxy. \nii\ line flux maps were derived similarly to \Ha\ ones (see section \ref{hakin}). However, since the \nii\ flux is rather weak, we used the \Ha\ line to constrain the line position.
We do not find any significant peak in the flux ratio maps, contrary to \citet{Wright:2009} who found an AGN in two of their six $z\sim 1.6$ galaxies.
This lack of AGN contribution is further supported by other indicators.
\begin{itemize}
 \item the quite low mean value of reddening ($E(B-V)\sim 0.3$ or $A_V \sim 1$) measured in our sample galaxies is against any dusty AGN alternative
 \item we do not detect any strong [Ne\,{\sc iii}] $\lambda$3869 or [Ne\,{\sc v}] $\lambda$3426 emission in the VIMOS composite spectrum which would be a clear signature of Sey2-like activity (e.g. \citealp{Zakamska:2003})
 \item no evidence for Sey2-like activity is found in X-ray data. Indeed, we have checked in the HEASARC archive that no X-ray counterpart was yet detected at less than 30\arcsec\ from our objects. Moreover, when existing, the counterparts are closer to brighter objects.
\end{itemize}
% This lack of AGN contribution is further supported by the VIMOS composite spectrum. Indeed we do not detect any strong [Ne\,{\sc iii}] $\lambda$3869 or [Ne\,{\sc v}] $\lambda$3426 emission which would be a clear signature of Sey2-like activity (e.g. \citealp{Zakamska:2003}). Moreover, the quite low mean value of reddening ($E(B-V)\sim 0.3$ or $A_V \sim 1$) measured in our sample galaxies is against any dusty AGN alternative.
%Moreover, VIMOS spectra strenghten this results since no neon line (3871\AA) is observed.
Note also that type 1 AGN are excluded from the parent VVDS sample on the basis of the presence of broad emission lines in the VIMOS spectrum.

The \nii/H$\alpha$ ratio can be used to estimate roughly the typical metallicity of our sample. Using the last calibrations proposed by \citet{Perez-Montero:2009}, we have estimated a rather low oxygen abundance $12+log(O/H)=8.37$, corresponding to half the solar metallicity (see \citealp{Queyrel:2009} for a full discussion on the mass-metallicity relation).

% 6548: 4.605e-5
% 6563: 6.416e-4
% 6584: 8.225e-5
% 6717: 8.296e-5
% 6731: 6.541e-5
% 6584/6563=0.13 - log(6584/6563)=-0.89
% (6717+6731)/6563=0.23 - log((6717+6731)/6563)=-0.63
% 12+log(O/H)=8.37 (PMC08)

\section{Results}

\subsection{Kinematical classification}
\label{classif}

\begin{figure}
\begin{center}
\resizebox{1.0\columnwidth}{!}{\includegraphics{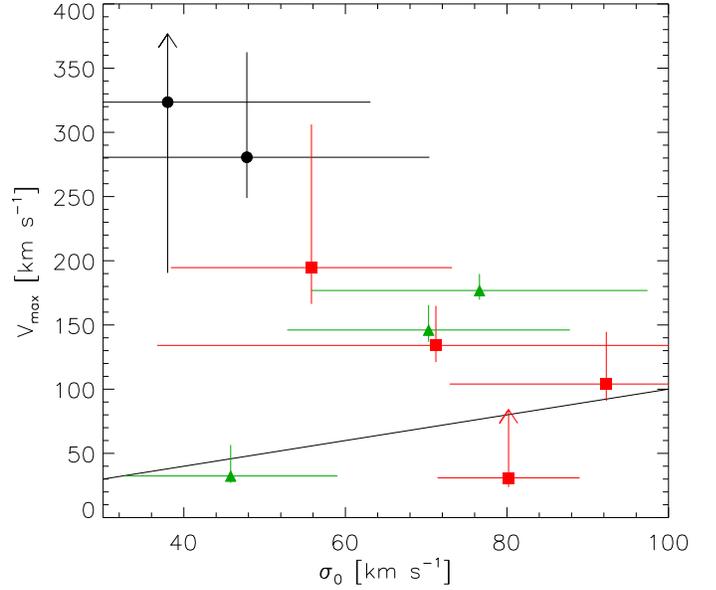}}
\end{center}
\caption{Maximum rotation velocity $V_{max}$ deduced from rotating disk model fitting as a function of ``1/error''-weighted mean local \vd\ corrected for beam smearing effects $\sigma_0$. Black dots, red squares, green triangles correspond respectively to objects classified as rotating disks (RD), perturbed rotators (PR) and merging systems (MS). The solid line represents $y=x$.}
\label{classification}
\end{figure}

%\begin{figure}
%\begin{center}
%\resizebox{1.0\columnwidth}{!}{\includegraphics{eps/masses_frac_rot_sig.eps}}
%\end{center}
%\caption{Black dots, red squares, green triangles correspond respectively to objects classified as rotating disks (RD), perturbed rotators (PR) and merging systems (MS).}
%\label{mass_frac}
%\end{figure}

\begin{figure*}
\begin{center}
%\resizebox{2.0\columnwidth}{!}{\includegraphics{eps_figures/maps_VVDS020116027model.eps}}
%\resizebox{2.0\columnwidth}{!}{\includegraphics{eps_figures/maps_VVDS020182331model.eps}}
%\resizebox{2.0\columnwidth}{!}{\includegraphics{eps_figures/maps_VVDS020147106model.eps}}
%\resizebox{2.0\columnwidth}{!}{\includegraphics{eps_figures/maps_VVDS020261328model.eps}}
%\resizebox{2.0\columnwidth}{!}{\includegraphics{eps_figures/maps_VVDS220596913model.eps}}
%\resizebox{2.0\columnwidth}{!}{\includegraphics{eps_figures/maps_VVDS220584167_Hmodel.eps}}
%\resizebox{2.0\columnwidth}{!}{\includegraphics{eps_figures/maps_VVDS220544103_Hmodel.eps}}
\resizebox{2.0\columnwidth}{!}{\includegraphics{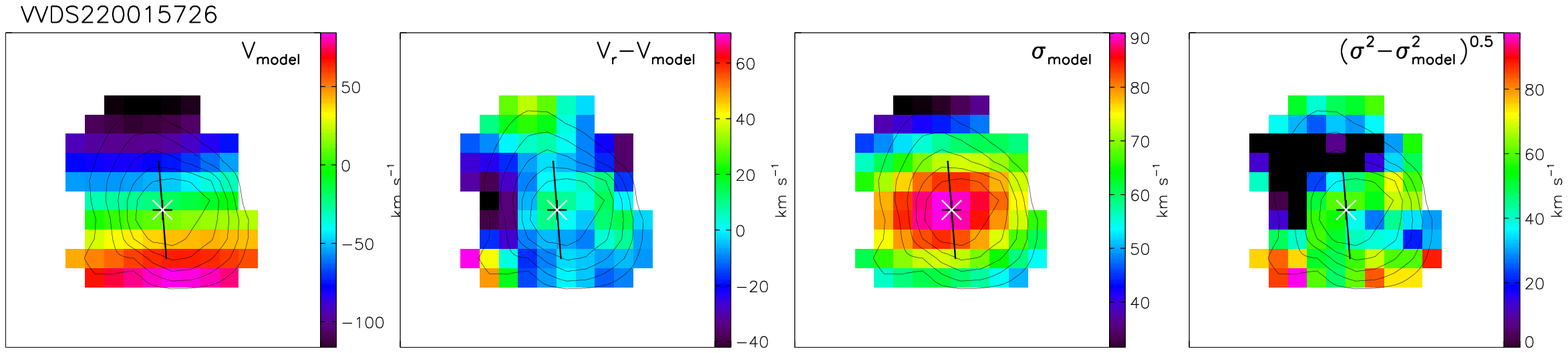}}
\resizebox{2.0\columnwidth}{!}{\includegraphics{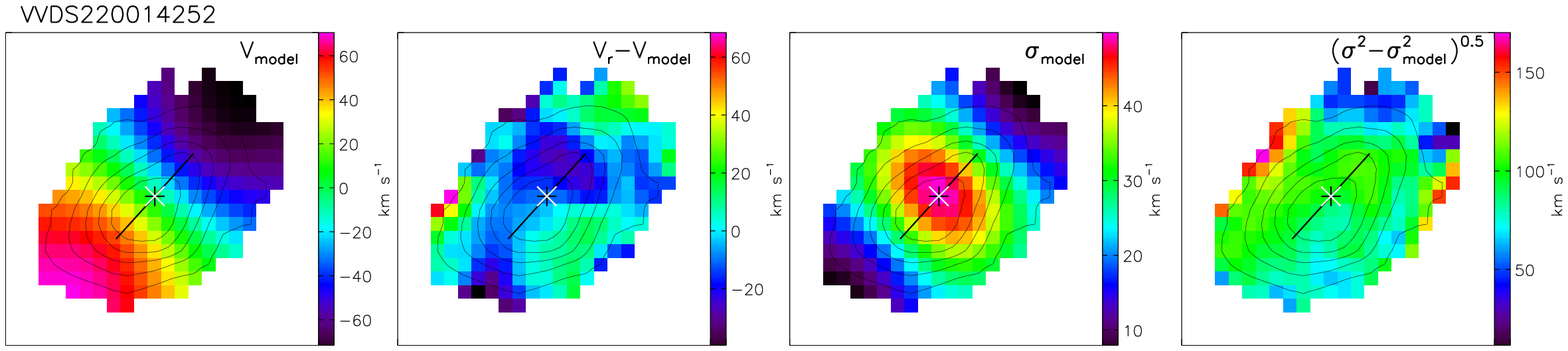}}
\end{center}
\caption{From left to right: Model \vf, \vf~residuals, model \vdm~and \vdm~quadratic residuals.
North is up, East is left. The center of the models is marked on each kinematical map with a black and white double cross. The solid line represents the orientation of the kinematical major axis and ends at half light radius measured on I-band images. The flux contours are overplotted.}
%XXX CONTOURS, MAJOR AXIS.}
%, the direction of North-East, the VVDS identification and VIMOS-based redshift are indicated for each galaxy.}.
\label{models}
\end{figure*}

It has already been observed by other authors studying $z>1$ star-forming galaxies with 3D-spectroscopy
\citep{Law:2007,Wright:2007,Wright:2009,Genzel:2006,Genzel:2008,Forster-Schreiber:2006,Cresci:2009} that the gaseous velocity dispersion of these objects is larger than at lower redshifts \citep{Yang:2008}.
%\citep{Law:2007,Wright:2007,Wright:2009,Genzeletal:2006,Genzeletal:2008,Forster-Schreiberetal:2006}, the gaseous velocity dispersion of these objects is larger than at lower redshifts \citep{Yangetal:2008}.
As shown from kinematical modeling, an unresolved velocity gradient cannot account for high \vd~values (\citealp{Epinat:2009}). The kinematical classification as performed by \citet{Flores:2006} on FLAMES/GIRAFFE data relies mainly on the ability to observe a signature of unresolved velocity gradient in the \vdm: rotating disks are defined as systems showing a regular velocity gradient and a central peak in the \vdm, where the actual velocity gradient is the steepest; systems with an offset \vd~peak from the center, or without peak, are defined as perturbed rotators; the galaxies with no regular velocity gradient and no \vd~central peak are classified as galaxies having complex kinematics.
%For high local gaseous dispersion systems
For systems with a high local velocity dispersion of the ionized gas, the \vd~argument is thus not valid anymore. In Figure \ref{models}, we show examples of rotating disk models for two objects from our sample illustrating this point. For VVDS220015726 (first line) and VVDS220014252 (second line), the model \vfs~(first column) show that the kinematics of these objects reasonably match rotating disks (the second column shows the residual \vfs). For VVDS220015726, the signature in the \vdm~(third column) correctly matches the unresolved velocity gradient signature since the \vdm~central signature (see Figure \ref{map2}) is well removed when subtracting quadratically the model (last column of Figure \ref{models}). For VVDS220014252 which has a higher \vd, the model \vdm\ absolutely does not account for the observed \vd\ as demonstrated by the resemblance between corrected (last column of Figure \ref{models}) and uncorrected (see Figure \ref{map2}) \vdm s.

The $\sim$0.65\arcsec~seeing continuum images from CFHT do not allow a morphological classification as accurate as performed by \citet{Neichel:2008}
%The $\sim$0.65\arcsec~seeing continuum images from CFHT do not allow an accurate morphology classification such as done by \citealp{Neicheletal:2008}
 from HST data on IMAGES sample. However, relying on line flux morphology, the kinematical maps derived from our SINFONI data and from attempts to fit a rotating disk model, we are able to derive a first order kinematical classification.

We defined rotating disks (RDs) as systems for which the rotation mass is higher than the dispersion mass (rotation dominated systems, see the two parameters $V_{max}/\sigma_0$ and $M_{\theta}/M_{\sigma}$ linked by equation \ref{mtms_vs} and given in Table \ref{table_mass}) as shown in Figure \ref{mrot_msig} and for which both \vf\ and \vdm\ are described by rotating disk models.
Mergers systems (MSs) are defined as systems with two spatially resolved components in the \Ha\ map and for which the \vf\ is not described accurately by rotating disk models.
Other systems are classified as perturbed rotators (PRs). They mainly show high \vd s, higher than around 60\kms, and their \vfs\ show some peculiarities not described by rotating disks.

%A clumpy flux distribution such as for VVDS220596913 associated with a \vf~that could be interpreted as two disks may be interpreted as close-by companions that are in a process of merging (ie. a merging system). It is also the case for VVDS020116027 and VVDS220544103. However, VVDS020116027 is quite different from the two other systems showing lower intrinsic \vd.

%High rotational velocity systems with a relatively low velocity dispersion and a central peak as for
%VVDS220584167 or VVDS220015726 have to be interpreted as rotating disks. These two galaxies have the highest $\chi^2$ because their \vfs~show the highest values.

%The other systems presenting peculiarities in their \vf~or in their \vdm~or with high \vd~are classified as perturbed rotators. Indeed, a rotating disk with a high intrinsic \vd~could account for thick star forming accreting disks.

Figure \ref{classification} shows a correlation between the maximum rotation velocity $V_{max}$ and the ``1/error''-weighted mean local \vd\ corrected for beam smearing effects $\sigma_0$, in particular for rotating disks and perturbed rotators.
The most massive galaxies (see Table \ref{table_mass}) are our two RDs.
This would imply that the most massive disks are stable earlier or that stable disks can already be formed at these redshifts. Random motions convert in organized motions (rotation) more rapidly for massive systems.

From our classification, we obtain three clear major merging systems, two clear rotating disks and four perturbed rotators that could be either perturbed by minor merging events or by continuous gas accretion.
The classification is summarized in column (10) of Table \ref{table_mass} and individual detailed comments on the kinematics and dynamics of each galaxy are provided in Appendix \ref{individual}.

%Three objects are compatible with dispersion dominated systems, however, it is worth noting that the two objects with very low maximum rotation velocities could be nearly face on objects.
%Since the maximum velocity is a mass indicator for rotating disks, 

%It is interesting to note that for our sample, we see a correlation between the maximum rotation velocity and the intrinsic local velocity dispersion $\sigma_0$ (Figure \ref{classification}). Since the maximum velocity is a mass indicator for rotating disks, this would imply that the most massive disks are stable earlier or that stable disks can already be formed at these redshifts. Random motions convert in organised motions (rotation) more rapidly for massive systems.

%The fact that the RDs rotation mass is preponderant underlines our dynamical classifiaction.

%\subsection{Properties of different kinematical types}

\subsection{The Tully-Fisher relation}
\label{tullyfisher}

\begin{table}
\caption{Magnitudes and maximum velocities} 

\begin{center}
\begin{tabular}{ccccc}
\hline
VVDS ID & K$_{AB}$ & B$_{AB}$ & $V_{max,~model}$ & $V_{max,~map}$ \\
   & mag & mag & \kms  & \kms \\
 (1)  & (2) & (3) & (4)  & (5) \\
\hline
020116027 & -22.7 $\pm$ 0.4 & -21.8 $\pm$ 0.2  &  32 $^{+ 29}_{-  5}$ &  23 $^{+ 21}_{-  4}$ \\
020182331 & -22.3 $\pm$ 0.2 & -21.7 $\pm$ 0.1  & 134 $^{+ 32}_{- 14}$ & 114 $^{+ 27}_{- 12}$ \\
020147106 & -22.8 $\pm$ 0.1 & -22.1 $\pm$ 0.1  &  30 $^{+\infty}_{-  7}$ &  21 $^{+\infty}_{-  5}$ \\
020261328 & -21.6 $\pm$ 1.0 & -21.1 $\pm$ 0.3  & 194 $^{+125}_{- 31}$ &  82 $^{+ 53}_{- 13}$ \\
220596913 & -23.2 $\pm$ 0.5 & -22.6 $\pm$ 0.1  & 177 $^{+ 13}_{-  7}$ & 153 $^{+ 11}_{-  6}$ \\
220584167 & -24.2 $\pm$ 0.5 & -23.1 $\pm$ 0.1  & 280 $^{+ 81}_{- 31}$ & 198 $^{+ 57}_{- 22}$ \\
220544103 & -23.2 $\pm$ 0.5 & -22.5 $\pm$ 0.2  & 146 $^{+ 19}_{-  9}$ & 108 $^{+ 14}_{-  6}$ \\
220015726 & -23.2 $\pm$ 0.5 & -22.1 $\pm$ 0.1  & 323 $^{+\infty}_{-135}$ & 189 $^{+\infty}_{- 79}$ \\
220014252 & -23.3 $\pm$ 0.5 & -22.6 $\pm$ 0.2  & 103 $^{+ 39}_{- 13}$ &  84 $^{+ 32}_{- 11}$ \\
\hline
\label{table_tf}
\end{tabular}

(1) Source VVDS identification number,
(2) AB absolute magnitude in rest frame $K$-band,
(3) AB absolute magnitude in rest frame $B$-band,
(4) Maximum velocity derived from the model,
(5) Maximum velocity derived from the maps (equation \ref{vmaxmap}).
\end{center}
\end{table}
%         &  1.2 $^{+ 1.7}_{- 0.5}$
%         &  5.8 $^{+ 1.0}_{- 1.9}$
%         &  1.7 $^{+ 0.1}_{- 0.6}$
%         &  0.6 $^{+ 1.8}_{- 0.4}$
%         &  8.5 $^{+ 2.7}_{- 4.1}$
%         & 12.1 $^{+10.9}_{- 5.0}$
%         &  5.1 $^{+ 2.3}_{- 3.1}$
%         &  6.2 $^{+ 5.7}_{- 1.6}$
%         &  6.1 $^{+ 1.0}_{- 3.6}$
% (4) Stellar mass,
% (5) Maximum velocity derived from the model,
% (6) Maximum velocity derived from the maps (equation \ref{vmaxmap}).

\begin{figure}
\begin{center}
\resizebox{1.0\columnwidth}{!}{\includegraphics{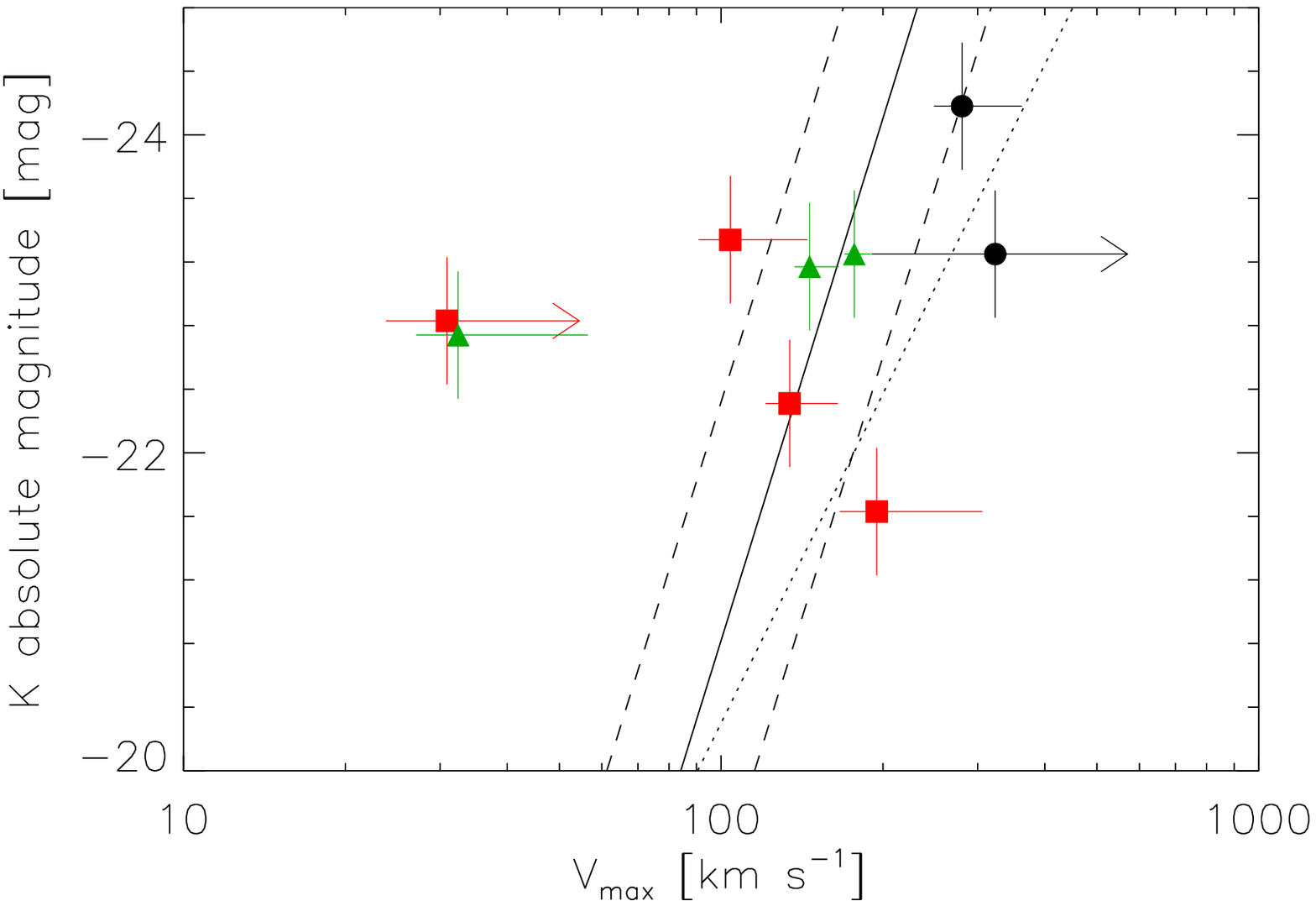}}\\
\resizebox{1.0\columnwidth}{!}{\includegraphics{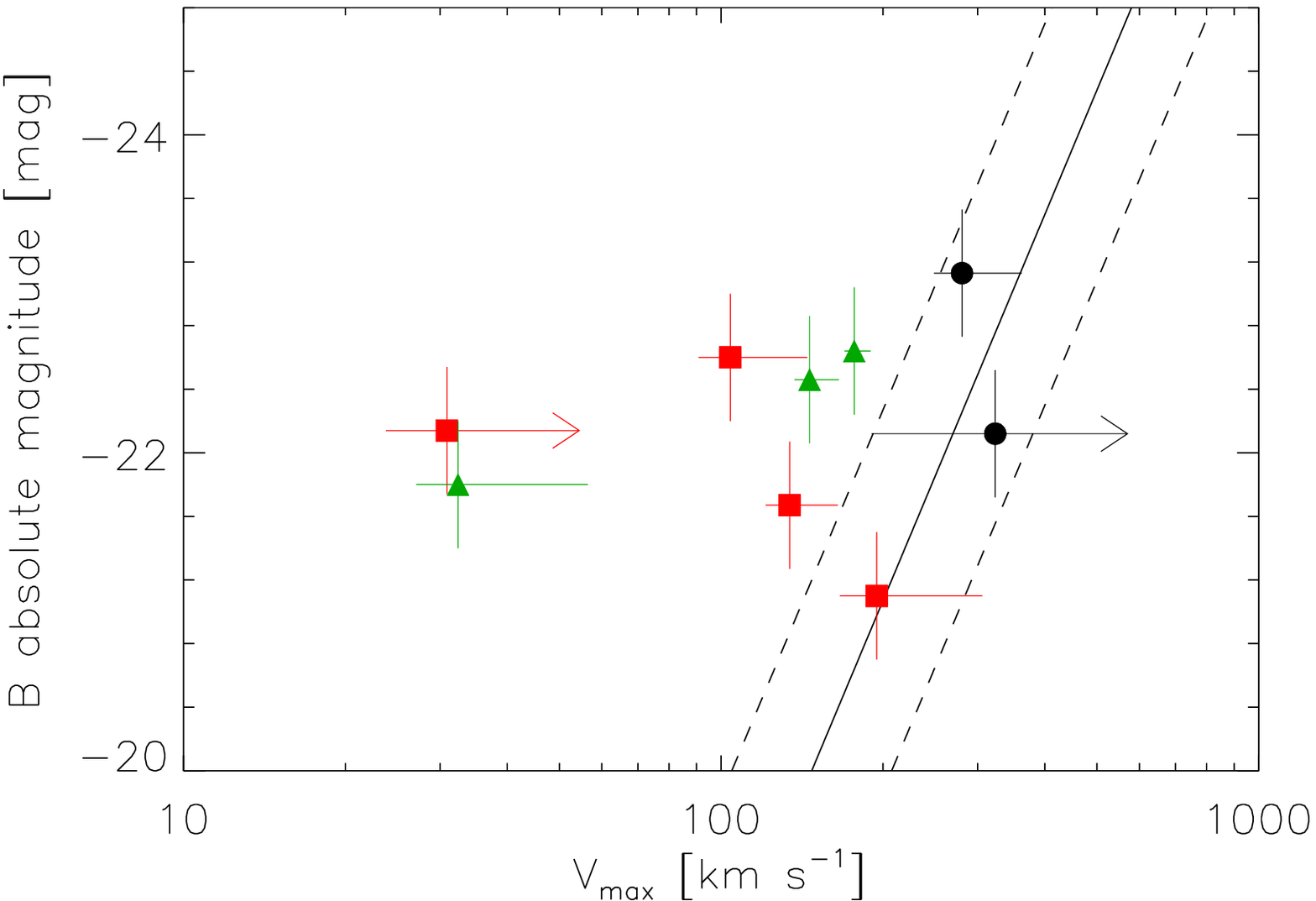}}\\
\resizebox{1.0\columnwidth}{!}{\includegraphics{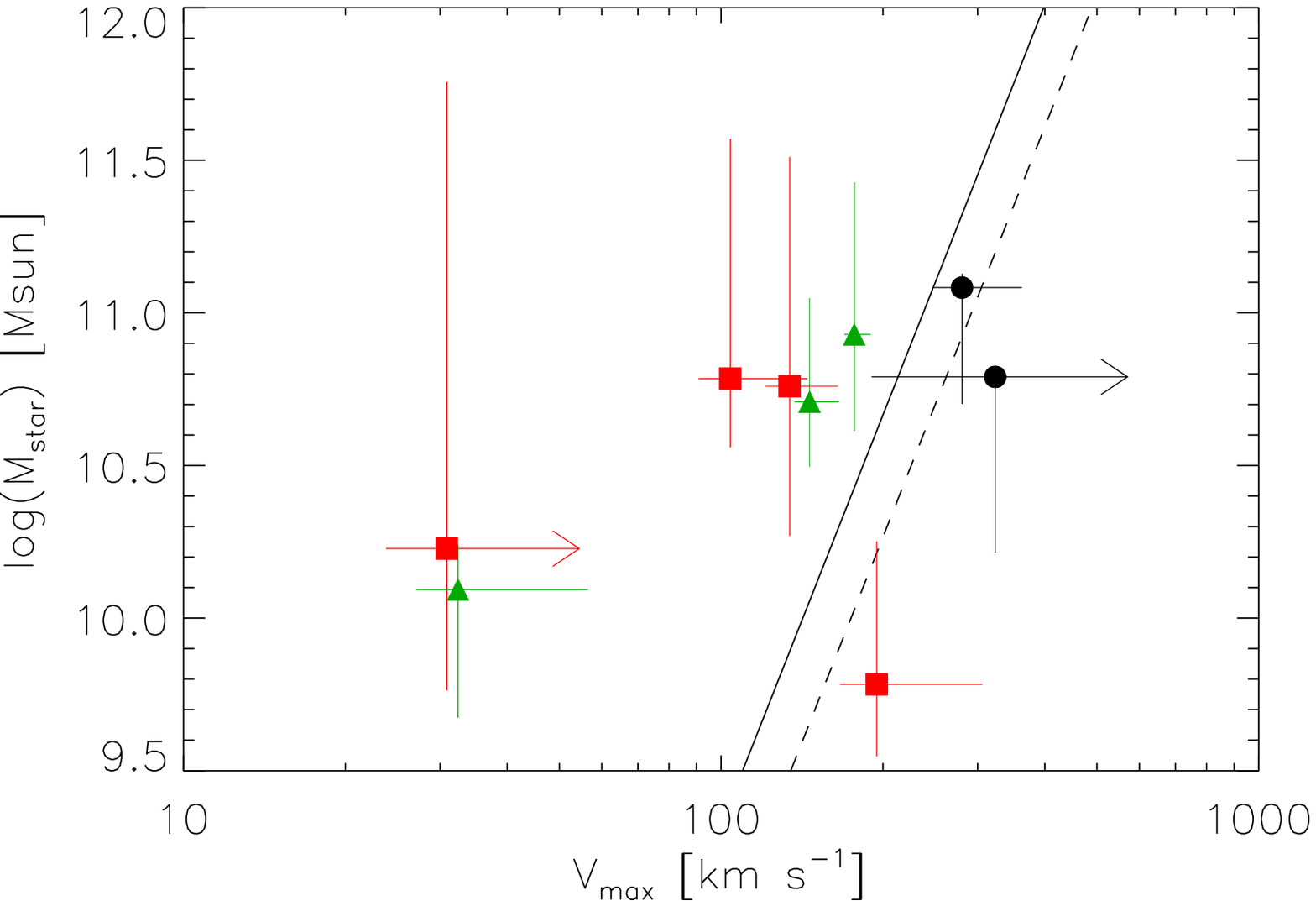}}
\end{center}
\caption{$K$-band (top), $B$-band (middle) and stellar mass (bottom) TFRs for our nine $z\sim1.4$ galaxies.
Black dots, red squares, green triangles correspond respectively to objects classified (see section \ref{classif}) as rotating disks (RD), perturbed rotators (PR) and merging systems (MS). The maximum velocity has been determined from the rotating disk fit modeling.
\textbf{Top:} the dotted line represents the local reference $K$-band TFR used by \citet{Puech:2008} derived from \Ha\ on a SDSS sub-sample.
\textbf{Top and middle:} the solid line represents respectively the local $K$-band and $B$-band TFRs derived from a local HI sample by \citet{Meyer:2008} whereas the dashed lines represent the dispersion of the HI galaxies around these local relations.
\textbf{Bottom:} the solid line represents the $z=0$ TFR derived by \citet{Bell:2001}. The dashed lines represent the $z=2.2$ TFR derived from SINS data by \citet{Cresci:2009}.
% (filled symboled) and directly from the \vf~(open symbols).
}
\label{tfkbm}
\end{figure}

The Tully-Fisher relation (hereafter TFR) \citep{Tully:1977} links the maximum rotation velocity to the actual flux or mass of a galaxy (via their logarithm). Extensive work has been produced in order to probe the evolution of the TFR with cosmic time in both $B$-band and $K$-band up to redshifts $z\sim1.4$ (e.g. \citealp{Fernandez-Lorenzo:2009} for references). Since it requires a large sample of galaxies (more than $\sim$50 galaxies), high redshift TFR were derived from long-slit spectroscopy samples, except the ones derived by
(i) \citet{Flores:2006} and \citet{Puech:2008}
%\citet{Flores:2006,Puechetal:2008}
from FLAMES/GIRAFFE IFUs data at $z\sim 0.6$;
(ii) \citet{Cresci:2009} from SINFONI data at $z\sim 2.2$ and
(iii) \citet{Swinbank:2006} from lensed galaxies at $z\sim 1$ observed with GMOS.
%In this section, we aim at initiating the TFR study at redshifts $z\sim1.4$, using SINFONI integral field spectroscopy. This study will be completed with the MASSIV sample which is beeing constituted with representative objects from VVDS catalogs.

In the frame of the present work, due to the small number of objects, we only look at the distribution of the different kinematical types accross the local TFR. We present the local TFR derived from HI data by \citet{Meyer:2008} (solid line) since they provide both $K$-band (Figure \ref{tfkbm}, top) and $B$-band (Figure \ref{tfkbm}, middle) relations.
\citet{Puech:2008}
%\citet{Puechetal:2008}
used a subsample of SDSS sample \citep{Pizagno:2007} to derive a local reference TFR. The latter has been derived using $V_{80}$ velocities instead of maximum velocities (from \Ha~kinematics) since they argue that their maximum velocity is statistically not measured
(see \citealp{Puech:2008}
%(see \citealp{Puechetal:2008}
and \citealp{Pizagno:2007} for details). We thus also plot this $K$-band reference TFR in Figure \ref{tfkbm}, top (dotted line).

For local samples with \Ha\ kinematical data, usually one has to exclude galaxies with low inclination because the uncertainty on their maximum velocity is high, but also galaxies with high inclination because of dust extinction in the disk that can bias the maximum velocity determination. %\citep{Epinat:2008b}.
For high redshift, as mentioned in section \ref{morphologies}, the inclination is poorly known, thus we expect to observe both face-on and edge-on systems that we are not able to identify.

Absolute magnitudes in the desired rest-frame band (using the appropriate filter response curve) and stellar masses with their associated error bars for our objects were obtained from the same spectral energy distribution modeling than described in section \ref{seds} using the GOSSIP software \citep{Franzetti:2008}. Stellar masses and the associated errors are provided in Table \ref{table_mass} whereas rest-frame $B$- and $K$-band absolute magnitudes and their associated errors determined from the PDF are provided in Table \ref{table_tf}. Uncertainties on $K$-band magnitudes are larger than for $B$-band. They result from the lack of any measurement of $K$ rest-frame ($\sim3.5\mu$m) to constrain efficiently the SED fitting.
%the GOSSIP spectral energy distribution (SED) modeling software \citep{Franzetti:2008} using the same data and procedure than described in section \ref{seds}.}
%
%We used as input for the SED fitting the multi-band photometric observations available in the VVDS fields, including $BVRI$ data from the CFHT, $UBVRZs$ data from the CFHT Legacy Survey, $J$- and $K$-bands data from SOFI at the NTT and from the UKIDSS survey, and the VVDS-Deep spectra. The photometric and spectroscopic data were fitted with a grid of stellar population models, generated using the PEGASE2 population synthesis code \citep{Fioc:1997}, assuming a set of ``delayed'' star formation histories (see \citealp{Gavazzi:2002} for details), and a \citet{Salpeter:1955} initial mass function. 
% together with the apparent magnitudes that have been computed back from apparent magnitudes.

%A large uncertainty on $K$-band magnitudes results from the lack of any measurements near $K$ rest frame ($\sim3.5\mu$m) to constrain efficiently the SED fitting. For VVDS-02h field, $K$-band magnitudes have been determined using several methods. We used the dispersion over these methods to have an estimate of the error made on the absolute magnitudes determination ($\pm 0.4$).

In Figures \ref{tfkbm} top and middle, we respectively show the $K$- and $B$-band TFR for our nine $z\sim1.4$ objects. The maximum velocities $V_{max}$ have been derived from the rotating disk models. $V_{max}$ is systematically higher when coming from the models that take into account the beam smearing effects (see Table \ref{table_tf}).
% (filled symbols) but also directly deprojected from the \vfs~(open symbols). 
On these Figures, to distinguish the three kinematical types we affected different colors and symbols (see caption).
%In the next paragraph, we comment the velocities obtained by the models.
We see that the RD galaxies follow accurately the local $B$-band TFR as well as the PR with the highest $V_{max}$. These galaxies show a better agreement with the SDSS $K$-band TFR rather than the HI one.
The two objects with no consequent rotation observed are the furthest from the local TFR (both $K$- and $B$-band). For the one classified as PR, this can be accounted for by the inclination. As for local galaxies, these galaxies have to be excluded from the TFR analysis because of the uncertainty on the inclination.
The other objects classified as PR or MS stand on a restricted area on the TFR and present some correlation (if we exclude the slowest rotator) parallel to the local TFR, both in $B$- and $K$-bands. The three kinematical types seem to behave as the three kinematical types of \citet{Puech:2008}.

Given the strong evolution in luminosity with redshift and the fact that the $B$-band rest-frame magnitude is sensitive to short bursts of star formation, we also consider in Figure \ref{tfkbm} (bottom) the stellar mass TFR which has the advantage of probing more closely the actual stellar mass build-up of galaxies. As for the absolute magnitude TFR, the two rotating disks follow the stellar mass TFR traced by \citet{Cresci:2009} between $z\sim 1.5$ and $z\sim 2.5$.

A larger sample will be needed in order to provide more advanced conclusions on TFR evolution.

%\subsection{Spectrophotometric properties}

%\subsection{Mass comparison}

%\begin{figure}
%%\vspace{-2cm}
%\begin{center}
%%\resizebox{1.0\columnwidth}{!}{\includegraphics{eps/mdyn_vs_mstel.eps}}
%\resizebox{1.0\columnwidth}{!}{\includegraphics{eps/masses.eps}}
%\end{center}
%%\vspace{2cm}
%\label{mdyn_mstel}
%\caption{Comparison between enclosed mass (y-axis) and stellar mass (x-axis) in log units. Black dots, red squares, green triangles correspond respectively to objects classified (see section \ref{classif}) as rotating disks (RD), perturbed rotators (PR) and merging systems (MS). The maximum velocity has been determined from the rotating disk fit modeling (filled symboled) and directly from the \vf~(open symbols).}
%\end{figure
\section{Discussion and conclusions}

%Need to discuss the presence of AGN. Wright et al. 2008 (arXiv: 0810.5599) noted that 2 of their discs have a weak AGN at their center. They use the spatially resolved NII/Ha ratio. We ABSOLUTELY NEED to do this map and test for the presence of AGN for the galaxies with NII (VVDS-020116027, VVDS-220014252, VVDS-220015726, VVDS-220584167, VVDS-220596913)

\begin{figure}
\begin{center}
\resizebox{1.0\columnwidth}{!}{\includegraphics{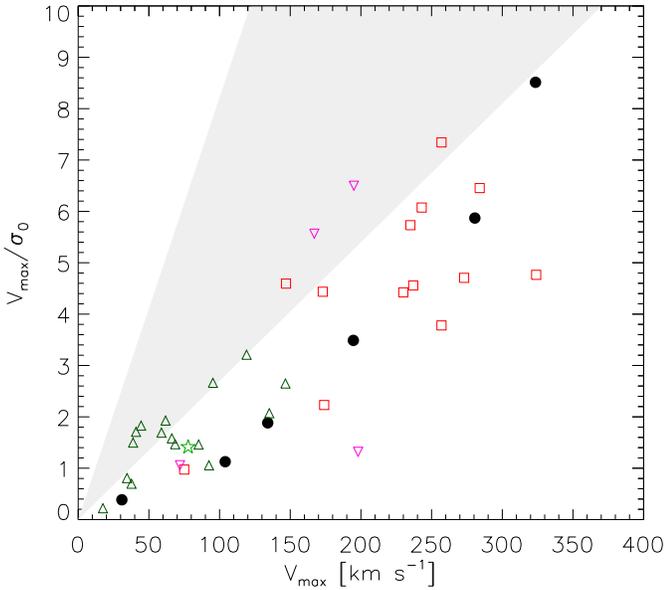}}
\end{center}
\caption{Ratio of the maximum rotation velocity $V_{max}$ (from modeling) with the mean local \vd\ corrected for beam smearing effects $\sigma_0$ as a function of the maximum rotation velocity. Black dots are our rotating disks (RD) and perturbed rotators (PR). The grey cone corresponds to the area covered by local rotating disk galaxies (GHASP sample, \citealp{Epinat:2009}). Galaxies from the literature are also plotted: dark green triangles are from \citet{Law:2007,Law:2009}, light green star is from \citet{Stark:2008}, pink up side down triangles are rotating disks from \citet{Wright:2007,Wright:2009} and red squares are rotating disks from the SINS sample \citet{Genzel:2008,Cresci:2009}.
The maximum rotation velocities and the mean local \vd\ for the galaxies from \citet{Law:2007,Law:2009} have been corrected for beam smearing using the recipes presented in \citet{Epinat:2009}, and a mean statistic inclination of 45\degr\ has been used to deproject the velocities in the galaxy plane.}
\label{v_vsig}
\end{figure}

Using high spatial resolution near infrared integral
field spectroscopy, we have produced a detailed analysis
of the kinematical maps of nine galaxies with $1.2 \leq z \leq 1.6$
selected from the VVDS. We have identified three major mergers
systems, two massive rotating disks with total
dynamical mass $>10^{11}~\msun$,
and four galaxies with an indication of rotation but with perturbed
velocity maps, two of them with high gaseous
velocity dispersion unseen in local isolated disks (\citealp{Epinat:2009}).
Our sample, although still small but comparable in
size to other samples in the literature, has the
advantage to be selected from an unbiased sample drawn
from the  $I$-band magnitude selected VVDS
and with [OII]3727 $EW>50$\AA.

The effects of beam smearing on the data (mean seeing around 0.65\arcsec\ corresponding to around 5 kpc at these redshifts) have been taken into account in the interpretation of the kinematical maps in particular to correct the velocity dispersion maps from unresolved velocity gradients, using model fitting methods as done by \citet{Forster-Schreiber:2006} or \citet{Wright:2007}.
%[Discuss beam smearing effects: how can they affect our data and interpretation ? Law et al. make a lot of comments on this, what can we say ?]

Finding three major merging systems, representing 30\% of
our sample, is a high occurence. We can compare this pair fraction to the measurements at lower
redshifts: \citet{de-Ravel:2009} find that at
$z\sim1$ about 10\% of galaxies with a
stellar mass $M_*>10^{10}~\msun$ are in pairs
likely to lead to a major merger, making the
merger rate in our higher redshift sample significantly
higher.
% [Cite Kartaltepe]
When taking into account typical merger timescales
of $\sim0.5-1$~Gyr, the high rate in our sample means that
over a $\sim3$~Gyr period virtually
all galaxies with stellar mass $>10^{10}~\msun$
will have undergone at least one major merger event.
Although the observed sample is small, this
result is comparable to the fraction of
mergers in other samples at similar \citep{Wright:2009} or higher
redshifts \citep{Shapiro:2008}.
%: \citet{Wright:2009} find that 50\% of their sample of six galaxies at similar redshifts can be classified as mergers.
In two of our merger systems, we can identify that
one of the galaxies is compatible with
a rotating disk, indicating that disk galaxies
are indeed involved in merging events at these
high redshifts.

The two massive rotating disks, with velocity dispersion
maps indicating stable disks,
confirm that massive rotating disks
with total dynamical mass $M>10^{11}~\msun$ are already
well assembled at $z\sim1.5$. The stability is underlined by the agreement of these galaxies with local Tully-Fisher relations obtained from local samples. These disks obviously need
to have been assembled at earlier cosmic times, therefore
requiring an early process of formation. These objects
are comparable to some of the massive disks observed at similar
redshifts by \citet{Wright:2009}, or at higher
redshifts by \citet{Forster-Schreiber:2006},
\citet{Genzel:2008} and \citet{Law:2009}.
However, we note that these disks are
only about 20\% of our sample.

Perturbed rotators are the dominant population in our
sample with four galaxies over nine. These objects show perturbed
velocity maps, two of them with
high velocity dispersion, with a $V/\sigma_0$ lower than for typical rotating disks (Figure \ref{v_vsig}), indicating that the gravitational support is not mainly due to rotation.
%[lower than observed in local galaxies (TBC)].
These objects
are clearly a separate class not observed at lower redshifts.
Their dynamical state is comparable to some of the galaxies observed by
\citet{Forster-Schreiber:2006}, \citet{Wright:2009},
or \citet{Law:2009}. At a resolution of a few kiloparsecs,
we argue that it is difficult to disentangle whether
the high velocity dispersion is produced by instabilities resulting from
gas accretion e.g. along the filaments of the cosmic web,
representing early-stage disks with a high cold gas fraction
which fragmented under self-gravity and collapsed to form
a starburst \citep{Immeli:2004b}, or by gas rich minor merger events
(e.g. \citealp{Semelin:2002}).
Indeed, both scenarios of continuous accretion of cold gas coming from cosmological filaments and frequent mergers of minor cold gas satellites may fuel the disk in fresh gas. If minor mergers (ranging from 10:1 to 50:1) are relatively significant, the relaxation processes will expulse the pre-existing stars from the disk to a spheroid structure or to a very thick disk not directly observable for high redshift galaxies. On the other hand, the formation of these structures would not perturb dramatically the kinematics of the disk, and thus would not be easily observable. Indirect signatures of the existence of a spheroid could be the stabilization of the gaseous disk, reducing the star formation out of merger phases and thus reducing the amount of sub-structures like \Ha\ or UV clumps in the disk. On the contrary, a very smooth accretion of diffuse gas coming from an extended halo or from the cosmological filaments will not be so efficient to form a stellar spheroid. Thus instabilities would be more numerous in the disks and possibly observable through deep imaging addressing the formation of clumps. For very minors mergers (e.g. mass ratio $>$ 100:1), the dwarf galaxies will be dislocated by the tidal field once they experience the gravitational field of the main galaxy. Thus, torn by tidal field, the accretion of very small companions will resemble very much to diffuse gas accretion. For a baryonic mass galaxy of $10^{10}~\msun$, these satellites correspond to mass lower than $10^8~\msun$. If they exist, due to the lack of spatial resolution, these galaxies are not observed, neither in observations nor in numerical simulations.

For two of the perturbed rotators
observed in our sample (VVDS020147106 and VVDS220014252),
the velocity maps are asymmetric and consistent
with a small companion in the process of merging. More generally, these perturbed rotators indicate that
disk build-up is fully on-going at the peak of cosmic star
formation history.

Star formation rates in our sample are high, with
an average of $\sim55$~\msunyr,
meaning that over the $\sim$1~Gyr between $z=1.6$ and $1.2$ more
than $5\times10^{10}~\msun$ would be formed if this
rate is sustained, comparable to the average dynamical mass.
This seems to indicate that any external accretion
should last only a limited time.

We have not found any evidence for type 2 AGN in our sample,
(type 1 were already excluded from our parent
galaxy sample).
The average \ntwo/H$\alpha$ and \stwo/\Ha\ ratios are typical of star-forming
galaxies \citep{Stasinska:2006}, and the \ntwo/H$\alpha$ 2D maps do not
show a central excess of \ntwo\ which could indicate an AGN.
Although we cannot exclude low luminosity or highly
obscured AGN, this seems to indicate that the AGN
and star formation phenomena may be only weakly connected
in the main star-forming galaxy population.
%[NOTE: SPECIFIC SECTION ON THIS SEARCH FOR AGN-type2: Say also we have excluded AGN-type 1 from our input sample].

The question of the assembly of galaxies via major dissipative
mergers or internal secular processes has been recently
highly debated in the literature. \citet{Genzel:2008} advocated
a secular process of assembly to form bulges and disks in
massive galaxies at $z\sim2$. \citet{Law:2009} concluded that the
high velocity dispersions they observe in most galaxies
at $z\sim2$ may be neither a `merger' nor a `disk', but rather
the result of instabilities related to cold gas accretion
becoming dynamically dominant. Our data seem rather to indicate
that several processes are acting at these epochs.
Among them, merging seems to play a key role. On the one hand, we
find a high 30\% rate of close pairs of galaxies expected to
merge in less than 1~Gyr, indicating that the hierarchical build up of
galaxies at the peak of star formation is fully in progress.
On the other hand, in our sample the dominant `perturbed rotators'
may include a significant fraction of galaxies with
minor mergers in progress or cold gas accretion along streams of the cosmic web, producing
a high velocity dispersion. Minor merging
may be an important phenomenon in the build-up of
galaxies, which will require higher spatial resolution
than currently available on 8-10m telescopes, even with adaptive
optics, to be confirmed.
In large numerical simulations, major and minor mergers
are indeed playing an important role \citep{Bournaud:2005}, in parallel to continuous gas accretion \citep{Semelin:2005,Dekel:2009}.
%a major role (refs) [more on this later].
Whether the two massive disks that we observe in our sample
are the result of secular processes with continuous accretion \citep{Genzel:2008} or
of anterior hierarchical build up by major and/or minor mergers
cannot be assessed as both processes take a relatively short time
(less than a 1~Gyr) to complete. These arguments will need to
be revisited from larger representative samples.

In conclusion, the statistics of this representative 
%our complete
sample, based on nine galaxies, show that there does not seem to be one single process driving the mass assembly in galaxies. Major mergers certainly play an
important role, while the contribution of minor mergers is likely
but will remain difficult to confirm. In addition, secular evolution
with accretion which drives gas and stars in the central regions of
galaxies remains a possibility to assemble bulges and disks early
in the life of the universe. Furthermore, the absence of AGN type 2 in
our sample indicates that the AGN phenomenon in high
redshift galaxies is at best a short lived event.
With the small samples observed to date, we cannot exclude
that in the life of galaxies these processes will act at
different epochs.

Whether the differences between our sample and other existing
samples are mainly due to different selection functions will need to
be investigated with larger well controlled samples.
With the on-going MASSIV program at the VLT, we
will be able to investigate these issues in more details. 

%%\input{maps_work.tex}

%\section*{Acknowledgments}
\begin{acknowledgements}
%We are very grateful to the VLT Observatory for accepting this programme for Science Verification Time.
%%We also thank for some helpful suggestions, ?? for a critical reading of the original version of the paper. 
%This work was supported by the Marie Curie Research Training Network {\it Euro3D; contract No. HPRN-CT-2002-00305}.
We wish to thank the ESO staff at Paranal Observatory and especially the SINFONI team at VLT for their support during observation. 
We thank the referee for the constructive comments which help to improve the quality of this paper. We thank David R. Law for kindly providing us before publication their velocity and velocity dispersion maps \citep{Law:2009}.
This work has been partially supported by the CNRS-INSU Programme National de Galaxies and  by the french ANR grant ANR-07-JCJC-0009.
Based on observations obtained with MegaPrime/MegaCam, a joint project of CFHT and CEA/DAPNIA, at the Canada-France-Hawaii Telescope (CFHT) which is operated by the National Research Council (NRC) of Canada, the Institut National des Science de l'Univers of the Centre National de la Recherche Scientifique (CNRS) of France, and the University of Hawaii. This work is based in part on data products produced at TERAPIX and the Canadian Astronomy Data Centre as part of the Canada-France-Hawaii Telescope Legacy Survey, a collaborative project of NRC and CNRS.
\end{acknowledgements}

\bibliographystyle{aa}
\bibliography{biblio}

\begin{appendix}
\section{Properties of individual galaxies}
\label{individual}

\begin{figure*}
\begin{center}
\resizebox{2.0\columnwidth}{!}{\includegraphics{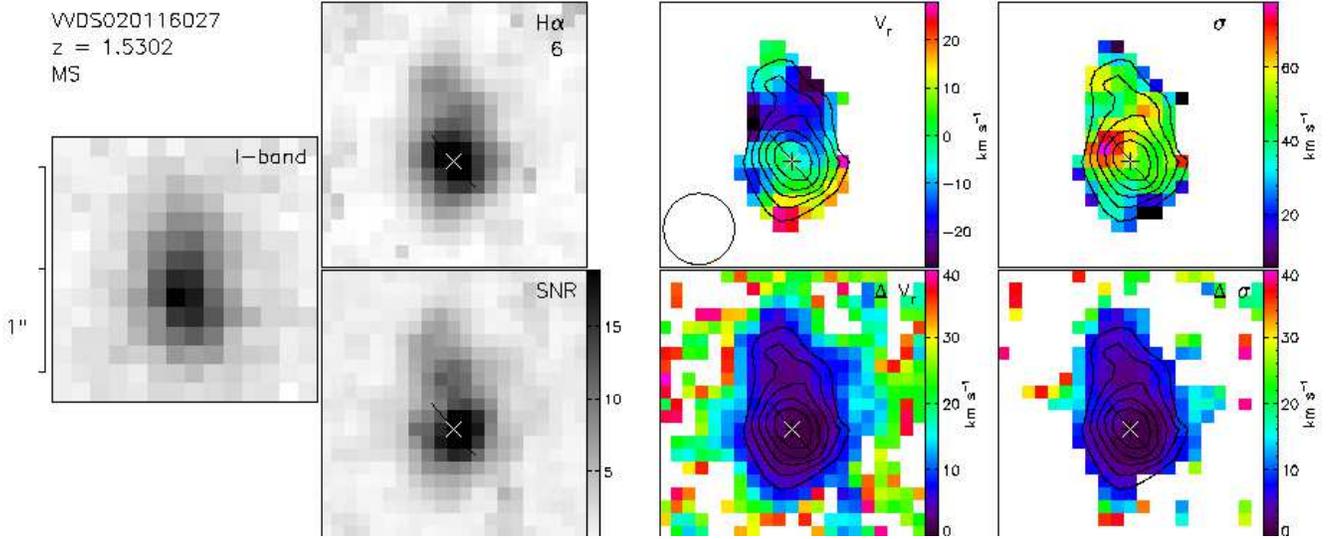}}
\end{center}
\caption{VVDS020116027 maps. From left to right: $I$-band CFHT legacy survey best seeing image, H$\alpha$ flux (top) and signal-to-noise (bottom) maps, H$\alpha$ velocity field (top) and associated errors (bottom), H$\alpha$ velocity dispersion map and associated errors (bottom) obtained from Gaussian fits to the SINFONI data cubes after smoothing spatially with a two-dimensional Gaussian of FWHM $=$ 2 pixels. Except the instrumental spectral PSF, no correction has been applied to compute the \vdm. Kinematical maps have been masked using the following criteria: (i) the line width must be larger than the one of the spectral PSF (the majority white pixels in the velocity dispersion error map), (ii) the uncertainty on the velocity must be less than 30\kms\ and (iii) the signal-to-noise ratio larger than $\sim2$. The $I$-band image and H$\alpha$ maps are color-coded with a linear scaling such that the values increase from light to dark. The integrated \Ha\ flux is quoted in each \Ha\ flux map in $10^{-17}$\ergscm. In error maps, values larger than the scale maximum are displayed using white.
% These data have been acquired with the 125$ \times $250mas sampling configuration of SINFONI. 
The angular size is indicated and the final seeing FWHM (including observational seeing and gaussian smoothing) is plotted as a circle on the \vf.
%of 1\arcsec (corresponding to $\sim 8$ kpc at the redshifts of the objects). 
North is up, East is left. The center used for kinematical analysis is marked on each kinematical map with a black and white double cross. The solid line represents the orientation of the kinematical major axis and ends at half light radius measured on I-band images. The flux contours are overplotted on the \vf, the \vdm\ and their associated error maps. The redshift derived from SINFONI data and the dynamical classification are quoted (RD for rotating disks, PR for perturbed rotators, MS for merging systems).}
%XXX CONTOURS, MAJOR AXIS.}
%, the direction of North-East, the VVDS identification and VIMOS-based redshift are indicated for each galaxy.}.
\label{map1}
\end{figure*}

\begin{figure*}
\begin{center}
\resizebox{2.0\columnwidth}{!}{\includegraphics{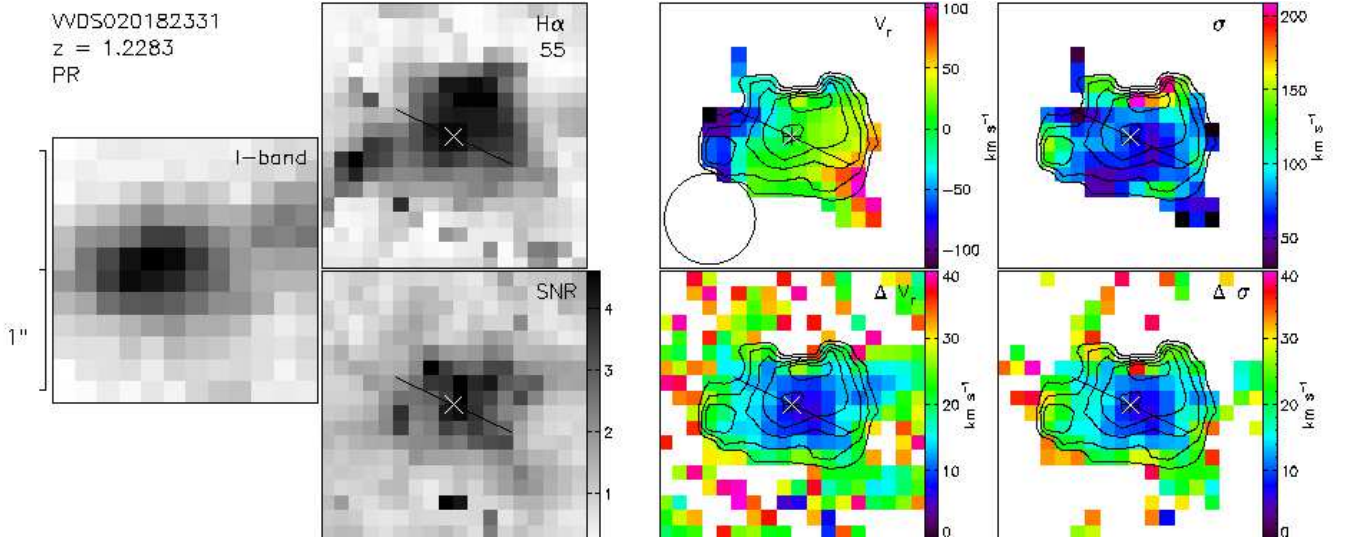}}
\end{center}
\caption{Similar as Figure \ref{map1} for VVDS020182331. The signal-to-noise threshold for this galaxy has been set to 1.5.}
\label{map2}
\end{figure*}

\begin{figure*}
\begin{center}
\resizebox{2.0\columnwidth}{!}{\includegraphics{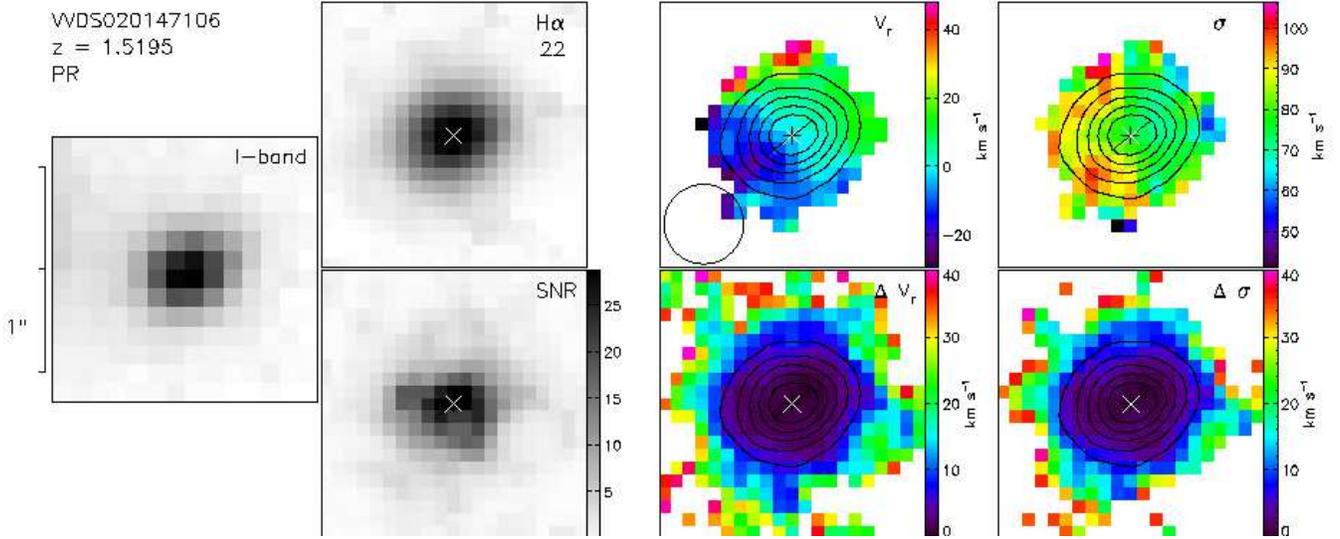}}
\end{center}
\caption{Similar as Figure \ref{map1} for VVDS020147106.}
\label{map3}
\end{figure*}

\begin{figure*}
\begin{center}
\resizebox{2.0\columnwidth}{!}{\includegraphics{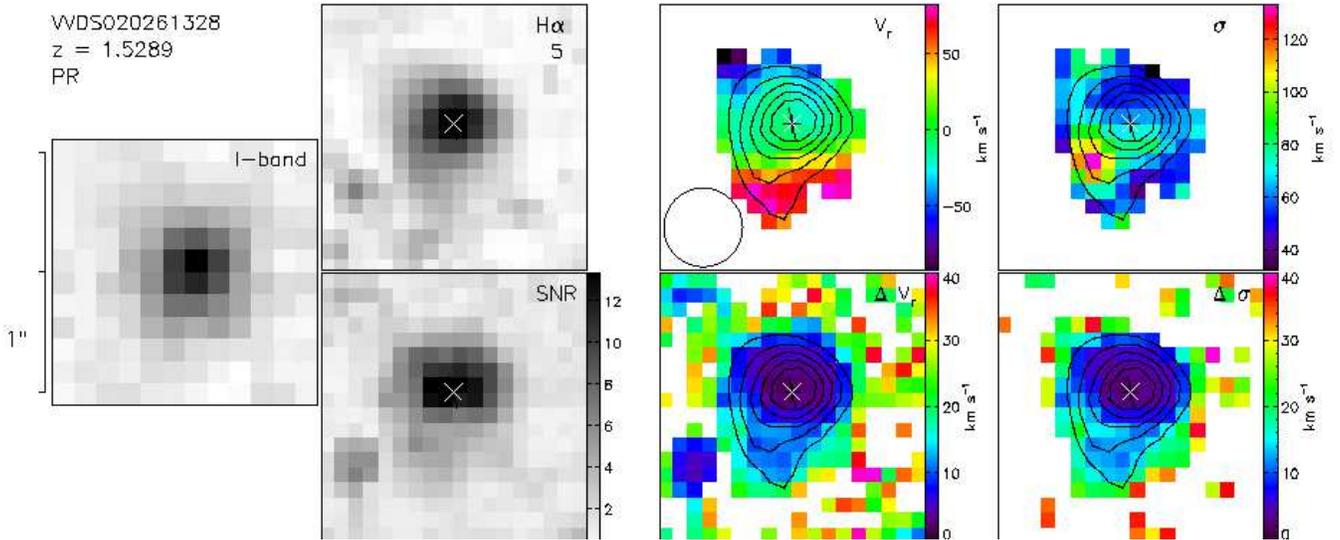}}
\end{center}
\caption{Similar as Figure \ref{map1} for VVDS020261328. The East feature in the signal-to-noise map correspond to a badly subtracted cosmic ray.}
\label{map4}
\end{figure*}

\begin{figure*}
\begin{center}
\resizebox{2.0\columnwidth}{!}{\includegraphics{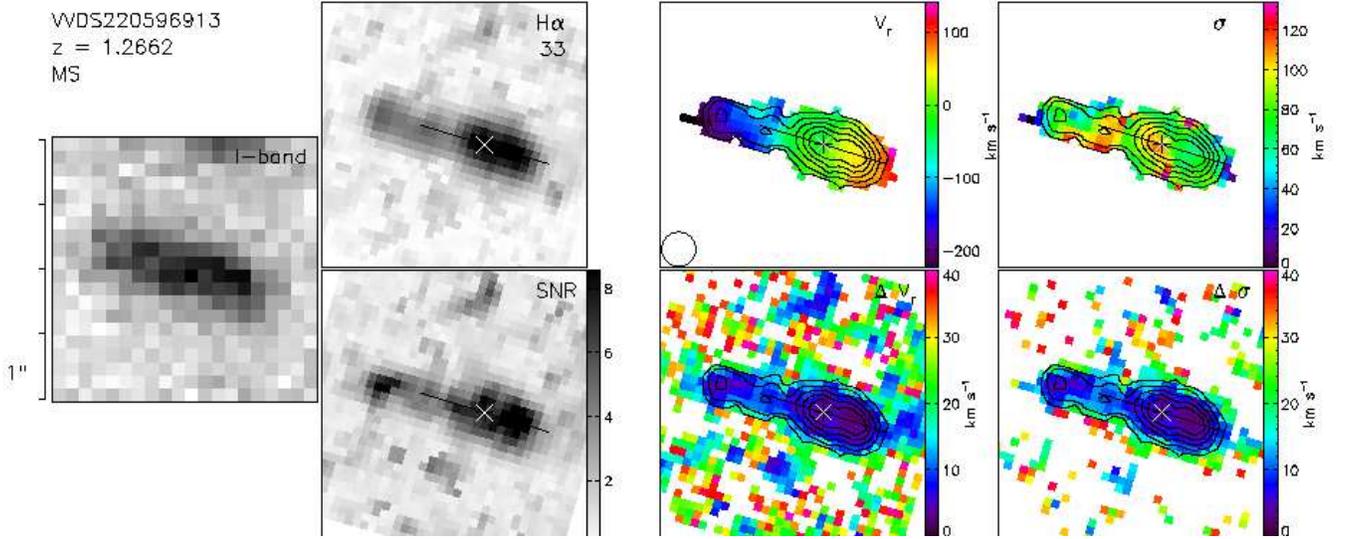}}
\end{center}
\caption{Similar as Figure \ref{map1} for VVDS220596913. The North feature is compatible with sky line residuals. $I$-band map is a CFH12K/CFHT image.}
\label{map5}
\end{figure*}

\begin{figure*}
\begin{center}
\resizebox{2.0\columnwidth}{!}{\includegraphics{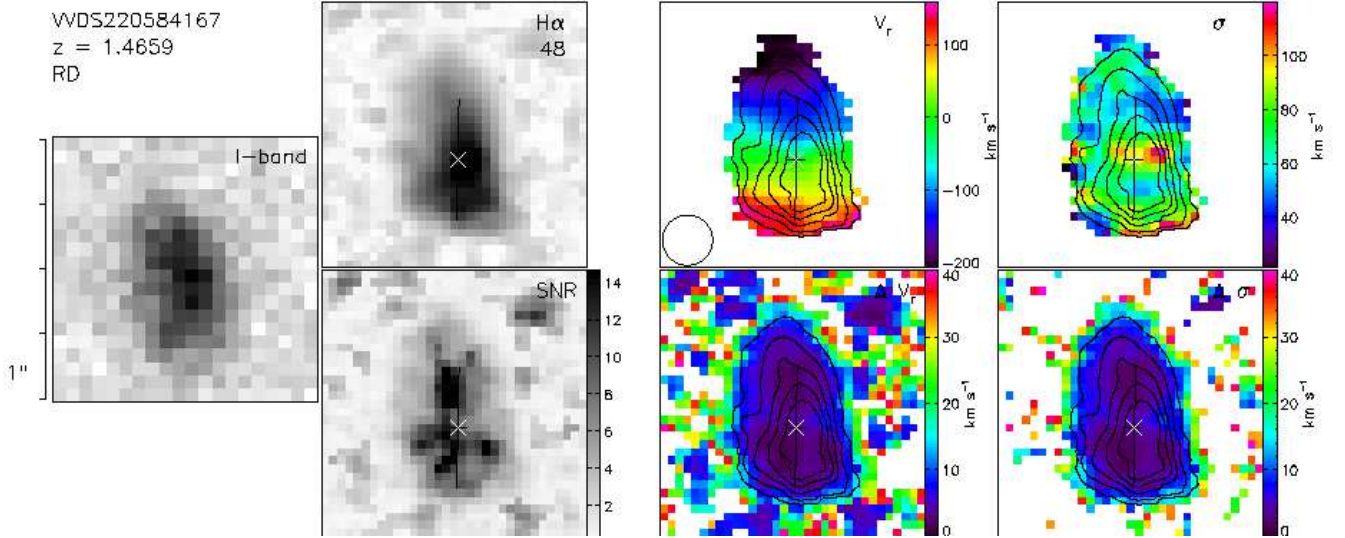}}
\end{center}
\caption{Similar as Figure \ref{map5} for VVDS220584167. The North-West feature is compatible with sky line residuals.}
\label{map6}
\end{figure*}

\begin{figure*}
\begin{center}
\resizebox{2.0\columnwidth}{!}{\includegraphics{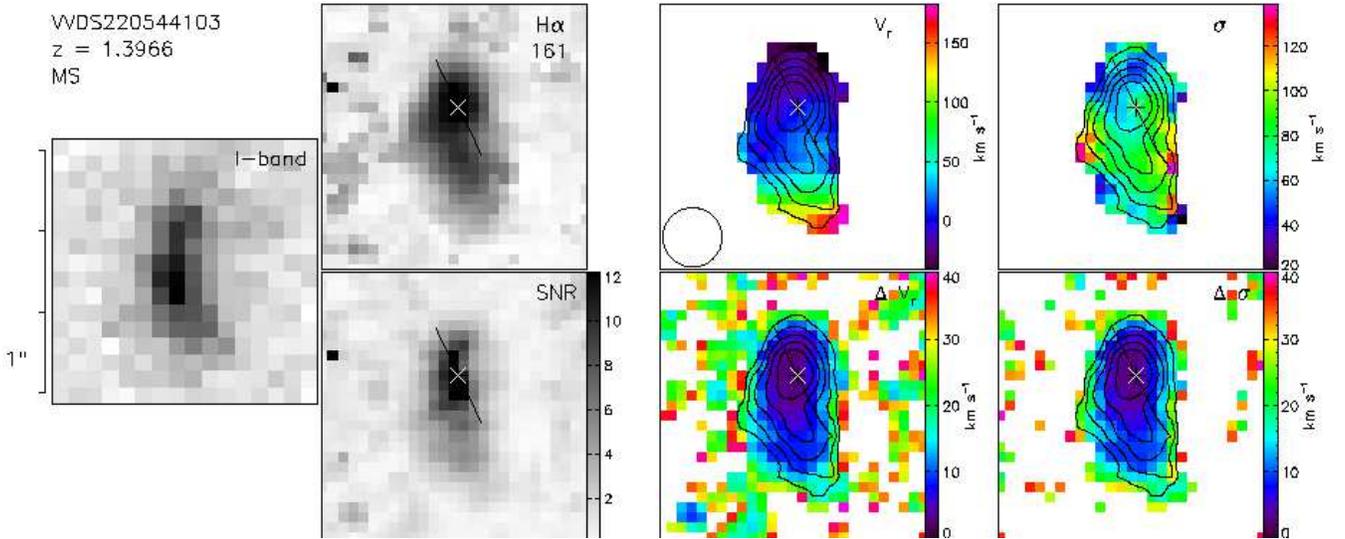}}
\end{center}
\caption{Similar as Figure \ref{map5} for VVDS220544103.}
\label{map7}
\end{figure*}

\begin{figure*}
\begin{center}
\resizebox{2.0\columnwidth}{!}{\includegraphics{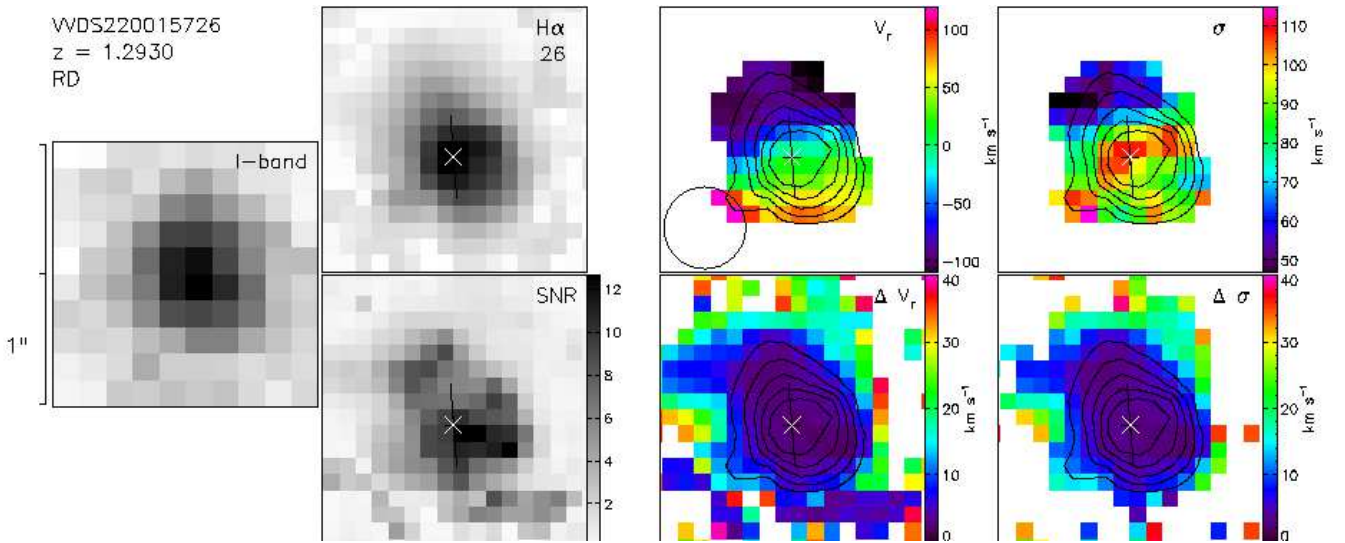}}
\end{center}
\caption{Similar as Figure \ref{map5} for VVDS220015726.}
\label{map8}
\end{figure*}

\begin{figure*}
\begin{center}
\resizebox{2.0\columnwidth}{!}{\includegraphics{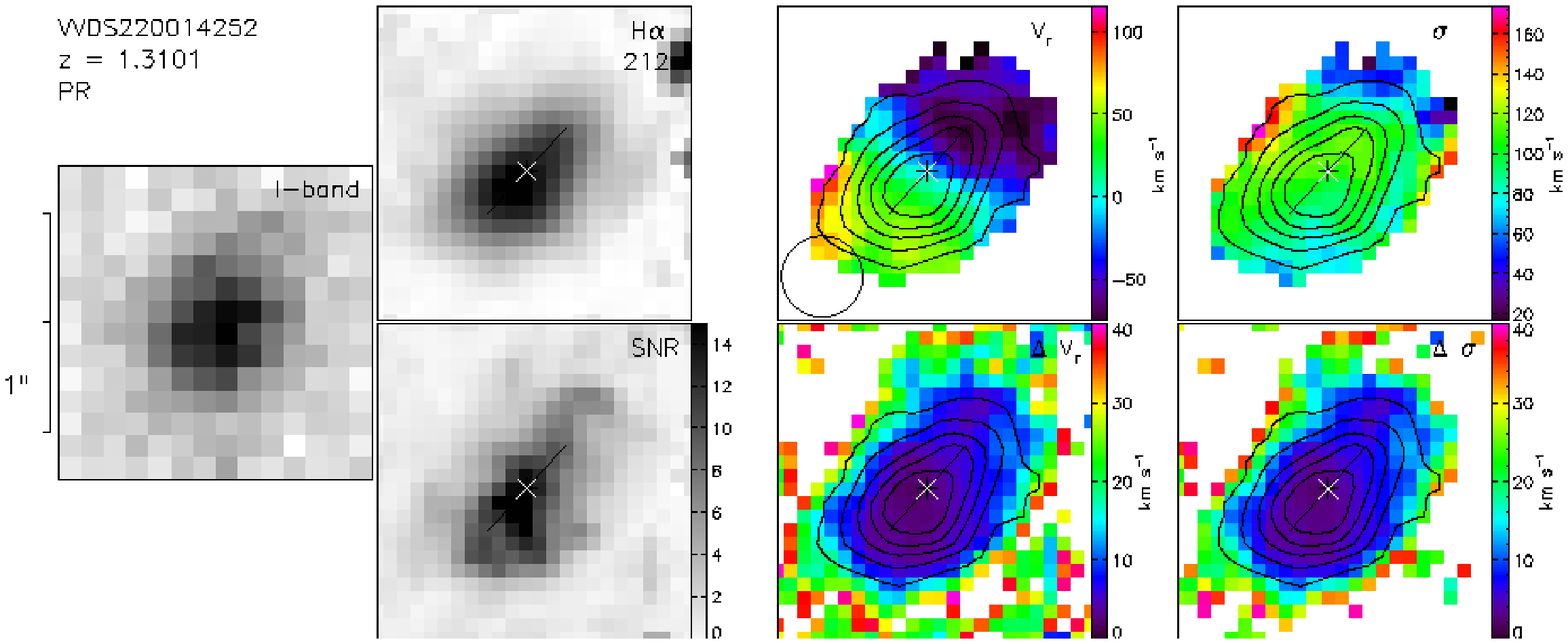}}
\end{center}
\caption{Similar as Figure \ref{map5} for VVDS220014252.}
\label{map9}
\end{figure*}

In this appendix, we present the morphological, kinematical and dynamical properties for each galaxy derived from 
%the results derived from the velocity maps for each galaxy. 
$I$-band images, \Ha\ flux maps, radial velocity maps, velocity dispersion maps and their associated error maps (Figures \ref{map1} to \ref{map9}). On each velocity map, the kinematical center used for the kinematical analysis is plotted with a black and white double cross.
The kinematical position angle is also displayed and ends at half light radius $r_{1/2}$ (Table \ref{table_mass}). The flux contours are overplotted on both \vfs\ and \vdms. In all figures, North is up and East is to the left. The kinematical parameters and the masses are stored in Tables \ref{table_kin} and \ref{table_mass}.
% \textbf{Error maps pixels with values larger than the scale maximum are displayed using white.}\\
On error maps, pixels with values out of the scale range are displayed using white.\\

%\subsection{Kinematical characterization of the 10 objects in the  $1.3 \leq  z  \leq 1.6$ sample}
%We used a kinematical classification inspired by the one defined by \citet{Flores:2006} for which a proper rotating disk is characterised by a $\sigma$-map showing a single peak near the center of the galaxy where the gradient of the rotation is the steepest; a perturbed rotation has a $\sigma$-map with a peak shifted off the center or does not show any peak; and complex kinematics if it is not a rotation. We had to this the information deduced from the analysis of the $\sigma$-map and of its residual after fitting the simple rotational model to the velocity field. \\
%

{\bf VVDS020116027}

This galaxy shows a faint and irregular velocity gradient. The \Ha\ flux map, the \vf\ and the \vdm\ suggest
%that this object is composed by
two components separated of about 6~kpc (0.75\arcsec). Indeed, the velocities decrease from South to North accross the main component but decrease of around 15\kms\ when reaching the faint component. The separation between the two components is characterized by a higher \vd\ that could indicate a merging event \citep{Amram:2007} or an unresolved gap in velocity between the two components.
%main component on the South shows a velocity gradient from positive to negative values compatible with a slowly rotating disk, or a nearly face on rotating disk. The northern component has less flux and positive velocities that could indicate a close small companion.
The velocity separation of the centers of these two components is close to 0\kms. The merging timescale would be around 0.4~Gyr (from \citealp{Kitzbichler:2008}). The faint component is not seen in the $I$-band CFHTLS image, but the system appears as distorted and elongated, leading to an inclination of the whole system of 50\degr.
The very low maximum rotation velocity (50\kms) may indicate that the object is almost not rotating ($V_{max}/\sigma_0=1.1$) or that the main component is seen nearly face-on.
% However, the low velocity dispersion (around 45\kms) of this object is surprising if we consider this object as a non rotating object.
%The overall \vf\ is compatible with a non rotating object or a nearly face-on object. 
%
%The maximum \vd~is around 70\kms~and peaks at 0.3\arcsec~at the North-East of the maximum of the \Ha~flux.\\
Fitting a rotating disk model for this system has been attempted both with or without masking the northern component. In both cases, a low $\chi^2$ is obtained but is due to the low velocities measured in this galaxy. The $\sigma$-peak around 70\kms\ located at 0.3\arcsec\ at the North-East of the maximum of the \Ha~flux is not accounted for by these models.
Due to both the low \vd\ and maximum rotation velocity, this object is one of the less massive of our sample ($5.5\times10^{10}\msun$).
%despite the fact that it does not have the smaller half light radius $r_{1/2}=4.0$~kpc.
%The very low maximum rotation velocity (50\kms) may indicate that the object is almost not rotating ($V_{max}/\sigma_0=1.1$) or that it is seen nearly face-on. However, the low velocity dispersion (around 45\kms) of this object is surprising if we consider this object as a non rotating object. From the $I$-band image, the inclination is estimated to 50\degr\ since the source is elongated. 
%Considering this object as a merging system is compatible with a lower inclination at least for one component.
This object is classified as a major merger, presumably in a premerging state since the SFR$_{H\alpha}$ is rather faint, indicating that the starburst event would not be yet engaged.\\
%We thus classified this object as merging system. The SFR of this galaxy is one of the lowest of our sample.\\
%
%XXX Isn't that surprising for a merger? As well as the low \vd?
%XXX Compute individual masses from Ha map morphology + kinematics
%
%XXX Use Ha map to compute the relative flux/mass?\\
%The enclosed mass would be at most $XXX\msun$.
%
%closest strong sky line :16610\AA, flux 1.326e+02 (16612.\AA, flux 1.326e+02)

{\bf VVDS020182331}

The flux of this galaxy is mostly concentrated in the central parts. This galaxy shows a perturbed \vf. A strong sky line residual (14605\AA) perturbs the line measurements and is responsible for both northern \vd\ and line flux peaks observed towards the North.
%It may also be responsible for the eastern flux and \vd~peaks and may bias velocity measurements.
The uncertainty on the velocity measurements is lower than 20\kms\ for the central arcsecond only (in diameter). In this central part, a rotating disk model is acceptable. The corrected \vd\ of this object is high (71\kms). We do not detect in \Ha\ the close-by object revealed in CFHTLS best seeing images, indicating that this object is a foreground or background galaxy or that it is a companion with no star formation with which VVDS020182331 may interact. The latter explanation could explain the perturbed rotation observed in VVDS020182331. This object is slightly more rotation supported than dispersion supported ($V_{max}/\sigma_0 = 1.9$) and has a dynamical mass of $4.1\times10^{10}\msun$. It is classified as a perturbed rotator.\\
%
%closest strong sky line :14605\AA, flux 2.166e+03
%
%Le champ de vitesses de cette galaxie n'est pas tr�s bien reproduit par le mod�le de disque en rotation. Toutefois, les cartes pr�sent�es pour cette galaxie sont entach�es d'une grande incertitude, en particulier sur la r�gion Nord Ouest, r�gion o� le champ de vitesses est maximum. En enlevant cette r�gion, le champ de vitesses devient comparable � celui obtenu pour VVDS020261328 qui est bien reproduit par le mod�le de disque en rotation. Un pic de dispersion de vitesses qui ne peut pas �tre expliqu� par le gradient de vitesse est observ� au Sud sur le petit axe. Le plateau semble ne pas �tre atteint sur la courbe de rotation.

{\bf VVDS020147106}

The \Ha\ flux is concentrated in the center. The size of the galaxy is twice the seeing (see Table \ref{runs}). This galaxy shows a faint velocity shear compatible with a face-on rotating disk. The \vdm\ does not show any peak but is higher on the eastern side (of about 20\kms). The northern edge shows higher velocities of about 30\kms. However the one pixel crown of the edges has higher uncertainties due to a low signal to noise ratio (lower than 4) and was masked to fit a rotating disk model. The fit indicates that a low velocity plateau (30\kms) is reached close to the center. This low maximum rotation velocity suggests that the inclination of this system is even lower than suggested from the $I$-band image or that this object is strongly dispersion dominated ($V_{max}/\sigma_0 = 0.4$). Due to compactness of this object ($r_{1/2}=1.2$ kpc) and compared to the total extent of \Ha\ emission ($R_{last}=6.4$ kpc), the dispersion mass is estimated to $27.4\times10^{10}\msun$, which is 15 times larger than the stellar mass ($1.7\times10^{10}\msun$). This suggests that this star-forming galaxy is embedded in a large dark matter halo or that it contains a large amount of gas. Even if it can not be excluded that this galaxy is a dispersion dominated spheroid
% characterized by a very low velocity gradient but with a large \vd
, this galaxy is classified as a perturbed rotating disk since it has a high \vd\ (80\kms) and shows rotation in its perturbed \vf. The asymmetries observed in the velocity maps could be accounted for by a minor merging event.\\
%The rotation mass is thus unreliable, but 
%
% given the uncertainties on the velocity measurements and the low velocity gradient, 
%
%closest strong sky line :16553.906\AA, flux 6.473e+02
%
%{\bf VVDS020461235}
%
%This object presents a complex kinematics. it is impossible to find a clear center for this clumpy galaxy. Its $\sigma$-map shows several peaks which seem to match with the peaks in the H$\alpha$ intensity map. The peak with the maximum H$\alpha$ flux shows a very high dispersion of $\sim 250 km/s$. Our simple model disk didn't succeed in reproducing its complex velocity field. This object appears to be more close to a system of mergers than to a perturbed rotation.
%BE: ce qui est √©tonnant c'est aussi que la dispersion de vitesse en dehors des pics est relativement faible.
%

{\bf VVDS020261328}

The \Ha\ flux map is peaked and faintly extends toward the South. The \vdm\ has a peak around 120\kms\ in the South-East. The \vf\ shows a clear but irregular velocity shear of about 100\kms\ that extends over 0.8\arcsec\ in radius. Fitting a rotating disk model suggests that only the central part of the galaxy is detected since the plateau is not reached. It also shows that the \vd\ peak cannot be explained by the beam smearing effects. For the same reasons as VVDS020147106, VVDS020261328 has a dynamical mass ($19.6\times10^{10}\msun$) 30 times larger than the stellar mass ($0.6\times10^{10}\msun$) suggesting
% that the gas amount is relatively low and 
that dark matter and cool gas dominate the gravitational potential of this galaxy. The SFR is the lowest of our sample (see Table \ref{tablesfr}).
This galaxy is classified as a pertubed rotator mainly rotation supported ($V_{max}/\sigma_0 = 3.5$).\\
%However, this peak could be not real since it is located in a low flux region of the galaxy.\\
%
%closest strong sky line :16610\AA, flux 1.326e+02 (16612.\AA, flux 1.326e+02)
%
%It is a rotating disk with a peak in the H$\alpha$-map located at the center of the object.
%%The velocity gradient seems to be rotated ($\pm 45 km/s$).
%The velocity gradient is about 100\kms~and only extends over 1.2\arcsec.
%Unfortunately, only the central part of the galaxy is visible and therefore the plateau has not been reached in this observation. The rotating disk model fits relatively well the observed inner gradient. $\sigma$-map peaks at a position slightly shifted from the H$\alpha$ peak, still visible in the residual map. The South-East $\sigma$ peak could be not real since the observation contains a sky line residual.\\
%L'inclinaison morphologique estim�e � $0$\degr~n'est pas r�aliste puisqu'un gradient de vitesse est observ� et  est plut�t bien d�crit par un mod�le de disque en rotation. Toutefois, ce gradient est faible, ce tablesfrqui indique que l'inclinaison est probablement tr�s faible. Une inclinaison de $10$\degr~a �t� utilis�e. Par ailleurs, la diff�rence entre les d�terminations cin�matique et morphologique de la position du grand axe n'est pas �tonnante pour un syst�me de faible inclinaison. La carte de dispersion de vitesses pr�sente un pic d�cal� par rapport au centre qui n'est pas correctement d�crit par le mod�le de disque en rotation. Ce pic n'est pas corr�l� avec un pic dans la carte de flux \Ha.

{\bf VVDS220596913}

The \Ha\ flux map shows two main components separated by 12.5~kpc (1.5\arcsec) that are not clearly distinguished in the $I$-band image. The brigthest component is located in the western side and is composed by two peaks of equal intensity. The eastern component is much fainter. It has a smooth velocity gradient and two peaks in the \vdm\ (around 100\kms). One corresponds to the brightest \Ha\ knot, and the other is more diffuse and corresponds to the transition region between the two components where the \Ha\ emission is the faintest (signal to noise ratio around 4). The high velocity dispersion between the two components can be a signature for merging as for Hickson compact group H31 \citep{Amram:2007}. The velocity shear of the faint component is misaligned by around 20\degr\ with respect to the one of the brigth component suggesting, as does the flux distribution, that this system is composed of two galaxies in the process of merging. The velocity separation of the two components is estimated to $\sim150$\kms. The merging timescale would be around 1.2~Gyr.
%However, the velocity field could also be well fitted by a large and unique rotating disk extending over 3\arcsec~(plateau not reached) and with an asymmetric flux distribution. The center used in that case is the brightest flux peak.\\
Although a large and unique rotating disk consituted by large clumps as seen in \citet{Bournaud:2008} cannot be excluded (leading to $V_{max}/\sigma_0=2.3$), fitting the brightest component alone gives better results in terms of $\chi^2$, suggesting that the main component is rotating with a maximum velocity of around 200\kms.
We fixed the center between the two flux peaks of the main component, and used an inclination of $55$\degr.
Assuming that both components have the same dimensions (half the whole system dimensions), we obtained a dynamical mass of around $8\pm2\times10^{10}\msun$ (dominated by rotation) for the bright component and $4\pm2\times10^{10}\msun$ (dominated by dispersion) for the faint one. This imply a total dynamical mass of $\sim12\times10^{10}\msun$ of the same order as the one computed considering a unique object ($\sim13\times10^{10}\msun$).
This object is classified as a major merger system.\\

{\bf VVDS220584167}

VVDS220584167 is the object that presents the most extended \Ha\ map of the sample. It has an elongated peak in the \vdm\ close to the center. The observation is affected by a strong sky line residual (16195\AA, see Figure \ref{profils2}) that induces large uncertainties on line measurements, mainly on the southern side.
% western apparent peak on the \vdm\ is an artefact due to a strong sky line residual (16195\AA, see Figure \ref{profils2}) as well as the southern outer region which presents a relatively high dispersion.
The \vf\ shows asymmetry and the galaxy presents a distorted morphology in the $I$-band image which makes possible the existence of a strong bar (see for comparison UGC 08937 in \citealp{Epinat:2009}). Moreover, the observed displacement between $I$-band and \Ha\ morphologies suggests a violent star formation event compatible with the high SFR of this galaxy.
The velocity field is reasonnably fit by a simple rotational disk model, and the central \vd\ peak is not accounted for by the model since it does not simulate the effect of a bar. VVDS220584167 is the most massive disk of the sample with a dynamical mass of $21.1\times10^{10}\msun$. It is also the most rotation-dominated object ($M_{\theta}/M_{\sigma}=18$ and $V_{max}/\sigma_0 = 5.9$).
%XXX This could be an artefact due to the sky line residuals that affect the southern side.
%XXX The fit should thus be unrealistic, in particular for that the maximum velocity is reached so far from the center. This would thus also explain that velocity dispersion peak cannot be explained by the unresolved velocity gradient. Depending on the interpretation of these anomalous features, this galaxy would be classified as rotating disk or perturbed rotator.
Thus this galaxy is classified as a rotating disk with a bar. \\

{\bf VVDS220544103}

The \Ha\ flux map of VVDS220544103 shows two peaks separated by around 6~kpc (0.75\arcsec). It also shows a distorted morphology both in the line flux map and in the $I$-band image. A faint velocity gradient ($\sim 30$\kms) is observed toward the main northern peak, whereas a strong velocity gradient ($\sim 120$\kms) is observed at the South of the fainter southern peak. The two gradients are in the same direction and the velocity field is continuous accross the whole object. The isovelocities contours are perpendicular to the distorted and elongated morphology. The strongest peaks in the \vd\ are located at the edges where the signal to noise ratio is low. Excluding these peaks, the \vdm\ shows an elongated peak close to the faint \Ha\ flux peak.
% (the dispersion in the residual velocity field is the highest of our sample). 
Fitting a simple rotating disk model to the whole system leads to a ratio $V_{max}/\sigma_0=2.1$. However, this object is hard to fit with such a model and the dynamical mass estimates just give an indication on its total mass.
From our data, it is not possible to separate two components and compute mass estimates for both. We estimated the total dynamical mass to $10.3\times10^{10}\msun$ using a unique rotating disk object hypothesis. However, one might guess that at least one component could be seen nearly face-on and thus have a larger maximum velocity and thus a larger total mass.
%The merging hypothesis could explain the total dynamical mass ($10.3\times10^{10}\msun$, computed from a unique rotating disk object) lower than the stellar mass ($5.1\times10^{10}\msun$, the highest of the sample). In particular one might guess that at least one component could be seen nearly face-on and thus have a larger maximum velocity.
The velocity separation between the two components is around 40\kms. The merging timescale for this system would be around 0.5~Gyr. This object could be two or more galaxies in a merging process and is classified as a merging system.\\

{\bf VVDS220015726}

%The redshift of VVDS220015726 makes that the \Ha~line lies very close to two strong sky lines . Our observation is thus polluted by strong sky residuals. These residuals may be responsible for the eastern high velocities but also for the \vd~peak located at the same position. It probably perturbs the whole velocity measurements but mostly the low signal regions (edges). The center measurements are quite secure.
The central regions give a clean signal, while the low signal regions are affected by sky residuals due to the proximity of two strong sky lines next to \Ha\ (15053\AA and 15056\AA).
The H$\alpha$ map is peaked at the center and it has also a peak at the same location in the \vdm\ ($\sim 100$\kms). The kinematical maps of this galaxy are well reproduced by the rotating disk model. In particular, this model shows that the central \vd\ is due to beam smearing effects (see Figure \ref{models}). The relatively high $\chi^2$ value of the fit is only due to the fact that this object shows the highest velocities. VVDS220015726 is clearly dominated by rotation ($V_{max}/\sigma_0=8.5$) since it is the fastest rotator of the sample ($V_{max}=323$\kms) but also the object with the lowest velocity dispersion ($\sigma_0=38$\kms). Its total dynamical mass ($16.7\times10^{10}\msun$) is compatible within the error bars to its stellar mass ($6.2\times10^{10}\msun$). This indicates that the halo contribution is low in the central parts of this galaxy. This object is classified as a rotating disk.\\
%
%closest strong sky lines :
%15052.882\AA 5.187e+02
%15055.541\AA 4.751e+03 one of the strongest of the whole spectrum (max=5.1e+03)
%
%Unfortunately, we have a high uncertainty on the inclination for this object.
%L'inclinaison morphologique estim�e � $0$\degr~n'est pas r�aliste puisqu'un gradient de vitesse est observ�. Ce gradient est plut�t bien reproduit par le mod�le de disque en rotation. La diff�rence entre les d�terminations cin�matique et morphologique de la position du grand axe s'explique par la faible inclinaison du syst�me. Une inclinaison de $10$\degr~a donc �t� utilis�e. Pourtant, le gradient de vitesse est relativement important et sugg�re que l'inclinaison est plut�t de l'ordre de $30$\degr. La vitesse maximale non d�projet�e de cette galaxie n'est donc exploitable qu'en tant que limite inf�rieure �tant donn�e l'incertitude sur l'inclinaison. Bien que le champ de vitesses soit plut�t bien d�crit par le mod�le de disque en rotation, le pic de dispersion de vitesses semble ne pas �tre d� au gradient de vitesse.

{\bf VVDS220014252}

The \Ha\ flux map of VVDS220014252 shows an elongated peak and diffuse emission in the outer parts, compatible with the $I$-band morphology. The peak does not perfectly match the center of external isophotes. The \Ha\ flux map also suggests the presence of an arm in the western side.
%The velocity gradient is aligned with the morphological major axis. 
The velocity field is the one of a rotating disk except in the eastern side where it shows unexpected high velocities.
A rotating disk model correctly fits the \vf. The fit is better (smaller $\chi^2$) when the center does not match the \Ha\ flux peak, but the center of external isophotes. Adopting an inclination of 56\degr, the deprojected maximum rotation velocity is quite low (103\kms). The model shows that the maximum rotation velocity is reached close to the kinematical center.
% so that the inner gradient is quite high ($48$\kms~$kpc^{-1}$). 
The North-West side of the \vf\ shows a bump and thus a small decrease.
Around this bump, profiles are broad and asymmetric. Broad profiles (more than 150\kms) are also observed at the North-East edge, with a rather good signal to noise ratio (larger than 5). The \vd\ is not peaked at the center of the galaxy but is high everywhere. Moreover, the mean \vd\ of this galaxy is the highest of the whole sample (92\kms). The model shows that the velocity dispersion is not due to beam smearing effects (see Figure \ref{models}).
The broad and asymmetric profiles suggests the possibility to have intrinsic double profiles that can be signatures of interactions \citep{Amram:2007}. Moreover, the kinematics is not dominated by rotation since $V_{max}/\sigma_0=1.1$ and the total dynamical mass ($10.2\times10^{10}\msun$) is dominated by the dispersion mass ($7.9\times10^{10}\msun$).
For these reasons, this galaxy showing the highest SFR of the sample is classified as a perturbed rotator possibly in the process of a minor merging event.

\end{appendix}

\end{document}